\documentclass[aps,reprint,nofootinbib,nobibnotes,notitlepage,superscriptaddress,onecolumn,prd,
 amsmath,amssymb
]{revtex4-2}

\usepackage[caption=false]{subfig}
\usepackage{braket}
\usepackage{float}
\usepackage{lipsum}
\usepackage{graphicx}% include figure files
\usepackage{dcolumn}% align table columns on decimal point
\usepackage{bm}% bold math
\usepackage{natbib}
\usepackage{hyperref}
\hypersetup{
	colorlinks = true,
    linkcolor = Red,
    urlcolor  = Red,
    citecolor = Red
}
\usepackage[skip=4pt plus1pt, indent=10pt]{parskip}

\usepackage{microtype}

\usepackage{verbatim}
\usepackage[amssymb]{SIunits}
\usepackage{tabularx}
\usepackage[dvipsnames]{xcolor}
\usepackage{wasysym}

\usepackage{tikz}
\usepackage[compat=1.1.0]{tikz-feynman}

\def\be{\begin{equation}}
\def\ee{\end{equation}}

\usepackage{color}
\definecolor{darkgreen}{RGB}{0,120,0}

\definecolor{darkgreen}{RGB}{0,120,0}

\newcommand{\delD}[1]{(2\pi)^3\delta_\mathrm{D}\left({#1}\right)}

\newcommand{\av}[1]{\left\langle{#1}\right\rangle} 

\newcommand{\vk}{\vec k}
\newcommand{\hk}{\hat{\vec k}}
\newcommand{\vK}{\vec K}

\newcommand{\vx}{\vec x}

\newcommand{\Si}{\mathsf{S}^{-1}}
\newcommand{\F}{\mathcal{F}}
\newcommand{\G}{\mathcal{G}}

\newcommand{\A}{\mathsf{A}}
\newcommand{\Ai}{\mathsf{A}^{-1}}

\newcommand{\T}{\mathsf{T}}

\newcommand{\Ci}{\mathsf{C}^{-1}}
\newcommand{\tCi}{\tilde{\mathsf{C}}^{-1}}

\newcommand{\hr}{\hat{\vec r}}

\newcommand{\hn}{\hat{\vec n}}

\newcommand{\hz}{\hat{\vec z}}

\newcommand{\hK}{\hat{\vec K}}
\newcommand{\tjo}[3]{\begin{pmatrix} {#1} & {#2} & {#3}\\ 0 & 0 & 0\end{pmatrix}}
\newcommand{\tj}[6]{\begin{pmatrix} {#1} & {#2} & {#3}\\ {#4} & {#5} & {#6}\end{pmatrix}}

\renewcommand{\vr}{\vec r}

\newcommand{\gnldotdot}{g_{\rm NL}^{\dot{\sigma}^4}}
\newcommand{\gnldotdel}{g_{\rm NL}^{\dot{\sigma}^2(\partial\sigma)^2}}
\newcommand{\gnldeldel}{g_{\rm NL}^{(\partial\sigma)^4}}
\newcommand{\fnl}{f_{\rm NL}^{\rm loc}}
\newcommand{\gnl}{g_{\rm NL}^{\rm loc}}
\newcommand{\taunl}{\tau_{\rm NL}^{\rm loc}}
\newcommand{\gcon}{g_{\rm NL}^{\rm con}}

\DeclareSymbolFont{toneletters}{T1}{\familydefault}{m}{it}
\SetSymbolFont{toneletters}{bold}{T1}{\familydefault}{bx}{it}

\DeclareMathSymbol\edth{\mathord}{toneletters}{"F0}

\usepackage{adjustbox}
\usepackage{dashbox}

\definecolor{darkgreen}{RGB}{0,120,0}
\newcommand{\resub}[1]{{#1}}%\color{darkgreen}{#1}}}

\def\beq{\begin{eqnarray}}
\def\eeq{\end{eqnarray}}

\let\vec\mathbf

\usepackage{empheq}

\def\mnras{\rm{MNRAS}}

\def\aap{\rm{A\&A}}

\def\apj{\rm{ApJ}}                 % Astrophysical Journal
\def\jcap{\rm{JCAP}}

\defcitealias{Philcox4pt2}{Paper 2}
\defcitealias{Philcox4pt3}{Paper 3}

\usepackage{xspace}

\newcommand{\papertwo}{\citetalias{Philcox4pt2}\xspace}
\newcommand{\paperthree}{\citetalias{Philcox4pt3}\xspace}

\begin{document}

\title{\texorpdfstring{\Large Searching for Inflationary Physics with the CMB Trispectrum:\\
\large
1. Primordial Theory \& Optimal Estimators}{Searching for Inflationary Physics with the CMB Trispectrum: 1. Primordial Theory \& Optimal Estimators}
%Constraining Single-Field Inflation, Local Interactions, Spinning Massive Particle Exchange, Chiral Physics, Gravitational Lensing, and Beyond
}
%\date{\today}
\setlength{\parskip}{0pt}

\author{Oliver~H.\,E.~Philcox}
\email{ohep2@cantab.ac.uk}
\affiliation{Simons Society of Fellows, Simons Foundation, New York, NY 10010, USA}
\affiliation{Center for Theoretical Physics, Columbia University, New York, NY 10027, USA}
\affiliation{Department of Physics,
Stanford University, Stanford, CA 94305, USA}

\begin{abstract} 
    \noindent The primordial four-point function encodes a wealth of information about the inflationary Universe. Despite extensive theoretical work, most models of four-point physics have never been compared to data. In this series, we conduct a detailed analysis of Cosmic Microwave Background temperature and polarization trispectra, searching for a wide variety of phenomena including local effects, self-interactions, curvatons, DBI inflation, gauge fields, solid inflation, scalar field exchange, spinning massive field exchange, chiral physics, point sources, and gravitational lensing. After presenting a suite of separable primordial templates, we derive thirteen quasi-optimal estimators that directly estimate the underlying template amplitudes. These are unbiased, minimum variance, mask-deconvolved, and account for correlations between templates (including with lensing). Each estimator can be efficiently implemented using spherical harmonic transforms, Monte Carlo methods, and optimization techniques, and asymptotes to standard forms in certain limits. In \papertwo, we implement these estimators in public code, and in \paperthree, use them to constrain primordial trispectra with \textit{Planck} data. This enables a wide variety of tests of inflation, including some of the first direct constraints on cosmological collider physics.
\end{abstract}

\maketitle
%\tableofcontents

\section{Introduction}

\setlength{\parskip}{4pt}

\noindent At the humongous energy scale of the early Universe, many new things can happen. In the standard paradigm, one assumes an inflationary period dominated by a single scalar field, $\varphi$, slowly rolling down some potential, whose quantum fluctuations source curvature perturbations, $\zeta$, in the post-inflationary Universe \citep[e.g.,][]{Guth:1980zm,Linde:1981mu}. Interactions of $\varphi$ with itself or with other particles generically alter the statistics of $\zeta$; a corollary is that the observed distribution of $\zeta$ can be used to place constraints on the phenomenology of inflation. 

A vast body of work exists theorizing the possible impacts of high-energy physics on the primordial curvature distribution (and other observables, such as isocurvature fluctuations and spectral distortions). One of the most exciting possibilities is that new physics can source \textit{non-Gaussianity}: non-negligible $(n\geq 3)$-point correlation functions that could be measured in late-Universe data. The precise order and shape of the correlators can yield information on the physical model that generates it, for example, a three-point function peaking in the squeezed limit indicates multi-field inflation \citep{Maldacena:2002vr}. 

To fully explore the wide zoology of inflationary non-Gaussianity, it is useful to have some systematic approach for categorizing the various $n$-point functions. A promising avenue for this is the Effective Field Theory (EFT) of Inflation \citep{Cheung:2007st,Senatore:2009gt,Senatore:2010wk,Weinberg:2008hq,Cabass:2022avo}, which probes the low energy consequences of generic inflationary theories subject to some symmetry assumptions, such as approximate scale-invariance. This has recently been coupled to analytic and numerical techniques such as the bootstrap formalism and the `Cosmological Flow' solver to yield efficient computation of inflationary non-Gaussianity in a wide range of physical regimes \citep{Arkani-Hamed:2018kmz,DuasoPueyo:2023kyh,Wang:2022eop,Cabass:2021fnw,Jazayeri:2022kjy,Pimentel:2022fsc,Pajer:2020wnj,Werth:2023pfl,Pinol:2023oux}. Another approach is the so-called `Cosmological Collider' picture \citep{Arkani-Hamed:2015bza,Lee:2016vti,Chen:2009zp,Flauger:2016idt}, which connects particle production during inflation to kinematic limits of the correlation functions in a (relatively) model-agnostic framework. The end-product of such approaches is a theoretical (or numerical) prediction for the correlators of $\zeta$ that can be compared to data.

What is the best-way to search for such signatures? Currently, our best hope is the Cosmic Microwave Background (CMB). On large-scales, the temperature and polarization anisotropies directly trace the primordial curvature perturbations, thus we may use the correlators of CMB $T$- and $E$-modes to constrain inflationary predictions.\footnote{At leading order, scalar physics does not generate $B$-modes. However, these can be used to trace tensor physics through primordial gravitational wave signatures. Whilst the characterization and estimation of higher-point tensor statistics is of great theoretical interest (see \citep[e.g.,][]{Duivenvoorden:2019ses,Philcox:2023xxk,Philcox:2024wqx} for the three-point function) it is beyond the scope of this work.} A similar game can be played in other data-sets including spectroscopic galaxy clustering and 21cm emission \citep[e.g.,][]{Cabass:2022wjy,Cabass:2022ymb,DAmico:2022gki,Meerburg:2016zdz,Floss:2022grj,Cabass:2024wob,Goldstein:2024bky,MoradinezhadDizgah:2017szk,MoradinezhadDizgah:2018ssw,Cabass:2018roz,Cabass:2022epm}. Since these surveys provide a three-dimensional view of primordial physics (in contrast to the two-dimensional slice seen in the CMB), they will eventually be the leading source of inflationary information \citep{Cabass:2022epm,Sailer:2021yzm}. At present, however, the primordial volume surveyed by three-dimensional surveys is considerably smaller than that of \textit{Planck} and contemporary CMB experiments (partly due to non-linearities and systematic effects), thus the CMB gives the strongest constraints on most primordial models.

Utilizing CMB data from WMAP, \textit{Planck} and beyond, many previous works have searched for non-Gaussian signatures from inflation  \citep[e.g.,][]{2014A&A...571A..24P,Planck:2015zfm,Planck:2019kim,Creminelli:2005hu,Senatore:2009gt,Fergusson:2010gn,2015arXiv150200635S,Marzouk:2022utf,2006JCAP...05..004C,Komatsu:2003iq,Liguori:2010hx,Babich:2004gb,Sohn:2024xzd,Cabass:2022wjy,Cabass:2022ymb,Cabass:2024wob,Feng:2015pva,Fujita:2013qxa,Komatsu:2001wu,Smidt:2010ra,2003MNRAS.341..623S,Munchmeyer:2014nqa,Shiraishi:2012rm,Philcox:2023xxk,Philcox:2024wqx,Philcox:2023ypl,PhilcoxCMB,Philcox:2024jpd,Hikage:2012bs,Kunz:2001ym,Vielva:2009jz,WMAP:2006bqn}. To date, there have been no robust detections. This is not necessarily a cause for nihilism, however, since (a) most previous analyses have been limited to phenomenological templates, which do not capture some regimes of interest, including the cosmological collider (though see \citep{Cabass:2024wob,Sohn:2024xzd}), and (b) almost all previous analyses have restricted to three-point functions. The official \textit{Planck} non-Gaussianity analyses provide a clear demonstration of this: four-point functions are discussed in only $13$ of the $187$ total pages \citep{2014A&A...571A..24P,Planck:2015zfm,Planck:2019kim}. 

\begin{figure}
\centering
\begin{minipage}{0.4\textwidth}
    % Contact diagram
    \begin{tikzpicture}
        \begin{feynman}
            \vertex (a1) at (-2.5, 1.5) {$\vk_1$};
            \vertex (a2) at (-2.5, -1.5) {$\vk_2$};
            \vertex (a3) at (2.5, 1.5) {$\vk_3$};
            \vertex (a4) at (2.5, -1.5) {$\vk_4$};
            \vertex (c) at (0, 0);
            
            \diagram* {
                (a1) -- [fermion] (c),
                (a3) -- [fermion] (c),
                (a2) -- [fermion] (c),
                (a4) -- [fermion] (c),
            };
        \end{feynman}
        \node at (0,-1.8) {\textbf{Contact}: $g_{\rm NL}$};
    \end{tikzpicture}
\end{minipage}
\hspace{3em}
\begin{minipage}{0.4\textwidth}
    % Exchange diagram
    \begin{tikzpicture}
        \begin{feynman}
            \vertex (b1) at (-2.5, 1.5) {$\vk_1$};
            \vertex (b2) at (-2.5, -1.5) {$\vk_2$};
            \vertex (b3) at (2.5, 1.5) {$\vk_3$};
            \vertex (b4) at (2.5, -1.5) {$\vk_4$};
            \vertex (m1) at (-0.7, 0);
            \vertex (m2) at (0.7, 0);
            \vertex (mid) at (0, -0.3) {$\vK$}; % Label for the middle line
            
            \diagram* {
                (b1) -- [fermion] (m1),
                (b3) -- [fermion] (m2),
                (b2) -- [fermion] (m1),
                (b4) -- [fermion] (m2),
                (m1) -- [photon] (m2),
            };
        \end{feynman}
        \node at (0,-1.8) {\textbf{Exchange}: $\tau_{\rm NL}$};
    \end{tikzpicture}
\end{minipage}
\caption{Cartoon illustrating the two types of primordial trispectra considered in this work. Contact diagrams (left) correspond to cubic inflationary interactions, such as that generated by the local transformation $\phi\to \phi+g_{\rm NL}\phi^3$ for some primordial scalar field $\phi$. The associated trispectra can be factorized into a sum of products of functions of the four momenta $\vk_i$, \textit{i.e.}\ $T_\zeta(\vk_1,\vk_2,\vk_3,\vk_4)\sim\sum_n \alpha^{(n)}(\vk_1)\beta^{(n)}(\vk_2)\gamma^{(n)}(\vk_3)\delta^{(n)}(\vk_4)$. Exchange diagrams (right) involve a pair of quadratic interactions corresponding to, for example, the exchange of a new particle with momentum $\vK\equiv \vk_1+\vk_2$. These can be factorized into two pieces connected by $\vK$, \textit{i.e.}\ $T_\zeta(\vk_1,\vk_2,\vk_3,\vk_4)\sim\sum_n \epsilon^{(n)}(\vk_1,\vk_2)\zeta^{(n)}(\vk_3,\vk_4)$, and are generally more difficult to constrain.}\label{fig: cartoon}
\end{figure}
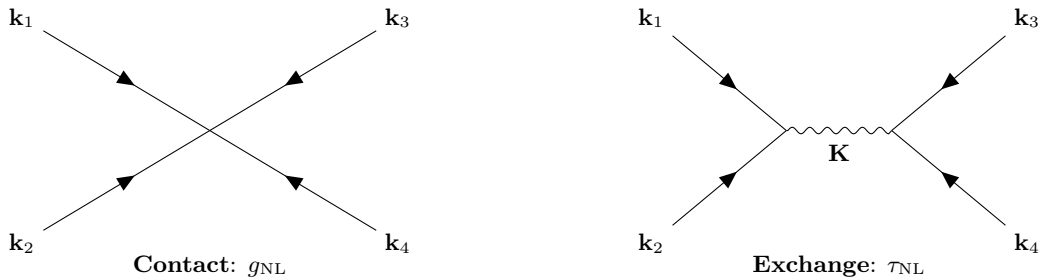

From a theoretical point of view, trispectra are no less interesting than bispectra. Many inflationary models predict large four-point non-Gaussianity without three-point functions; moreover, four-point functions are sensitive to both primordial scattering and exchange processes (which correspond to cubic and quadratic interactions, as sketched in Fig.\,\ref{fig: cartoon}). Measuring trispectra allows for the amplitudes of particular terms in the primordial Lagrangian to be immediately constrained (without requiring the breaking of time-translation invariance), and can reveal new interactions that cannot be probed at lower orders, such as parity-violation and graviton exchange. Some progress has been made towards this goal already; \citep{2015arXiv150200635S} present a detailed analysis of EFT trispectrum templates in WMAP (which was later applied to \textit{Planck} \citep{Planck:2015zfm,Planck:2019kim}), and a number of works have constrained the simplest `local' four-point function \citep{Fergusson:2010gn,2014A&A...571A..24P,Planck:2015zfm,Planck:2019kim,Smidt:2010ra,Feng:2015pva,Munshi:2010bh,Marzouk:2022utf}, though only \citep{Marzouk:2022utf} includes polarization. Much, however, remains to be done.

In this series of works, we will perform a detailed analysis of primordial physics in the CMB four-point function, filling an important void in observational studies of inflation. In particular, we will consider the following (non-exhaustive) categories of trispectra:
\begin{itemize}
    \item \textbf{Local effects}: These can be sourced by light scalar fields in inflation beyond the inflaton, optionally with additional symmetry restrictions \citep{Senatore:2010wk}. They can be described by the non-Gaussianity parameters $\gnl$ and $\taunl$ (which constrain, for example, curvaton models and ekpyrosis \citep{Sasaki:2006kq,Lehners:2013cka}), and are the most well-known and oft-constrained of all trispectrum parameters.
    \item \textbf{Self-interactions}: These can be modeled using the EFT of inflation \citep{Cheung:2007st,Senatore:2010wk}, which sets out the leading contributions to the inflationary Lagrangian allowed by symmetry, and their corresponding trispectra (for both single- and multi-field inflation). They may also be generated by integrating-out very massive fields in the primordial Lagrangian, and can be related to models such as DBI inflation \citep[e.g.,][]{Bartolo:2003jx,Silverstein:2003hf,Arroja:2009pd}.
    \item \textbf{Massive particle exchange}: Depending on the mass and spin of the exchanged field, novel signatures can be formed in the soft limits of the four-point function, which are predicted by the `Cosmological Collider' literature \citep[e.g.,][]{Chen:2009zp,Arkani-Hamed:2015bza,Lee:2016vti,Flauger:2016idt,Wang:2022eop,Pimentel:2022fsc,Kumar:2019ebj,Reece:2022soh,Arkani-Hamed:2015bza,Alexander:2019vtb,Jazayeri:2023xcj,McCulloch:2024hiz,Liu:2019fag,Meerburg:2016zdz,Wang:2019gbi,Tong:2022cdz,Sohn:2024xzd,Pimentel:2022fsc,Cabass:2024wob,Chen:2016uwp,Chen:2018xck,Lu:2019tjj,Wang:2020ioa,Bodas:2020yho,Jazayeri:2022kjy,Kim:2019wjo,Lu:2021wxu,Cui:2021iie,Qin:2022lva,Werth:2023pfl,Chen:2022vzh,Xianyu:2023ytd,Pinol:2023oux,Chakraborty:2023qbp,Craig:2024qgy,Yin:2023jlv,Baumann:2011nk,Assassi:2012zq,Baumann:2017jvh,Arkani-Hamed:2018kmz,Cabass:2022rhr,Bordin:2018pca,Bordin:2019tyb,Jazayeri:2023kji,Chen:2018sce,Green:2023ids,Noumi:2012vr}. Here, we will focus on the collapsed limit of the trispectrum, which exhibits interesting signatures for masses around the Hubble scale, practically acting as a particle collider at energies $H\lesssim 10^{14}\,\mathrm{GeV}$.
    \item \textbf{Phenomenological Templates}: We will additionally consider a number of more generic templates, such as featureless primordial shapes and direction-dependent scalar trispectra \citep[e.g.,][]{Fergusson:2010gn,Shiraishi:2016mok,Shiraishi:2013oqa}. These can be mapped to a number of ultraviolet models such as axion-gauge field couplings and solid inflation.
    \item \textbf{Late-Time Trispectra}: Non-linear physics in the late Universe, such as gravitational lensing and point sources, provide an additional source of non-Gaussianity. Though non-primordial, these are important contributors to the observed CMB trispectrum and their neglection can lead to spurious detections of inflationary physics \citep[e.g.,][]{Lewis:2006fu,Hanson:2010rp,Planck:2018lbu,2014A&A...571A..24P,Philcox:2025lxt}. Here, we constrain the overall point source and lensing amplitudes $t_{\rm ps}$ and $A_{\rm lens}$, noting that the corresponding estimators are very similar to those of primordial local effects and particle exchange respectively.
\end{itemize}
For reference, we provide a brief summary of these models in Tab.\,\ref{tab: models}. 

In each case of the above cases, we will develop, implement, test, and apply estimators to constrain the characteristic model amplitudes, which can then be related to microphysical primordial parameters. To this end, we will develop quartic estimators that allow close-to-optimal measurement of the amplitudes, following a procedure outlined in \citep{2015arXiv150200635S} (itself based on the Komatsu-Spergel-Wandelt estimator \citep{Komatsu:2003iq}). Given data $d$ and a target trispectrum $T$ with amplitude $A$, these have the schematic form
\beq
    \widehat{A}[d] \sim \sum_{i_1i_2i_3i_4}\frac{\partial T^{i_1i_2i_3i_4}}{\partial A}\bigg(d_{i_1}d_{i_2}d_{i_3}d_{i_4}-6\av{d_{i_1}d_{i_2}}d_{i_3}d_{i_4}+3\av{d_{i_1}d_{i_2}}\av{d_{i_3}d_{i_4}}\bigg),
\eeq
subtracting off the Gaussian contributions, and labeling pixels and fields by $i=1,\cdots,N_{\rm pix}$. These have a number of useful properties namely: they are \textbf{direct} estimators of the underlying amplitude (with no need for binning); given a suitable weighting scheme, they are \textbf{minimum variance}; they are \textbf{unbiased} by the mask, beam, correlations between templates and (if desired) late-time effects; they are \textbf{efficient} to compute, with the rate-limiting steps scaling as $\mathcal{O}(N_{\rm pix}\log N_{\rm pix})$ (instead of the na\"ive $\mathcal{O}(N_{\rm pix}^4)$ scaling). As we will see below, construction of such estimators can be somewhat painful; as such, the majority of the models listed above have not been analyzed in detail before.

Due to the scale of this project, we split our results into three papers. In the first paper (this work), we give a general overview of primordial four-point non-Gaussianity, both serving as a literature review and a detailed description of each primordial trispectra, as well as its relation to inflationary models. Given these templates, we define an optimal estimator for each and show how it can be efficiently implemented using various theoretical and computational tricks. \papertwo presents an efficient implementation of these estimators in the \href{https://github.com/oliverphilcox/PolySpec}{\textsc{PolySpec}} code and performs extensive validation of all parts of the analysis. Finally, in \paperthree, we apply the formalism to the latest \textit{Planck} data, placing constraints on every model discussed in this work and their corresponding microphysical parameters.

\vskip 8pt
The remainder of this paper is as follows. In \S\ref{sec: templates} we discuss the rich landscape of four-point inflationary physics and specify the various primordial templates that will be used throughout this paper. In \S\ref{sec: estimators}, \resub{we give a general overview} of our trispectrum estimators before presenting their specific forms for contact-, exchange-, and late-time trispectra in \S\ref{sec: estimators-contact},\,\S\ref{sec: estimators-exchange}\,\&\,\S\ref{sec: estimators-late-time} respectively. We discuss the relation of our estimators to standard forms in \S\ref{sec: estimator-comparison}, before considering their optimization in \S\ref{sec: optim}. We conclude with a summary and discussion of future work in \S\ref{sec: discussion}. Appendices \ref{app: direc-simp}, \ref{app: angle-simplifications}, \ref{app: forecasts}, \ref{app: lensing} \& \ref{app: analytic-fisher} provide a number of useful results relating to the templates and their practical implementation. For clarity, we box the definitions of all primordial trispectrum templates, and signify each estimator with a dashed box; these are implemented in the \textsc{PolySpec} code described in \papertwo.

% Contact diagram

\section{Primordial Templates}\label{sec: templates}

\noindent Four-point physics during inflation endows the gauge-invariant curvature fluctuation, $\zeta$, with a primordial trispectrum, $T_\zeta$. In this work, we consider two broad classes of trispectra: contact and exchange shapes, as sketched in Fig.\,\ref{fig: cartoon}. Generically, these can be expressed in the following Fourier-space forms
\beq\label{eq: contact-exchange-tzeta}
    \av{\zeta(\vk_1)\zeta(\vk_2)\zeta(\vk_3)\zeta(\vk_4)}_{\rm contact} &\equiv& T_\zeta(\vk_1,\vk_2,\vk_3,\vk_4)\,\delD{\vk_1+\vk_2+\vk_3+\vk_4}\\\nonumber
    \av{\zeta(\vk_1)\zeta(\vk_2)\zeta(\vk_3)\zeta(\vk_4)}_{\rm exchange} &\equiv&\int_{\vK}T_\zeta(\vk_1,\vk_2,\vk_3,\vk_4;\vK)\,\delD{\vk_1+\vk_2-\vK}\delD{\vk_3+\vk_4+\vK},
\eeq
where the exchange trispectra depend explicitly on an internal momentum $\vK$. Contact interactions are sourced by 2$-$2 scattering processes (or their de Sitter analog), whilst exchange forms usually involve the exchange of some intermediate particle with momentum $\vK$. Via homogeneity and isotropy, each trispectrum depends only on scalar quantities, \textit{i.e.}\ $\{k_i\equiv |\vk_i|$, $\vk_i\cdot\vk_j$, $\vk_i\cdot\vK$, $\vk_i\times\vk_j\cdot\vK\}$, where the final quantity breaks parity symmetry \citep[e.g.,][]{Coulton:2023oug}, and only the first two are present for contact interactions. Below, we specify the trispectrum shapes associated with a variety of physics models. In \paperthree, we measure their amplitudes from \textit{Planck} data and compare them to literature constraints.

\begin{table}
    \centering
    \begin{tabular}{c|c||c|c|c|c}
      \textbf{Amplitude} & \textbf{Eq.} & \textbf{Model} & \textbf{Type} & \textbf{Functions} & \textbf{Estimator}\\\hline\hline
      $\gnl$ & \eqref{eq: gNL-shape} & Local (cubic) & Contact & $P(\vr), Q(\vr)$ & \S\ref{subsec: loc-estimator}\\
      $g_{\rm NL}^{\rm con}$ & \eqref{eq: con-shape} & Constant shape & Contact & $R(\vr)$ & \S\ref{subsec: con-estimator}\\
      $g_{\rm NL}^{\dot{\sigma}^4,\dot{\sigma}^2(\partial\sigma)^2,(\partial\sigma)^4}$ & \eqref{eq: EFTI-shapes} & Self-interactions (EFT of Inflation) & Contact & $A(\vr,\tau), B(\vr,\tau), {}_{\pm1}C(\vr,\tau)$ & \S\ref{subsec: efti-estimators}\\\hline
      $\taunl$ & \eqref{eq: tauNL-shape} & Local (quadratic) & Exchange & $P(\vr), Q(\vr)$ & \S\ref{subsec: loctau-estimator}\\
      $\tau_{\rm NL}^{n, \rm even}$ & \eqref{eq: dn-even-def} & Parity-even direction-dependent local & Exchange & $P_{n\nu}(\vr)$, $Q(\vr)$ & \S\ref{subsec: direc-estimator}\\
      $\tau_{\rm NL}^{n, \rm odd}$ & \eqref{eq: dn-odd-def} & Parity-odd direction-dependent local & Exchange & $P_{n\nu}(\vr)$, $Q(\vr)$ & \S\ref{subsec: direc-estimator}\\
      $\tau_{\rm NL}^{n_1n_3n}$ & \eqref{eq: dnnn-def} & Generalized direction-dependent local & Exchange & $P_{n\nu}(\vr)$, $Q(\vr)$ & \S\ref{subsec: direc-estimator}\\
      $\tau_{\rm NL}^{\rm heavy}(s,\mu_s)$ & \eqref{eq: heavy-exchange-template} & Spin-$s$ heavy particle exchange & Exchange & $P_{s\mu}^{(-3/2\mp i\mu_s)}(\vr)$ & \S\ref{subsec: direc-estimator}\\
      $\tau_{\rm NL}^{\rm light}(s,\nu_s)$ & \eqref{eq: light-exchange-template} & Spin-$s$ intermediate/light particle exchange & Exchange & $P_{s\mu}^{(-3/2+\nu_s)}(\vr)$ & \S\ref{subsec: collider-estimator}\\\hline
      $t_{\rm ps}$ & \eqref{eq: point-source-tspec} & Unclustered point sources & Contact & ${}_0U^T(\hn)$ & \S\ref{subsec: point-source-estimator}\\
      $A_{\rm lens}$ & \eqref{eq: lens-trispectrum-def} & Weak gravitational lensing & Exchange & ${}_sU^X(\hn)$, ${}_sV^X_{\pm}(\hn)$ & \S\ref{subsec: lensing-estimator}
    \end{tabular}
    \caption{Summary of the trispectrum models considered in this work. In the first three columns we list the model amplitudes, the template definitions and the physical models generating them; further details can be found in \S\ref{sec: templates}. We consider two types of trispectra: `contact' and `exchange' (cf.\,Fig.\,\ref{fig: cartoon}) with the latter peaking in collapsed Fourier-space configurations, \textit{i.e.}\ $|\vk+\vk'|\ll k,k'$ (for two external legs $\vk,\vk'$). In \S\ref{sec: estimators-contact}\,\&\,\ref{sec: estimators-exchange} we derive separable KSW-like estimators for each trispectrum template, which involve a set of filter functions; the functions required are given in the penultimate column. The last two templates correspond to late-time effects (point sources and CMB lensing), whose covariance with the primordial templates must be carefully accounted for. Whilst these templates span a range of physical scenarios of interest (and can be projected onto many physical models, for example curvatons, colliders and solid inflation), they are by no means exhaustive.}
    \label{tab: models}
\end{table}

\subsection{Contact Trispectra}\label{subsec: contact-templates}
\subsubsection{Local Interactions}\label{subsec: contact-local-theory}
\noindent The simplest model of primordial non-Gaussianity induces a quadratic or cubic modulation to the Gaussian curvature, $\zeta_{\rm G}$:
\beq\label{eq: local-zeta-perturbation}
    \zeta_{\rm G}(\vx) \to \zeta_{\rm G}(\vx)+\frac{3}{5}\fnl\left(\zeta_{\rm G}^2(\vx)-\av{\zeta_{\rm G}^2}\right)+\frac{9}{25}\gnl\left(\zeta_{\rm G}^3(\vx)-3\av{\zeta^2_G}\zeta_{\rm G}(\vx)\right), 
\eeq
whose magnitude depends on the coupling strengths $\fnl$ and $\gnl$ (with numerical coefficients arising from the conversion to curvature from Bardeen potential). Whilst these cannot be generated in vanilla single-field inflation due to the consistency condition \citep{Maldacena:2002vr}, the situation changes in the presence of additional light scalars during inflation and the $\gnl$ term can become dominant in some regimes, including via the presence of additional approximate symmetries such as charge conservation ($\mathbb{Z}_2$) or supersymmetry \citep{Senatore:2010wk,Suyama:2013nva}. At leading order, $\fnl$ generates a curvature bispectrum due to the contraction of $\zeta_{\rm G}^2$ with two linear legs; analogously, $\gnl$ generates a curvature trispectrum due to the contraction of $\zeta_{\rm G}^3$ with three linear legs. In Fourier-space, this leads to
\beq\label{eq: fNL-shape}
    \av{\zeta(\vk_1)\zeta(\vk_2)\zeta(\vk_3)}' &\supset& \frac{6}{5}f^{\rm loc}_{\rm NL}P_\zeta(k_1)P_\zeta(k_2)+\text{2 perms.}
\eeq
\vspace{-1.6em}
\beq\label{eq: gNL-shape}
\boxed{\av{\zeta(\vk_1)\zeta(\vk_2)\zeta(\vk_3)\zeta(\vk_4)}'_c \supset \frac{54}{25}g^{\rm loc}_{\rm NL}P_\zeta(k_1)P_\zeta(k_2)P_\zeta(k_3)+\text{3 perms.},}
\eeq
where the prime indicates that we drop the momentum-conserving Dirac delta functions. Here, we are interested in the trispectrum contribution, which is (a) in the contact form, (b) explicitly separable in $k_i$, (c) roughly scale-invariant, with $T_\zeta \sim A_\zeta^3k^{3(n_s-4)}$ (assuming the standard definitions of $P_\zeta(k) \equiv A_\zeta k^{-3}(k/k_{\rm pivot})^{n_s-1}$ and $A_\zeta \equiv 2\pi^2A_s$). The $\gnl$ coefficient is, \textit{a priori}, unconstrained, though the assumption of perturbativity requires $|\gnl|A_\zeta\lesssim 1$ thus $|\gnl|\lesssim 10^{10}$. This parameter has been constrained in a number of previous studies including \citep{Creminelli:2005hu,Fergusson:2010gn,2015arXiv150200635S,2014A&A...571A..24P,Planck:2015zfm,Planck:2019kim,Feng:2015pva,Vielva:2009jz,Smidt:2010ra,Hikage:2012bs,Sekiguchi:2013hza,Desjacques:2009jb,Leistedt:2014zqa,Giannantonio:2013uqa}.

One physical realization of the local model is in curvaton scenarios \citep[e.g.,][]{Sasaki:2006kq,Bartolo:2003jx}, where the inflaton $\varphi$ is coupled to a light scalar $\chi$. If the curvaton has a subdominant energy density during inflation but decays to radiation before the redshift of primordial nucleosynthesis, a large non-Gaussianity can be generated. In the simplest adiabatic models, this corresponds to to the local parameters
\beq\label{eq: curvaton}
    \fnl &=& \frac{5}{6}\left(\frac{3}{2r_D}-2-r_D\right), \qquad 
    \gnl = \frac{25}{54}\left(-\frac{9}{r_D}+\frac{1}{2}+10r_D+3r_D^2\right)
\eeq
for curvaton decay fraction $r_D \equiv 3\rho_\chi/[\rho_\chi+4\rho_{\rm rad}]$ at the epoch of decay into radiation. Note that $\fnl$ and $\gnl$ are of the same order in these models (for small $r_D$), though some mechanisms can be added to enhance $\gnl$ \citep[e.g.,][]{Byrnes:2010em,Huang:2013yla}. Alternatives to inflation, such as the ekpyrotic/cyclic scenario, can also map to these templates. As discussed in \citep{Lehners:2013cka}, the two-field kinetic conversion model predicts 
\beq\label{eq: ekpyrosis-model}
    \fnl = \pm 5+ \frac{3}{2}\kappa_3\sqrt{\epsilon},\qquad  \gnl = (-40+\frac{5}{3}\kappa_4+\frac{5}{4}\kappa_3^2)\epsilon,
\eeq
where $\kappa_{3}, \kappa_4$ and $\epsilon\gtrsim 50$ describe the shape of the scalar field potential. Since this formalism (and its extensions \citep[e.g.,][]{Fertig:2015ola}) generically predicts non-zero $\gnl$, it could be confirmed or ruled out by future experiments. Additional physical sourcing includes various light-field models such as modulated reheating and thermal inflation \citep{Suyama:2013nva}, inflationary vector fields \citep{Dimopoulos:2006ms,Valenzuela-Toledo:2009bzd} and the supersymmetric multi-field model discussed in the next section.

Some works consider an additional `constant' primordial trispectrum shape \citep[e.g.,][]{Fergusson:2010gn}, defined such that the scale-invariant trispectrum $\sim k^{9}T_\zeta(k)$ is featureless in the de Sitter ($n_s=1$) limit. This can be represented by the following (trivially separable) form
\beq\label{eq: con-shape}
    \boxed{\av{\zeta(\vk_1)\zeta(\vk_2)\zeta(\vk_3)\zeta(\vk_4)}'_c \supset \frac{216}{25}g_{\rm NL}^{\rm con}\left[P_\zeta(k_1)P_\zeta(k_2)P_\zeta(k_3)P_\zeta(k_4)\right]^{3/4},}
\eeq
with characteristic amplitude $g_{\rm NL}^{\rm con}$. This was constrained in \citep{Fergusson:2010gn} using WMAP data, and has the perturbativity bound $|g_{\rm NL}^{\rm con}|\lesssim 10^{10}$.\footnote{Note that we have slightly altered the template of \citep{Fergusson:2010gn} to allow for departures from exact scale invariance, \textit{i.e.}\ $n_s\neq 1.$} Although this does not represent any specific inflationary model, it is a useful diagnostic for identifying generic primordial signatures and is thus included in our cosmic census.

\subsubsection{Self-Interactions and the Inflationary EFT}\label{subsubsec: EFTI}
\noindent The simplest models of inflation predict that the primordial Universe contains only a single light scalar degree of freedom, with negligible self-interactions \citep{Guth:1980zm,Linde:1981mu}. In this limit, the curvature perturbations are Gaussian (modulo non-linear reheating effects) and the remainder of this paper is largely irrelevant. 

\vskip 4pt
\paragraph{Single-Field EFT}
The EFT of inflation provides a rigorous manner in which to describe departures from the above picture, through the enumeration of all possible low-energy interactions consistent with the (assumed) softly symmetries of inflation \cite[e.g.,][]{Weinberg:2008hq,Cheung:2007st,Senatore:2010wk,Senatore:2009gt,Khosravi:2012qg,Shiu:2011qw,Fasiello:2011fj}. For single-field inflation, these can be described in terms of the Goldstone boson $\pi$, which encodes the slightly broken time-translation symmetry, and is related to primordial curvature via $\zeta = -H\pi+\cdots$. Up to fourth order in $\pi$, the single-field action can be written \citep{Cheung:2007st}
\beq
    S^{\rm EFT}_\pi &=& \int d^4x\sqrt{-g}\,\bigg\{-M_{\rm Pl}^2\dot{H}(\partial_\mu\pi)^2+2M_2^4\left[\dot{\pi}^2+\dot{\pi}^3-\dot{\pi}\frac{(\partial_i\pi)^2}{a^2}+(\partial_\mu\pi)^2(\partial_\nu\pi)^2+\cdots\right]\\\nonumber
    &&\qquad\qquad\qquad\,-\,\frac{M_3^4}{3!}\left[8\dot{\pi}^3+12\dot{\pi}^2(\partial_\mu\pi)^2+\cdots\right]+\frac{M_4^4}{4!}\left[16\dot{\pi}^4+\cdots\right]+\cdots\bigg\},
\eeq
neglecting metric fluctuations (\textit{i.e.}\ working in the decoupling limit). Here, $H$ is the Hubble parameter, $M_{\rm Pl}$ is the Planck mass, and $\{M_n\}$ are coupling constants, whose sizes are bounded by perturbativity and unitarity. The first term is the usual kinetic piece ($\sim \dot{\pi}^2-(\partial_i\pi)^2$), whilst the remaining three quantify self-interactions of $\pi$, with a structure fixed by the non-linear realization of diffeomorphisms. Usually, one rewrites this expression by noting that the $M_2^4\dot{\pi}^2$ term modifies the dispersion relation of the Goldstone mode \citep[e.g.,][]{Cheung:2007st,Senatore:2009gt,Planck:2019kim}
\beq\label{eq: EFTI-action}
    S^{\rm EFT}_\pi &=& \int d^4x\sqrt{-g}\,\bigg\{-\frac{M_{\rm Pl}^2\dot{H}}{c_s^2}\left[\dot{\pi}^2-c_s^2\frac{(\partial_i\pi)^2}{a^2}\right]\\\nonumber
    &&\qquad\qquad\qquad\,+\,M_{\rm Pl}^2
    \dot{H}\left(1-\frac{1}{c_s^2}\right)\left[\dot{\pi}^3\left(1+\frac{2\tilde{c}_3}{3c_s^2}\right)-\dot{\pi}\frac{(\partial_i\pi)^2}{a^2}+(\partial_\mu\pi)^2(\partial_\nu\pi)^2+\cdots\right]\\\nonumber
    &&\qquad\qquad\qquad\,+\,\frac{\tilde{c}_3M_{\rm Pl}^2\dot{H}}{c_s^2}\left(1-\frac{1}{c_s^2}\right)\left[\dot{\pi}^2(\partial_\mu\pi)^2+\cdots\right]+\frac{M_4^4}{4!}\left[16\dot{\pi}^4+\cdots\right]+\cdots\bigg\},
\eeq
for sound speed $c_s$ and cubic coupling $\tilde{c}_3$ with the definitions
\beq
c_s^{-2} = 1-\frac{2M_2^4}{M_{\rm Pl}^2\dot{H}}, \qquad \tilde{c}_3 =-c_s^2\frac{M_3^4}{M_2^4}.
\eeq

From \eqref{eq: EFTI-action}, we may assess the various non-Gaussian signals allowed by single-field inflation, making the standard symmetry assumptions. At third order, the self-interacting terms are given by $\dot{\pi}^3$ and $\dot{\pi}(\partial_i\pi)^2$, which source bispectra with amplitudes
\beq
    f_{\rm NL}^{\dot{\pi}(\partial\pi)^2}= \frac{85}{324}\left(1-\frac{1}{c_s^{2}}\right),\qquad f_{\rm NL}^{\dot{\pi}^3}=\frac{10}{243}\left(1-\frac{1}{c_s^{2}}\right)\left(\tilde{c}_3+\frac{3}{2}c_s^2\right).
\eeq
These have been constrained from WMAP and \textit{Planck} CMB anisotropies \citep{Senatore:2009gt,2014A&A...571A..24P,Planck:2015zfm,Planck:2019kim} as well as BOSS galaxy clustering \citep{Cabass:2022wjy,DAmico:2022gki}, via their overlap with the canonical equilateral and orthogonal templates. At fourth order, there are three contributions: $\dot{\pi}^4$, $\dot{\pi}^2(\partial_i\pi)^2$ and $(\partial_i\pi)^2(\partial_j\pi)^2$, each of which sources a different curvature trispectrum. However, from the structure of \eqref{eq: EFTI-action}, it is clear that the second and third terms are always accompanied by the cubic operators $\dot{\pi}^3$ or $\dot\pi(\partial_\mu\pi)^2$ that will dominate the non-Gaussianity. As such, the $\dot{\pi}^2(\partial_i\pi)^2$ and $(\partial_i\pi)^2(\partial_j\pi)^2$ operators are generically small in single-field inflation (recalling that the characteristic scale of $f_{\rm NL}$ is a factor of $A^{-1/2}_\zeta\sim 10^5$ smaller than $g_{\rm NL}$, and all current bounds on $f_{\rm NL}$ are consistent with zero). In contrast, the amplitude of $\dot{\pi}^4$ is independent of three-point interactions (due to its dependence on the hitherto unconstrained energy scale $M_4$), and could thus be large \citep[e.g.,][]{Senatore:2010jy,Cheung:2007st,2015arXiv150200635S}. This sources a primordial trispectrum with amplitude \citep[e.g.,][]{2015arXiv150200635S,Fergusson:2010gn}
\beq
    g_{\rm NL}^{\dot{\pi}^4}&=&\frac{25}{288}\frac{M_4^4}{H^4}A_\zeta c_s^3,
\eeq
which can be constrained from data, as we discuss below.

\vskip 4pt
\paragraph{Multi-Field EFT} In the above discussion we concluded that the canonical single-field EFT generates only a single trispectrum shape. By modifying the assumptions of \eqref{eq: EFTI-action}, we can generate additional, and potentially large, trispectra. One such extension is to assume that primordial perturbations are not sourced by the gauge boson, $\pi$, but by an additional light scalar field, $\sigma$, after horizon crossing, with $\zeta = (2A_\zeta)^{1/2}\sigma/H+\cdots$ \citep[e.g.,][]{Senatore:2010wk,Khosravi:2012qg,2015arXiv150200635S,Gong:2016qmq}. This leads to the multi-field EFT action:
\beq
    S^{\rm EFT}_\sigma = \int d^4x\sqrt{-g}\left\{\frac{1}{2}(\partial_\mu\sigma)^2+\frac{1}{\Lambda_1^4}\dot{\sigma}^4+\frac{1}{\Lambda_2^4}\dot{\sigma}^2(\partial_i\sigma)^2+\frac{1}{\Lambda_3^4}(\partial_i\sigma)(\partial_j\sigma)^2+\frac{\mu^4}{\Lambda^4}\sigma^4+\cdots\right\},
\eeq
where $\{\Lambda_n\}$ and $\mu$ are a set of characteristic amplitudes. Here, we have ignored cubic terms in the action, which could be suppressed via symmetry constraints (e.g., $\mathbb{Z}_2$ invariance for $\sigma\to-\sigma$). The first three terms are invariant under shifts $\sigma(\vx,t)\to \sigma(\vx,t)+A$, whilst the fourth (which could be sourced in supersymmetry theories) is not. Notably, the shift-symmetric interactions $\dot{\sigma}^4$, $\dot{\sigma}^2(\partial_i\sigma)^2$, and $(\partial_i\sigma)^2(\partial_j\sigma)^2$ are of the same form as those appearing in the single-field action \eqref{eq: EFTI-action}, but their amplitudes are now unconstrained (since our model contains more than just a single inflationary `clock'). Overall, the four vertices source non-Gaussianity with the following amplitudes (with shapes discussed below):
\beq\label{eq: gnl-LI}
    \gnldotdot A_\zeta &=& \frac{25}{768}\frac{H^4}{\Lambda_1^4}, \qquad \gnldotdel A_\zeta = -\frac{325}{6912}\frac{H^4}{\Lambda_2^4}, \qquad \gnldeldel A_\zeta = \frac{2575}{20736}\frac{H^4}{\Lambda_3^4}, \qquad \gnl \approx -\frac{50}{27}\frac{\mu^4}{\Lambda^4}N_e,
\eeq
assuming $N_e$ $e$-folds of inflation (where $\gnldotdot\equiv g_{\rm NL}^{\dot\pi^4}$). We note that the local shape can also be sourced by non-linearities in the relation of $\zeta$ and $\sigma$. Additional physical assumptions lead to relations between these parameters; for example, asserting Lorentz invariance implies that $\Lambda_1^4=-2\Lambda_2^4=\Lambda_3^4=\Lambda_{\rm LI}^4$ (all sourced by a $\Lambda_{\rm LI}^{-2}(\partial_\mu\sigma)^2(\partial_\nu\sigma)^2$ interaction) \citep[e.g.,][]{Senatore:2010wk}. The presence of additional particles also leads to exchange interactions; these are discussed in \S\ref{subsec: primordial-exchange}.

\vskip 4pt
\paragraph{Modified Kinetic Terms} Sizable four-point non-Gaussianity can also be generated by modifying the inflationary background, for example via non-standard kinetic terms. Canonical examples are $P(X,\phi)$ theories \citep[e.g.,][]{Babich:2004gb}, which are built around the following action for a light scalar field $\phi$:\footnote{Multi-field generalizations are also possible \citep[e.g.,][]{Mizuno:2009mv,Gao:2009gd}, though we do not consider them here.} 
\beq
    S^{P(X,\phi)}_{\phi} &=& \frac{1}{2}\int d^4x\sqrt{-g}\bigg\{M_{\rm Pl}^2R+2P(X,\phi)\bigg\},
\eeq
for some function $P(X,\phi)$ of $X\equiv-(1/2) g^{\mu\nu}\partial_\mu \phi\partial_\nu\phi$, where $P(X,\phi) = X-V(\phi)$ recovers the non-interacting single-field action. Notable subclasses of this are $K$-inflation \citep{Armendariz-Picon:1999hyi} and DBI inflation \citep{Alishahiha:2004eh,Silverstein:2003hf,Langlois:2008qf,Langlois:2008wt}, with the latter specified by
\beq
    %P_{\rm K-inflation}(X,\phi) &=& K(\phi)X+L(\phi)X^2+\cdots\\\nonumber
    P_{\rm DBI}(X,\phi) &=& -\frac{1}{f(\phi)}\left[\sqrt{1-2f(\phi)X}-1\right]-V(\phi)
\eeq
where the warp factor $f(\phi)$ and the potential $V(\phi)$ are informed by string theory. The dynamics of these models are principally set by the sound speed:
\beq
    c_s^2 \equiv \frac{P_{,X}}{P_{,X}+2XP_{,XX}};
\eeq
setting $c_s\ll 1$ implies that certain higher-order terms in the inflationary EFT expansion of \eqref{eq: EFTI-action} are non-negligible (and thus the previous bounds on single-field trispectra need not apply). The trispectra of $P(X,\phi)$ models can be extracted using their interaction-picture Hamiltonian density:
\beq
    \mathcal{H}_I^{(4)}= a^3\beta_1\dot{\alpha}^4+a\beta_2\dot{\alpha}^2(\partial\alpha)^2+\frac{1}{a}\beta_3(\partial\alpha)^4,
\eeq
where $\alpha$ is the perturbed inflaton $\delta\phi$ in the interaction picture, $A_\zeta =H^4/(2c_s\dot\phi^2P_{,X})$, and the constants are related to the $P(X,\phi)$ function by \citep{Arroja:2009pd}
\beq
    &&\beta_1 = P_{,XX}\left(1-\frac{9}{8}c_s^2\right)-\dot{\phi}^2P_{,XXX}\left(1-\frac{3}{4}c_s^2\right)+\frac{1}{8}\frac{\dot{\phi}^6c_s^2}{P_{,X}}P^2_{,XXX}-\frac{1}{24}\dot{\phi}^4P_{,XXXX},\\\nonumber
    &&\beta_2 = -\frac{1}{2}P_{,XX}\left(1-\frac{3}{2}c_s^2\right)+\frac{1}{4}\dot{\phi}^2c_s^2P_{,XXX}, \qquad 
    \beta_3 = -\frac{c_s^2}{8}P_{,XX}.
\eeq
Such models source contact trispectra with the exact same structure as in the previous section;\footnote{That generic $P(X,\phi)$ models can be expressed in terms of the same $\dot{\sigma}^4$, $\dot{\sigma}^2(\partial_i\sigma)^2$ and $(\partial_i\sigma)^2(\partial_j\sigma)^2$ interactions found in the multi-field EFT is expected; these are the only leading-order terms permitted by symmetry.} their amplitudes can be recast in terms of the non-Gaussianity parameters
% \beq
%     \gnldotdot A_\zeta = -\frac{25}{768}\frac{H^4}{c_sP^4_{,X}}\beta_1, \qquad \gnldotdel A_\zeta = -\frac{325}{6912}\frac{H^4}{c_s^3P_{,X}^4}\beta_2, \qquad \gnldeldel A_\zeta = \frac{2575}{20736}\frac{H^4}{c_s^5P_{,X}^4}\beta_3,
% \eeq
\beq
    \gnldotdot A_\zeta = -\frac{25}{768}\frac{H^4c_s}{P^2_{,X}}\beta_1, \qquad \gnldotdel A_\zeta = \frac{325}{6912}\frac{H^4}{c_sP_{,X}^2}\beta_2, \qquad \gnldeldel A_\zeta = -\frac{2575}{20736}\frac{H^4}{c_s^3P_{,X}^2}\beta_3,
\eeq
reinterpreting the results of \citep{Arroja:2009pd} in terms of the templates presented in \citep{2015arXiv150200635S}. These could be large, even in the absence of a measurable bispectrum \citep{Arroja:2008ga,Arroja:2009pd,Chen:2006dfn,Mizuno:2010by,Chen:2009bc,Huang:2013oya,Chen:2013aj}.\footnote{$P(X,\phi)$ models also source exchange trispectra (\S\ref{subsec: primordial-exchange}) with $A_\zeta\tau_{\rm NL}\sim H^4/(P_{,X}^3c_s^7)$ (and $\tau_{\rm NL}\sim 1/c_s^4$ in DBI inflation). As discussed in \citep{Arroja:2009pd}, their functional form is complex, but does not violate the Maldacena consistency relation. The same is true for the ghost models considered below.} As such, the non-Gaussianity amplitudes directly trace the structure of the inflationary kinetic term.

In DBI inflation, the above equations simplify considerably leading to
% In DBI inflation, $\beta_1 = 1/(2c_s^7\dot{\phi}^2)$, $\beta_2=1/(4c_s^3\dot{\phi}^2)$, $\beta_3=-1/(8c_s\dot{\phi}^2$ with $A_\zeta = H^4/(2\dot{\phi}^2)$; this leads to
\beq\label{eq: gnl-DBI}
    \left.\gnldotdot\right|_{\rm DBI} = -\frac{25}{768}\frac{1}{c_s^4}, \quad \left.\gnldotdel\right|_{\rm DBI} =  \frac{325}{13824}\frac{1}{c_s^2}, \quad \left.\gnldeldel\right|_{\rm DBI} = \frac{2575}{82944}\frac{1}{c_s^2}  
\eeq
at lowest order in the inverse sound speed \citep{Planck:2019kim,Arroja:2009pd,Chen:2009bc,Chen:2013aj}. Assuming $c_s\ll1$, only the first term is relevant; this implies that the DBI action is a special case of the single-field EFT action \eqref{eq: EFTI-action} with $\tilde{c}_3$ and $M_4$ set by $c_s$. In more generic $P(X,\phi)$ models, the other terms can dominate, thus we conclude that single-field inflation \textit{can} generate the three new shapes described above, albeit with modified kinetic couplings.

\vskip 4pt
\paragraph{Ghost Inflation} An additional option is to consider single-field inflation in the presence of a ghost condensate \citep{Arkani-Hamed:2003juy}, allowing the inflaton field to acquire a time-dependent vacuum expectation value. This is realized within the EFT of inflation if the previously ignored higher-derivative term $(\partial^2_i\pi)^2$ dominates over the quadratic $(\partial_i\pi)^2$ interaction, such that the quadratic action becomes  \citep{Senatore:2009gt,Cheung:2007st}
\beq
    S_\pi^{\rm ghost} = \int d^4x\sqrt{-g}\Bigg\{\left[2M_2^4\dot{\pi}^2-\frac{\tilde{M}^2}{2}\frac{(\partial^2_i\pi)^2}{a^4}\right]+\text{interactions}\Bigg\}.
\eeq
In this limit, the Goldstone dispersion relation becomes quadratic (and hence strongly non-relativistic), and interactions involving spatial derivatives are correspondingly enhanced. This can generate a large equilateral bispectrum via the $\dot{\pi}(\partial_i\pi)^2$ and $\partial^2\pi(\partial_i\partial^2\pi)$ interactions with $f_{\rm NL}\sim A_\zeta^{1/5}$ \citep{Cheung:2007st,Senatore:2009gt}. This was largely ruled out by \textit{Planck} constraints on the three-point function \citep[e.g.,][]{Planck:2019kim}. If there exist some mechanism of suppressing cubic interactions such as a $\mathbb{Z}_2$ symmetry (under $\pi\to-\pi, t\to -t$), the above terms are forbidden, thus it is natural to expect that the ghost formalism will lead to large contact non-Gaussianity from the $(\partial_i\pi)^2(\partial_j\pi)^2$ term (and higher-order generalizations). This is described in \citep{Izumi:2010wm,Huang:2010ab,Izumi:2011di} with additional parity-odd contributions discussed in \citep{Cabass:2022rhr,Cabass:2022oap}. Though a detailed treatment is beyond the scope of this work (since the corresponding trispectrum involves templates beyond $(\partial_i\pi)^2(\partial_j\pi)^2$), the conclusion is that such derivative interactions can be large in the presence of ghosts, and only partially captured by the $\gnldeldel$ shape.

\vskip 4pt
\paragraph{Inflationary Templates} From the above discussion, it is clear that single- and multi-field models of inflation can source primordial trispectra from quartic inflationary interactions, the simplest of which involve the Goldstone mode $\pi$ or a generic light scalar field $\sigma$. For each interaction, one can compute the corresponding curvature trispectrum via an `in-in' (or Schwinger-Keldysch \citep[e.g.,][]{Chen:2017ryl}) formalism; this results in the following shapes \citep{Fergusson:2010gn,2015arXiv150200635S} (see also \citep{Bartolo:2013eka} for a extended set of templates):
\beq\label{eq: EFTI-shapes}
    \av{\zeta(\vk_1)\zeta(\vk_2)\zeta(\vk_3)\zeta(\vk_4)}'_c &\supset& \frac{221184}{25}\gnldotdot A_\zeta^3\frac{1}{k_1k_2k_3k_4k_T^5}\\\nonumber
    &&\,-\,\frac{27648}{325}\gnldotdel A_\zeta^3\left(\frac{k_T^2+3(k_3+k_4)k_T+12k_3k_4}{k_1k_2(k_3k_4)^3k_T^5}(\vk_3\cdot\vk_4)+\text{5 perms.}\right)\\\nonumber
    &&\,+\,\frac{165888}{2575}\gnldeldel A_\zeta^3\left(\frac{2k_T^4-2k_T^2\sum_i k_i^2+k_T\sum_i k_i^3+12k_1k_2k_3k_4}{(k_1k_2k_3k_4)^3k_T^5}\right.\\\nonumber
    &&\qquad\qquad\,\left.\times\,(\vk_1\cdot\vk_2)(\vk_3\cdot\vk_4)+\text{2 perms.}\right),
\eeq
for $k_T\equiv k_1+k_2+k_3+k_4$. This assumes $n_s\approx 1$ and normalizes the templates such that $T_\zeta = (216/25)A_\zeta^3/k^9$ for $k_i=k$ and $\vk_i\cdot\vk_j=-k^2/3$ if $i\neq j$.\footnote{For the remainder of this work, we will allow for $n_s\neq 1$ by replacing $A_\zeta$ by (suitably symmetrized) $k^3P_\zeta(k)$, as in \citep{2015arXiv150200635S}.} Though not immediately obvious, such shapes can be written in a contact form using Schwinger-type representations as integrals over conformal time $\tau$ (with $\tau=0$ encoding the end of inflation):
\begin{empheq}[box=\fbox]{align}\label{eq: EFTI-schwinger} 
    \av{\zeta(\vk_1)\zeta(\vk_2)\zeta(\vk_3)\zeta(\vk_4)}'_c &\supset \frac{9216}{25}\gnldotdot A_\zeta^3\int_{-\infty}^0d\tau\,\tau^4\left(\prod_{i}\frac{e^{k_i\tau}}{k_i}\right)\\\nonumber
    &\,-\,\frac{13824}{325}\gnldotdel A_\zeta^3\int_{-\infty}^0d\tau\,\tau^2\frac{(1-k_3\tau)(1-k_4\tau)}{k_1k_2(k_3k_4)^3}(\vk_3\cdot\vk_4)e^{\sum_i k_i\tau}+\text{5 perms.}\\\nonumber
    &\,+\,\frac{82944}{2575}\gnldeldel A_\zeta^3\int_{-\infty}^0d\tau\,\left(\prod_{i}\frac{(1-k_i\tau)e^{k_i\tau}}{k_i^3}\right)(\vk_1\cdot\vk_2)(\vk_3\cdot\vk_4)+\text{2 perms.}
\end{empheq}

Upon discretizing the $\tau$ integral,\footnote{This itself is non-trivial, and will be discussed in \S\ref{sec: optim}.} each shape is an sum of terms separable in $k_i$, analogous to the $\gnl$ shape. Although generated by independent mechanisms, these shapes may not always be physically distinguishable; for example, \citep{2015arXiv150200635S} found that $\gnldotdel$ is almost $99\%$ correlated with a linear combination of $\gnldotdot$ and $\gnldeldel$ for a \textit{Planck}-like experiment. This is confirmed by the Fisher forecasts presented in Appendix \ref{app: forecasts}. In \paperthree, we will place constraints on the various $g_{\rm NL}$ amplitudes both independently and in concert, following the work of \citep{2015arXiv150200635S,Planck:2015zfm,Planck:2019kim} for WMAP and \textit{Planck} (none of which include polarization). These analyses would not be possible without the above separability.

\subsection{Exchange Trispectra}\label{subsec: primordial-exchange}
\subsubsection{Local Interactions}

\noindent At lowest order, exchange trispectra involve an intermediary field and a pair of cubic interactions, as shown in Fig.\,\ref{fig: cartoon}. A simple manifestation is the local transformation of \eqref{eq: local-zeta-perturbation}, which induces a coupling of the form $\av{\fnl\zeta_{\rm G}^2(\vx_1)\times \fnl\zeta_{\rm G}^2(\vx_3)\times \zeta_{\rm G}(\vx_3)\times \zeta_{\rm G}(\vx_4)}$ at second order in $\fnl$. In Fourier-space, we find
\beq\label{eq: tauNL-shape}
    \boxed{\av{\zeta(\vk_1)\zeta(\vk_2)\zeta(\vk_3)\zeta(\vk_4)}'_c \supset  \taunl P_\zeta(k_1)P_\zeta(k_3)P_\zeta(K)+\text{11 perms.},}
\eeq
where $\vK\equiv \vk_1+\vk_2$ and $\taunl=\left(\frac{6}{5}\fnl\right)^2$. As for the $\gnl$-shape, this is roughly scale-invariant and explicitly separable; however, it features explicit dependence on $K$ (\textit{i.e.}\ the $s$-channel momentum). This shape peaks in the collapsed regime with $K\ll k_1,k_3$ and $k_1\approx k_2$, $k_3\approx k_4$, unlike for the contact trispectra, which are enhanced when all $k_i$ (or all-but-one $k_i$) are similar in magnitude. According to the consistency relations \citep[e.g.,][]{Maldacena:2002vr}, these shapes cannot be produced within canonical single-field inflation, making them a smoking gun for new primordial physics.

As in \S\ref{subsec: contact-local-theory}, the $\taunl$ shape can be generated if there are multiple light fields in inflation \citep[e.g.,][]{Suyama:2013nva}. A simple example is the curvaton scenario \eqref{eq: curvaton}, which yields 
\beq
    \taunl=\left(\frac{3}{2r_D}-2-r_D\right)^2,
\eeq
in the simplest adiabatic models with decay fraction $r_D$ \citep{Bartolo:2003jx,Sasaki:2006kq}. Furthermore, let us consider a model where the Gaussian inflationary fluctuations $\zeta_{\rm G}$ are modulated by an additional light scalar field, $\sigma(\vx)$, whose power spectrum matches that of $\zeta_{\rm G}$:
\beq
    \zeta_{\rm G}(\vx)\to\left[1+A_{\sigma}\sigma(\vx)\right]\zeta_{\rm G}(\vx),
\eeq
where $A_{\sigma}$ is some coupling strength. If $\sigma$ is uncorrelated with $\zeta_{\rm G}$ we will form a trispectrum with amplitude $\taunl =A^2_{\sigma}$, without forming a bispectrum. This demonstrates the \textit{Suyama-Yamaguchi} inequality \citep{Suyama:2007bg}, which relates squeezed bispectra and collapsed trispectra: $\taunl\geq \left(\frac{6}{5}\fnl\right)^2$. As shown in \citep{Smith:2011if,Rodriguez:2013tlz,Green:2023ids}, this holds under quite general conditions and can be related to positivity bounds. Notbaly, this is saturated in the curvaton model, as well as many other simple scenarios (such as the ekpyrotic set-up discussed above). In this series, we search directly for the $\taunl$ amplitude, which has been previously constrained using WMAP \citep{Komatsu:2010hc,Smidt:2010ra} and \textit{Planck} \citep{2014A&A...571A..24P,Feng:2015pva} temperature anisotropies, as well as \textit{Planck} polarization \citep{Marzouk:2022utf}, though rarely in combination with other shapes.

\subsubsection{Direction-Dependent Templates and Parity Sensitivity}\label{subsubsec: direction-templates}

\paragraph{Phenomenological Templates} There are many ways to generalize the local exchange trispectrum of \eqref{eq: tauNL-shape}, and thus constrain more complex inflationary interactions that are not described by the local and EFT of inflation shapes discussed above. In this section, we focus on the \textit{direction-dependent} templates introduced in \citep{Shiraishi:2013oqa,Shiraishi:2016mok}. For the bispectrum, these arise as a modification to \eqref{eq: fNL-shape}:
\beq\label{eq: direction-dependent-Bk}
    \av{\zeta(\vk_1)\zeta(\vk_2)\zeta(\vk_3)}'&\supset& \frac{6}{5}\sum_{n\geq 0}f_{\rm NL}^{n}P_\zeta(k_1)P_\zeta(k_2)\mathcal{L}_n(\hk_1\cdot\hk_2)+\text{2 perms.},
\eeq
where $\mathcal{L}_n$ is a Legendre polynomial and $f_{\rm NL}^n$ are the non-Gaussianity amplitudes (equivalent to $(5/6)c_n$ in the notation of \citep{Shiraishi:2013oqa}). Here, $n=0$ reproduces the local shape (with $f_{\rm NL}^0=\fnl$), whilst the higher-order terms allow for dependence on the angle between the short- and long-mode (noting that the bispectrum peaks when $k_1\ll k_2,k_3$). \citep{Shiraishi:2013oqa} constructed a similar set of templates for the trispectrum: 
\begin{empheq}[box=\fbox]{align}\label{eq: dn-even-def} \av{\zeta(\vk_1)\zeta(\vk_2)\zeta(\vk_3)\zeta(\vk_4)}'_c &\supset \frac{1}{6}\sum_{n\geq 0}\tau_{\rm NL}^{n, \rm even}\left[\mathcal{L}_n(\hk_1\cdot\hk_3)+(-1)^n\mathcal{L}_n(\hk_1\cdot\hK)+\mathcal{L}_n(\hk_3\cdot\hK)\right]\\\nonumber
    &\,\times\,P_\zeta(k_1)P_\zeta(k_3)P_\zeta(K)+\text{23 perms.},
\end{empheq}
depending on the $\tau_{\rm NL}^{n,\rm even}$ amplitudes (equivalent to $6d_n^{\rm even}$ in the former work).\footnote{Strictly, we generalize the previous template by adding an extra factor $(-1)^{n}$ in the second term, which allows the template to more fully capture the behavior of various inflationary models, including the parity-breaking $\gamma F\tilde{F}$ gauge-field coupling. We thank Maresuke Shiraishi for suggesting this.} As before, $n=0$ reproduces the local model (with $\tau_{\rm NL}^0 = \taunl$), whilst larger $n$ encodes the geometry of the collapsed tetrahedra, specified by the angles between the two short legs $\vk_1$ and $\vk_3$ and the diagonal $\vK$. An analogous template can be used to represent the collapsed limits of \textit{parity-violating} models of inflation \citep{Shiraishi:2016mok}:
\begin{empheq}[box=\fbox]{align}\label{eq: dn-odd-def}
    \av{\zeta(\vk_1)\zeta(\vk_2)\zeta(\vk_3)\zeta(\vk_4)}'_c &\supset -\frac{i}{6}\sum_{n\geq 0}\tau_{\rm NL}^{n, \rm odd}\left[\mathcal{L}_n(\hk_1\cdot\hk_3)+(-1)^n\mathcal{L}_n(\hk_1\cdot\hK)+\mathcal{L}_n(\hk_3\cdot\hK)\right](\hk_1\times\hk_3\cdot\hK)\\\nonumber
    &\,\times\,P_\zeta(k_1)P_\zeta(k_3)P_\zeta(K)+\text{23 perms.}
\end{empheq}
(with $\tau_{\rm NL}^{n,\rm odd}\equiv 6\,d_n^{\rm odd}$ previously); here, the parity asymmetry (which has been realized in a number of inflationary models, including gauge fields and chiral gravitational wave exchange \citep[e.g.,][]{Lue:1998mq,Gluscevic:2010vv,CyrilCS,Shiraishi:2013kxa,Bartolo:2014hwa,Bartolo:2015dga,Shiraishi:2016mok,Moretti:2024fzb,Bartolo:2020gsh,Salvarese,Bartolo:2018elp,Bartolo:2017szm}) corresponds to a sign-flip under $\vk_i\to -\vk_i,\vK\to-\vK$ ensured by the triple product. 

Although these template have found significant theoretical use \citep[e.g.,][]{Bartolo:2015dga,Bartolo:2014hwa,Shiraishi:2013vja,Shiraishi:2013oqa,Naruko:2014bxa,Bartolo:2012sd,Dimastrogiovanni:2010sm,Shiraishi:2016mok,Shiraishi:2016hjd}, they do not capture all possible angular dependence of the collapsed trispectrum, since there are no terms involving all three of $\hk_1$, $\hk_3$ and $\hK$ (e.g., $\mathcal{L}_2(\hk_1\cdot\hK)\mathcal{L}_2(\hk_3\cdot\hK)$). To this end, we introduce a new suite of direction-dependent trispectrum templates:
\begin{empheq}[box=\fbox]{align}\label{eq: dnnn-def}        
    \av{\zeta(\vk_1)\zeta(\vk_2)\zeta(\vk_3)\zeta(\vk_4)}'_c &\supset \frac{1}{2}\sum_{n_1n_3n}\tau_{\rm NL}^{n_1n_3n}\left[\sum_{m_1m_3m}\tj{n_1}{n_3}{n}{m_1}{m_3}{m}Y_{n_1m_1}(\hk_1)Y_{n_3m_3}(\hk_3)Y_{nm}(\hK)\right]\\\nonumber
    &\,\times\,P_\zeta(k_1)P_\zeta(k_3)P_\zeta(K)+\text{23 perms.},
\end{empheq}
specified by the $\tau_{\rm NL}^{n_1n_3n}$ amplitudes for $n,n_1,n_3\geq 0$ satisfying triangle conditions. Here, the angular factor encodes the most general isotropic angular dependence (via tripolar spherical harmonics \citep{Khersonskii:1988krb}), utilizing the Wigner $3j$ symbol indicated by parentheses. The template amplitudes can be explicitly defined via the relation
\beq
    \tau_{\rm NL}^{n_1n_3n} &\equiv& \frac{1}{4}\int d\hk_1\,d\hk_3\,d\hK\,\sum_{m_1m_3m}\tj{n_1}{n_3}{n}{m_1}{m_3}{m}Y^*_{n_1m_1}(\hk_1)Y^*_{n_3m_3}(\hk_3)Y^*_{nm}(\hK)\\\nonumber
    &&\quad\,\times\,\left(\lim_{|\vk_1+\vk_2|\to0}\frac{\av{\zeta(\vk_1)\zeta(\vk_2)\zeta(\vk_3)\zeta(\vk_4)}'_c}{P_\zeta(k_1)P_\zeta(k_3)P_\zeta(K)}\right).
\eeq
Furthermore interchange symmetry implies, $\tau_{\rm NL}^{n_3n_1n}=(-1)^{n_1+n_3}\tau_{\rm NL}^{n_1n_3n}$, which reduces the number of free coefficients. It is straightforward to show that a single $\tau_{\rm NL}^{N, \rm even}$ term in \eqref{eq: dn-even-def} sources the following coefficients:
% \beq\label{eq: dn-even-dnnn}
%     \tau_{\rm NL}^{nn0} = (-1)^n\tau_{\rm NL}^{n0n} = \tau_{\rm NL}^{0nn} = (-1)^n\frac{(4\pi)^{3/2}}{\sqrt{2n+1}}\frac{\tau_{\rm NL}^{n, \rm even}}{3},
% \eeq
\beq\label{eq: dn-even-dnnn}
    \left.\tau_{\rm NL}^{n_1n_3n}\right|_{N,\rm even} = \frac{1}{3}(-1)^N\frac{(4\pi)^{3/2}}{\sqrt{2N+1}}\tau_{\rm NL}^{N, \rm even}\bigg[\delta^{\rm K}_{n_1N}\delta^{\rm K}_{n_3N}\delta^{\rm K}_{n0}+(-1)^N\delta^{\rm K}_{n_1N}\delta^{\rm K}_{n_30}\delta^{\rm K}_{nN}+\delta^{\rm K}_{n_10}\delta^{\rm K}_{n_3N}\delta^{\rm K}_{nN}\bigg].
\eeq
Similarly, one may compute \resub{the $\tau_{\rm NL}^{n_1n_3n}$ coefficients} sourced by a single $\tau_{\rm NL}^{N, \rm odd}$ contribution to \eqref{eq: dn-odd-def}. As derived in Appendix \ref{app: direc-simp}, we find
% \beq\label{eq: dn-odd-dnnn}
%     \tau_{\rm NL}^{nn'1} = (-1)^{N}\tau_{\rm NL}^{n1n'} = \tau_{\rm NL}^{1nn'} &=& \sqrt{2}(4\pi)^{3/2}\sqrt{(2n+1)(2n'+1)}(-1)^N\tjo{n}{1}{N}\tjo{n'}{1}{N}\begin{Bmatrix}n & 1 & n'\\ 1 & N & 1\end{Bmatrix}\frac{\tau_{\rm NL}^{n, \rm odd}}{3}
% \eeq
% \beq\label{eq: dn-odd-dnnn}
%     \tau_{\rm NL}^{n_1n_3n} &=& \frac{1}{3}\sum_{mm'}\bigg[\delta^{\rm K}_{n_1m}\delta^{\rm K}_{n_3m'}\delta^{\rm K}_{n1}+(-1)^N\delta^{\rm K}_{n_1m}\delta^{\rm K}_{n_31}\delta^{\rm K}_{nm'}+\delta^{\rm K}_{n_11}\delta^{\rm K}_{n_3m}\delta^{\rm K}_{nm'}\bigg]\\\nonumber
%     &&\,\times\,\sqrt{2}(4\pi)^{3/2}\sqrt{(2m+1)(2m'+1)}(-1)^N\tjo{m}{1}{N}\tjo{m'}{1}{N}\begin{Bmatrix}n & 1 & m'\\ 1 & N & 1\end{Bmatrix}\tau_{\rm NL}^{N, \rm odd}
% \eeq
\beq\label{eq: dn-odd-dnnn}
    \left.\tau_{\rm NL}^{n_1n_3n}\right|_{N,\rm odd}&=&\frac{\sqrt{2}}{3}(4\pi)^{3/2}(-1)^N\tau_{\rm NL}^{N,\rm odd}\sum_{m,m'=N\pm 1}\sqrt{(2m+1)(2m'+1)}\tjo{1}{m}{N}\tjo{1}{m'}{N}\begin{Bmatrix} 1 & m & m'\\ N & 1 & 1\end{Bmatrix}\nonumber\\
    &&\,\times\,\bigg[\delta^{\rm K}_{n_1m}\delta^{\rm K}_{n_3m'}\delta^{\rm K}_{n1}+(-1)^{N}\delta^{\rm K}_{n_1m}\delta^{\rm K}_{n_31}\delta^{\rm K}_{nm'}+\delta^{\rm K}_{n_11}\delta^{\rm K}_{n_3m}\delta^{\rm K}_{nm'}\bigg],
\eeq
where the curly brackets indicate a Wigner $6j$ symbol.
% where $m,m'=N\pm 1$, and we have used the relation
% \beq
%     i(\hk_1\times\hk_3\cdot\hK) &=& \sqrt{6}\left(\frac{4\pi}{3}\right)^{3/2}\sum_{m_1m_3m}\tj{1}{1}{1}{m_1}{m_3}{m}Y_{1m_1}(\hk_1)Y_{1m_3}(\hk_3)Y_{1m}(\hK),
% \eeq
% polypolar harmonic orthonormality, and the definition of the Wigner $6j$ symbol. 
Here, the local model of \eqref{eq: tauNL-shape} corresponds to $\tau_{\rm NL}^{000}=(4\pi)^{3/2}\taunl$ and the simplest parity-odd shape is specified by $\tau_{\rm NL}^{111}=(-\sqrt{2}/3)(4\pi)^{3/2}\tau_{\rm NL}^{0,\rm odd}$. Unlike the $\tau_{\rm NL}^{n,\rm even/odd}$ templates, the \eqref{eq: dnnn-def} trispectrum is explicitly separable in $\vk_{1,2}$ and $\vk_{3,4}$; this will be used in \S\ref{sec: estimators} to facilitate efficient trispectrum estimators.

The $\tau_{\rm NL}^{n_1n_3n}$ bifurcate into two sets: even $n_1+n_3+n$, which encode parity-even trispectra, and odd $n_1+n_3+n$, which describe parity-odd trispectra. Detecting any of the second set would be a strong indicator of parity-violating physics in primordial interactions.\footnote{Such physics can also source contact trispectra \citep[e.g.,][]{Cabass:2022rhr,Coulton:2023oug}, though their physical form is often complex.} This distinction will also be arise in the CMB trispectrum (defined in \S\ref{sec: estimators}); due to statistical isotropy, parity-even (parity-odd) physics appears in correlators with even (odd) $\ell_1+\ell_2+\ell_3+\ell_4$. Overall, the $\tau_{\rm NL}^{n_1n_3n}$ coefficients completely describe (isotropic) direction-dependent collapsed trispectrum, and, to our knowledge, have not been previously estimated from data (except for measurements of $\tau_{\rm NL}^{0,\rm odd},\tau_{\rm NL}^{1,\rm odd}$ from the parity-odd four-point function of \textit{Planck} and galaxy surveys \citep{Philcox:2022hkh,PhilcoxCMB,Philcox:2023ypl}). 

\vskip 4pt
\paragraph{Gauge Fields} Despite their phenomenological nature, the above templates can be generated by various ultraviolet models of inflation, and are often found to accompany some form of anisotropy in the power spectrum \citep[e.g.,][]{Bartolo:2015dga,Bartolo:2014hwa,Shiraishi:2013vja,Shiraishi:2013oqa,Naruko:2014bxa,Bartolo:2012sd,Dimastrogiovanni:2010sm,Shiraishi:2016mok,Bartolo:2017szm,Bartolo:2018elp}. They may also be sourced by early-Universe magnetic fields (both helical and non-helical) \citep[e.g.,][]{Shiraishi:2012sn,Shiraishi:2013vha,Shiraishi:2013vja,Planck:2015zrl,Shiraishi:2012rm,Caprini:2014mja,Shaw:2009nf,Trivedi:2011vt,Trivedi:2013wqa} and the exchange of chiral gravitational waves \citep[e.g.,][]{CyrilCS,Bartolo:2020gsh,Moretti:2024fzb,Salvarese}; these can have different phenomenology and would be an interesting topic for future study. 

As described in \citep{Bartolo:2015dga}, one sources of direction-dependent trispectra is a coupling between the pseudoscalar inflaton $\phi$ and a $U(1)$ gauge field $A_{\mu}$. This is described by the action:
\beq\label{eq: action-gauge}
    S_{\phi,A_\mu} &=& \int d^4x\sqrt{-g}\bigg\{-\frac{1}{2}(\partial_\mu\phi)^2-V(\phi)+I^2(\phi)\left(-\frac{1}{4}F^{\mu\nu}F_{\mu\nu}+\frac{\gamma}{4}\tilde{F}^{\mu\nu}F_{\mu\nu}\right)\bigg\},
\eeq
where $I(\phi)$ sets the the interaction strength, with $I(\phi)\propto a^{-2}$ inducing nearly scale-invariant correlators. Here, $F_{\mu\nu}\equiv\partial_\mu A_\nu-\partial_\nu A_\mu$ is the electromagnetic tensor and $\tilde{F}^{\mu\nu}\equiv\varepsilon^{\mu\nu\alpha\beta}F_{\alpha\beta}$ is its Hodge dual, which appears in a Chern-Simons interaction with coupling parameter $\gamma$.\footnote{The $\gamma=0$ case has been considered often in the literature \citep[e.g.,][]{Maleknejad:2012fw,Watanabe:2010fh,Dulaney:2010sq,Fujita:2013qxa}. Furthermore, in the limit of $I^2(\phi)\to 1$, $\gamma I^2(\phi)\to\phi$, \eqref{eq: action-gauge} describes axion inflation (with friction ensuring slow-roll), though the stability of such models is unclear \citep[e.g.,][]{Peloso:2022ovc,vonEckardstein:2023gwk}} Cubic couplings of the form $\delta\phi A_\mu A^\mu$ can source a variety of correlators, including an anisotropic power spectrum and a direction-dependent bispectrum. Furthermore, a loop diagram generates trispectra of the form (combining results from \citep{Shiraishi:2013oqa,Shiraishi:2013vja,Bartolo:2015dga,Shiraishi:2016mok}):
\beq\label{eq: gauge-field-Tk}
    \left.\av{\zeta(\vk_1)\zeta(\vk_2)\zeta(\vk_3)\zeta(\vk_4)}_c' \right|_{\gamma=0}&\supset& 24N_e^2|g_\ast|f^2(\gamma)\bigg[\mu_{13}^2+\mu_1^2+\mu_3^2-\mu_{13}\mu_1\mu_3\bigg]P_\zeta(k_1)P_\zeta(k_3)P_\zeta(K)+\text{23 perms.}\\\nonumber
    \left.\av{\zeta(\vk_1)\zeta(\vk_2)\zeta(\vk_3)\zeta(\vk_4)}_c' \right|_{\gamma\gg 1}&\supset& 24N_e^2|g_\ast|f^2(\gamma)\\\nonumber
    &&\,\times\,\bigg[\big\{-\mu_{13}+(\mu_1-\mu_3)(1-\mu_{13})-\mu_1\mu_3\big\}-i\left(\mu_{13}-\mu_1+\mu_3-1\right)\left(\hk_1\times\hk_3\cdot\hK\right)\bigg]\\\nonumber
    &&\,\times\,P_\zeta(k_1)P_\zeta(k_3)P_\zeta(K)+\text{23 perms.}+\left.\av{\zeta(\vk_1)\zeta(\vk_2)\zeta(\vk_3)\zeta(\vk_4)}_c' \right|_{\gamma=0}
\eeq
defining $\mu_{ij} \equiv \hk_i\cdot\hk_j$ and $\mu_i\equiv \hk_i\cdot\hK$. Here, we have assumed $N_e\approx 60$ $e$-folds of inflation with $I(\phi)\propto a^{-2}$ (but constant after inflation). In the penultimate line we separate out the imaginary parity-odd contribution, assuming $\gamma>0$ wlog. The trispectrum amplitude is controlled by $g_\ast\ll1$, which sets the anisotropy in the two-point function and encodes microphysical parameters with
\beq   
    g_\ast \approx -\frac{48N_e^2}{\epsilon}\frac{\rho_E^{\rm vev}}{\rho_\phi}f(\gamma), \qquad f(\gamma)=\begin{cases}1 & (\gamma=0)\\
    \frac{e^{4\pi|\gamma|}}{32\pi|\gamma|^3}& (\gamma\gg 1)\end{cases}.
\eeq
Here, $\epsilon$ is the slow-roll parameter, $\gamma$ is the Chern-Simons coupling strength in \eqref{eq: action-gauge}, and $\rho_E^{\rm vev}/\rho_\phi$ is the fractional energy density in the background electric field component of $A_\mu$ during inflation (which has characteristic value $3/(4\pi^3)(H/M_{\rm Pl})^{2}f(\gamma)(N_{\rm tot}-N_e)$ \citep{Bartolo:2015dga}, and can be generated from infrared stochastic fluctuations \citep{Bartolo:2012sd,Bartolo:2014xfa}). 

The exchange trispectrum of \eqref{eq: gauge-field-Tk} projects directly onto our non-Gaussianity parameters. Explicitly, we find the $\gamma=0$ coefficients:
\beq
    &&\tau_{\rm NL}^{000} = \frac{128}{3}\mathcal{A}(\gamma), \qquad \tau_{\rm NL}^{220}=\tau_{\rm NL}^{202}=\tau_{\rm NL}^{022}=\frac{64}{3\sqrt{5}}\mathcal{A}(\gamma), \qquad \qquad \tau_{\rm NL}^{222}=\frac{16\sqrt{14}}{45}\mathcal{A}(\gamma),
\eeq
defining $\mathcal{A}(\gamma) \equiv (4\pi)^{3/2}N_e^2|g_\ast|f^2(\gamma)$. For $\gamma\gg1$, additional terms are sourced:
\beq
    &&\tau_{\rm NL}^{110} = -\tau_{\rm NL}^{101}=\tau_{\rm NL}^{011}=\frac{64}{\sqrt{3}}\mathcal{A}(\gamma), \qquad \tau_{\rm NL}^{112}=-\tau_{\rm NL}^{121}=\tau_{\rm NL}^{211}=-16\sqrt{\frac{2}{3}}\mathcal{A}(\gamma)\\\nonumber
    &&\tau_{\rm NL}^{111} = 16\sqrt{2}\mathcal{A}(\gamma), \qquad \tau_{\rm NL}^{221} = -\tau_{\rm NL}^{212}=\tau_{\rm NL}^{122}=16\sqrt{\frac{2}{5}}\mathcal{A}(\gamma),
\eeq
where the top and bottom lines are parity-even and parity-odd respectively. This can be compared to the approximate forms used in \citep{Shiraishi:2016mok}:
\beq
    \tau_{\rm NL}^{0, \rm even} = \frac{1}{2}\tau_{\rm NL}^{2,\rm even} \approx  0.89\times 48N_e^2|g_\ast|f^2(\gamma), \quad \tau_{\rm NL}^{1,\rm even} \approx \delta_{\gamma\gg 1}\times(-0.7)\times 48N_e^2|g_\ast|f^2(\gamma),
\eeq
where higher-order angular components were replaced with their averages, to avoid the need for more complex templates \citep{Bartolo:2015dga}.\footnote{The $\tau_{\rm NL}^{1,\rm even}$ piece has not been previously considered in the literature, due to the differing sign convention in the templates.} If the Chern-Simons coupling is large (such that $f(\gamma)\gg 1$), the constraining power of the trispectrum far exceeds that of the power spectrum or bispectrum; this is due to an enhancement by $[N_ef(\gamma)]^n$ relative to the $(4-n)$ point function.\footnote{Explicitly, the three-point function can be written in terms of \eqref{eq: direction-dependent-Bk} with $f_{\rm NL}^0 = 2f_{\rm NL}^2 = (40/9)N_e|g_\ast|f(\gamma)$.} Though the templates themselves are generic, it is important to note that the relations between coefficients will depend on the model in question, for example the time-dependence of the $I^2(\phi)$ coupling.

\vskip 4pt
\paragraph{Solid Inflation} Another model capable of generating direction-dependent trispectra is `solid inflation' \citep{Endlich:2012pz,Gruzinov:2004ty}. In this formalism, the primordial Universe contains an isotropic combination of three anisotropic (in field-space) scalar fields, $\hat{\phi}^I$, which can have interesting phenomenological consequences, such as extended anisotropic periods of inflation and blue-tilted tensor spectra \citep[e.g.,][]{Endlich:2013jia,Bartolo:2014xfa}. Formally, the field-dependent action for solid inflation can be written
\beq
    S_{\rm solid} &=& \int d^4x\sqrt{-g}\,F[\hat{X},\hat{Y},\hat{Z}]+\cdots\\\nonumber
    \hat{X} &=& \mathrm{tr}\,\hat{B}, \quad \hat{X}^2\hat{Y}=\mathrm{tr}\,\hat{B}^2, \quad \hat{X}^3\hat{Z} = \mathrm{tr}\,\hat{B}^3,
\eeq
where $\hat{B}^{IJ} = g^{\mu\nu}\hat{\phi}^I_\mu\hat{\phi}^J_\nu$ and $F$ is some scalar function. This arises from SO(3) invariance and shift symmetry, assuming that the vacuum expectations of the three fields are orthonormal, such that the background state is isotropic and the inflationary `clock' is carried by the metric (giving a limit not described by the EFT of inflation). 

At lowest order in slow-roll, solid inflation features an anisotropic sound-speed for the phonon excitation modes (defined by $\pi^I = \hat{\phi}^I-\av{\hat{\phi}^I}$) with longitudinal and tangential components
\beq
    c_L^2 \approx \frac{1}{3}+\frac{8}{9}\frac{F_{,Y}+F_{,Z}}{XF_{,X}}, \qquad c_T^2 \approx \frac{3}{4}(1+c_L^2)
\eeq
(assuming $XF_{,XX}\ll F_{,X}$), whence the quadratic action can be written \citep{Endlich:2012pz}
\beq
    S_{\rm solid} &=& \int d^4x\sqrt{-g}\,\left(-\frac{1}{3}F_{,X}X\right)\left[\dot{\pi_i}^2-c_T^2(\partial_i\pi_j)^2-(c_L^2-c_T^2)(\partial_i\pi_i)^2\right]+\cdots,
\eeq
which sources the power spectrum $k^3P_\zeta(k) = H^2/(4M_{\rm Pl}^2\epsilon c_L^5)$. At third-order, the Lagrangian is schematically 
\beq
    \mathcal{L}_3\supset M^2_{\rm Pl}a^3H^2F_{,Y}/F\left[(\mathrm{tr}\,\pi_{i,j})^3+\cdots\right],
\eeq
where the ellipses denote other terms of the same order. As noted in \citep{Bartolo:2014xfa}, this generates bispectra matching \eqref{eq: direction-dependent-Bk} with $f_{\rm NL}^{2} = (50/3)F_{,Y}/(9F\epsilon c_L^2) \gg f_{\rm NL}^0$, featuring a large enhancement for small sound speeds (as in DBI models, but with an extra $1/\epsilon$ enhancement due to the differing amplitudes of third-order terms). Naturally, such models will also yield direction-dependent trispectra. Whilst the full trispectrum calculation has not been performed (to our knowledge), we can bound the collapsed limit of the trispectrum via the generalized Suyama-Yamaguchi relation \citep{Smith:2011if}:
\beq
    \left.\av{\zeta(\vk_1)\zeta(\vk_2)\zeta(\vk_3)\zeta(\vk_4)}'_c\right|_{K\ll k_1,k_3} \gtrsim \int_{\vK'\vK''}\frac{\partial\av{\zeta(\vk_1)\zeta(\vk_2)}}{\partial\zeta_L(\vK')}\av{\zeta_L(\vK')\zeta_L(\vK'')}\frac{\partial\av{\zeta(\vk_3)\zeta(\vk_4)}}{\partial\zeta_L(\vK'')},
\eeq
where $\zeta_L$ is a long-wavelength fluctuation, which can be treated as a background mode.\footnote{This is strictly a lower limit; as shown in \S\ref{subsubsec: collider}, all polarizations of the exchange particle contribute to the trispectrum, but only the longitudinal mode contributes to the bispectrum, thus we are neglecting additional terms in this approximation.} Following \citep{Bartolo:2014xfa}, the action of a long mode is equivalent to a modification of the longitudinal sound-speed, with $c_L^2\to c_L^2+8F_{,Y}/(9F\epsilon)(1-3\cos^2\theta)$ for relative angle $\theta$, leading to 
\beq
    &&\left.\av{\zeta(\vk_1)\zeta(\vk_2)\zeta(\vk_3)\zeta(\vk_4)}'_c\right|_{K\ll k_1,k_2} \gtrsim \frac{\partial\av{\zeta(\vk_1)\zeta(\vk_2)}}{\partial c_L^2}\frac{\partial\av{\zeta(\vk_3)\zeta(\vk_4)}}{\partial c_L^2}P_\zeta(K)\left(\frac{\partial c_L^2}{\partial\zeta_L}\right)^2\\\nonumber
    &&\Rightarrow \av{\zeta(\vk_1)\zeta(\vk_2)\zeta(\vk_3)\zeta(\vk_4)}'_c \gtrsim \frac{1}{8}\left(\frac{40}{9}\frac{F_{,Y}}{F}\frac{1}{\epsilon c_L^2}\right)^2\left[\mathcal{L}_2(\hk_1\cdot\hK)\mathcal{L}_2(\hk_3\cdot\hK)\right]P_\zeta(k_1)P_\zeta(k_3)P_\zeta(K)+\text{23 perms.}
\eeq
with a saturation of the Suyama-Yamaguchi bound. This does not project well onto the $\tau_{\rm NL}^{n, \rm even}$ templates of \citep{Shiraishi:2013oqa}, but can be easily expressed in our generalized $\tau_{\rm NL}^{n_1n_3n}$ basis with non-zero coefficients
\beq
    \tau_{\rm NL}^{220} &=& -7\sqrt{\frac{7}{10}}\tau_{\rm NL}^{222} =\frac{1}{3}\sqrt{\frac{7}{2}}\tau_{\rm NL}^{224} \gtrsim (4\pi)^{3/2}\frac{16\sqrt{5}}{81}\left(\frac{F_{,Y}}{F}\frac{1}{\epsilon c_L^2}\right)^2.
\eeq
As such, we can measure the anisotropic sound-speed $c_L^2$ from the direction-dependent trispectrum; this is analogous to using $\tau^{\rm loc}_{\rm NL}$ to constrain $f^{\rm loc}_{\rm NL}$, but for the anisotropic $f_{\rm NL}^2$ mode of the direction-dependent bispectrum. Since $c_L\gg c_T$, this signature is almost uncorrelated with the standard local template, \textit{i.e.}\ it is not indirectly constrained by $\taunl$ measurements.

\subsubsection{Massive Spinning Particles}\label{subsubsec: collider}
\noindent
The final set of primordial templates we will consider are motivated by the `cosmological collider' program \citep[e.g.,][]{Arkani-Hamed:2015bza,Lee:2016vti,Chen:2009zp,Flauger:2016idt}, which links curvature correlators to inflationary particle interactions. Whilst there exists a huge wealth of literature discussing the theoretical and computational aspects of this formalism \citep[e.g.,][]{Chen:2009zp,Arkani-Hamed:2015bza,Lee:2016vti,Flauger:2016idt,Wang:2022eop,Pimentel:2022fsc,Kumar:2019ebj,Reece:2022soh,Arkani-Hamed:2015bza,Alexander:2019vtb,Jazayeri:2023xcj,McCulloch:2024hiz,Liu:2019fag,Meerburg:2016zdz,Wang:2019gbi,Tong:2022cdz,Sohn:2024xzd,Pimentel:2022fsc,Cabass:2024wob,Chen:2016uwp,Chen:2018xck,Lu:2019tjj,Wang:2020ioa,Bodas:2020yho,Jazayeri:2022kjy,Kim:2019wjo,Lu:2021wxu,Cui:2021iie,Qin:2022lva,Werth:2023pfl,Chen:2022vzh,Xianyu:2023ytd,Pinol:2023oux,Chakraborty:2023qbp,Craig:2024qgy,Yin:2023jlv,Baumann:2011nk,Assassi:2012zq,Baumann:2017jvh,Arkani-Hamed:2018kmz,Cabass:2022rhr,Bordin:2018pca,Bordin:2019tyb,Jazayeri:2023kji,Chen:2018sce,Green:2023ids,Noumi:2012vr}, there has been little connection to observational data (apart from a number of forecasts \citep[e.g.,][]{Bordin:2019tyb,Meerburg:2016zdz,Floss:2022grj,Lee:2020ebj,MoradinezhadDizgah:2018ssw}) until very recently (see \citep{DuasoPueyo:2023kyh,Sohn:2024xzd,Cabass:2024wob} for some exciting recent work). These models are of particular interest here, since the resulting $\zeta$ correlators can display unique signatures in the squeezed and collapsed limits, such as angular dependence (akin to that of \S\ref{subsubsec: direction-templates}) and oscillatory features, neither of which can be sourced in the vanilla EFT of inflation models (\S\ref{subsubsec: EFTI}), due to the assumed symmetries of the inflationary background \citep{Maldacena:2002vr}. Below, we briefly outline the underlying concepts and define associated a set of collider trispectrum templates.

Much like scattering amplitudes can be used to probe new particles in terrestrial colliders, inflationary correlators can be used to infer the existence of new inflationary fields in the cosmological collider. \resub{This primarily involves} exchange diagrams, where, heuristically speaking, a pair of inflatons (or some other light scalar sourcing curvature perturbations) creates a new particle, $\sigma$, which then decays into another pair of inflatons.\footnote{This picture is at best schematic; the exchange field can also interact with itself and then couple linearly to the Goldstone mode \citep[e.g.,][]{Chen:2009zp}. Moreover, there is no notion of temporality; in the in-in picture, we evaluate all external fields at the end of inflation.} The physics of this interaction are defined by three properties: (1) the mass, $m$, of the exchange particle $\sigma$; (2) the (bosonic) spin, $s$, of the exchange particle $\sigma$; (3) the symmetries of the inflationary background (e.g., de Sitter without boost breaking).\footnote{In this work, we assume (approximate) conformal invariance, setting the propagation speed of the inflaton to $c_\pi = 1$ as in \citep{Arkani-Hamed:2015bza}. When $c_\pi\ll 1$, the results are phenomenologically similar, though the allowed level of non-Gaussianity can be much larger \citep{Lee:2016vti} (and the templates come with a different complex phase).} The production rate of $\sigma$ is set by the Boltzmann factor ($\sim e^{-m\pi/H}$); to avoid undue suppression, we usually consider masses comparable to the Hubble scale, which can be made natural through loop corrections and supersymmetry \citep[e.g.,][]{Meerburg:2016zdz}. There are four main regimes of interest \citep[e.g.,][]{Lee:2016vti}:
\begin{itemize}
    \item \textbf{Heavy} (principal series): $m^2/H^2 > (s-1/2)^2$ (or $m^2/H^2>9/4$ for $s=0$).\footnote{The differences between $s=0$ and $s>0$ arise since we assume scalars to be minimally coupled, not conformally coupled (whence $m=\sqrt{2}H$). The $s=0$ regime is often known as `quasi-single-field' inflation \citep[e.g.,][]{Chen:2009we,Chen:2009bc,Chen:2009zp,McAneny:2019epy,Noumi:2012vr,Sefusatti:2012ye}.} This sources correlators with characteristic oscillatory features due to particle production. These are parametrized by $\mu_s\equiv \sqrt{m^2/H^2-(s-1/2)^2}$ with $\mu_s>0$ (with $\mu_0 \equiv \sqrt{m^2/H^2-9/4}$).
    \item \textbf{Intermediate} (complementary series): $s(s-1) < m^2/H^2 < (s-1/2)^2$ (or $0< m^2/H^2<9/4$ for $s=0$). This sources correlators with enhanced squeezed limits relative to the massless case, with local shapes recovered for $m=s=0$. These are parametrized by $\nu_s\equiv \sqrt{(s-1/2)^2-m^2/H^2}=i\mu_s$ with $\nu_s\in(0,1/2)$ (and $\nu_0 \equiv \sqrt{9/4-m^2/H^2}$, with $\nu_0\in(0,3/2)$).
    \item \textbf{Conformal}: $m^2 = s(s-1)H^2$ (or $m^2=9H^2/4$ for $s=0$). This lies between the heavy and intermediate regimes, and is specified by $\nu_s = \mu_s = 0$.
    \item \textbf{Light}: $m^2/H^2 < s(s-1)$. This violates the \textit{Higuchi bound} (or unitarity), but can be realized by breaking special conformal transform symmetries \citep{Higuchi:1986py,Bordin:2018pca}. This is relevant only for $s \geq 2$, and parametrized by $\nu_s\in(1/2,s-1/2)$.\footnote{Note that some previous authors have used $\nu_s=\nu_0$ for light particles (replacing $(s-1/2)^2 \to 9/4$). This simply corresponds to changing the mass parameter, which is allowed in spaces with broken boost symmetries.}
\end{itemize}
Other scenarios are also possible, such as discrete partially massless states \citep{Baumann:2017jvh} and supersymmetric
fermion contributions \citep{Alexander:2019vtb}; these are beyond the scope of this work. 

\vskip 4pt
\paragraph{Fields with Mass \& Spin} To understand the cosmological collider trispectra, it is useful to first consider the exchange particles. In de Sitter space, a spin-$s$ field with $s$ polarization directions can be written\footnote{We assume all indices of $\sigma$ are spatial; only these are needed in the inflationary EFT with de Sitter isometries \citep[e.g.,][]{Lee:2016vti}.}
\beq
    \sigma_{i_1\cdots i_s}(\vk,\tau) \equiv \sum_{\lambda=-s}^s\sigma^{(\lambda)}_{i_1\cdots i_s}(\vk,\tau) =\sum_{\lambda=-s}^s\sigma^{(\lambda)}(\vk,\tau)\varepsilon^{(\lambda)}_{i_1\cdots i_s}(\hk)
\eeq
where we sum over the $(2s+1)$ helicity states with polarization tensors $\varepsilon^{(\lambda)}$ (where $\lambda$ represents the angular momentum component around $\hk$). These satisfy the relations \citep[e.g.,][]{Lee:2016vti,Baumann:2017jvh}
\beq\label{eq: polarization-tensors}
    \hat{q}^{i_1}\cdots \hat{q}^{i_s}\varepsilon^{(\lambda)}_{i_1\cdots i_s}(\hk)\propto e^{i\lambda\varphi}\mathcal{L}^\lambda_s(\cos\vartheta), \qquad \hat{n}^{i_1}\cdots \hat{n}^{i_s}\varepsilon^{(s)}_{i_1\cdots i_s}(\hk)=e^{is\varphi}, \qquad \hat{k}^{i_1}\cdots \hat{k}^{i_n}\varepsilon^{(\lambda)}_{i_1\cdots i_s}(\hk) = 0\quad (n<s-\lambda),
\eeq
where $(\vartheta,\varphi)$ are the polar and azimuthal angles of an arbitrary vector $\hat{\vec q}$ relative to $\hk$, $\hn$ is a null vector, $\mathcal{L}_s^\lambda$ is an associated Legendre polynomial, and the last equation ensures that the basis functions are transverse. As shown in \citep{Arkani-Hamed:2015bza}, symmetries of the de Sitter background yield the asymptotic scaling (late in inflation) \citep{Lee:2016vti,Assassi:2012zq,MoradinezhadDizgah:2018ssw,Arkani-Hamed:2015bza}:
\beq
    \lim_{\tau\to0}\sigma_{i_1\cdots i_s}(\vx,\tau) = \sigma^{+}_{i_1\cdots i_s}(\vx)(-\tau)^{\frac{3}{2}+i\mu_s-s}+\sigma^{-}_{i_1\cdots i_s}(\vx)(-\tau)^{\frac{3}{2}-i\mu_s-s}
\eeq
motivating the above mass parameters $\mu_s$. For intermediate and light particles, $\mu_s$ is imaginary and the dynamics of $\sigma$ are dominated by the most slowly decaying mode, which scales as $(-\tau)^{3/2-\nu_s}$ with $\nu_s\equiv i\mu_s>0$. For heavy particles, we find an oscillatory amplitude, since massive particles are produced by the expansion of space (though with vanishing frequency if $\mu_s\gg 0$); this leads to the aforementioned oscillations in the trispectrum.

Symmetry constraints also set the two-point function of the spin-$s$ field $\sigma$, which plays a key role in the inflationary correlators. As shown in \citep{Lee:2016vti}, the various polarization states are orthogonal with late-time power spectra:\footnote{Here and throughout, we drop the additional `local' contributions to the two-point functions \citep{Lee:2016vti}, as they are largely degenerate with inflaton self-interactions and do not produce oscillatory features.}
\beq\label{eq: sigma-2pt}
    &&\lim_{\tau,\tau'\to0}\av{\left(\epsilon^{i_1}\cdots\epsilon^{i_s}\sigma^{(\lambda)}_{i_1\cdots i_s}(\vk,\tau)\right)\left(\tilde{\epsilon}^{j_1}\cdots\tilde{\epsilon}^{j_s}\sigma^{(\lambda)}_{j_1\cdots j_s}(\vk',\tau')\right)^*}'\\\nonumber
    &=&\frac{(H\tau\tau')^{3/2-s}}{4\pi H}e^{i\lambda(\varphi-\varphi')}\left[\frac{(2s-1)!!\,s!}{(s-\lambda)!(s+\lambda)!}W_\lambda(s,\mu_s)\Gamma(-i\mu_s)^2\left(\frac{k^2\tau\tau'}{4}\right)^{i\mu_s}+\text{c.c.}\right]\\\nonumber
    &&W_{\lambda}(s,\mu_s)\equiv \frac{\Gamma(1/2+s-i\mu_s)\Gamma(1/2+\lambda+i\mu_s)}{\Gamma(1/2+s+i\mu_s)\Gamma(1/2+\lambda-i\mu_s)},
\eeq
for $\Delta \equiv 3/2+i\mu_s$, where we have introduced null vectors $\epsilon,\tilde{\epsilon}$, at azimuthal angles $\varphi,\varphi'$ to $\hk$. The phenomenology of this matches that obtained from the asymptotic scalings: heavy fields exhibit oscillations in $\mu_s\log k$, whilst intermediate and light fields decay monotonically, with only the second term in the expansion contributing. In full, the two-point function is a sum over all $(2s+1)$ helicity states $\lambda$, \resub{each of which has a power spectrum} specified by $W_\lambda(s,\mu_s)$ (and a normalization factor). This weighting is set by invariance under special conformal transforms; if one relaxes these assumptions, different helicity states are allowed independent amplitude. This is realized in the light particle model of \citep{Bordin:2018pca,Bordin:2019tyb} where a preferred foliation is introduced to avoid the Higuchi bound. In this case, the de Sitter isometries are broken, and each helicity state can be assigned an independent sound speed, $c_\lambda$, with different choices altering the angular structure of the correlators.

\vskip 4pt
\paragraph{Particle Interactions} Given the new field $\sigma$, one may compute exchange bispectra and trispectra using an effective action for $\sigma$ and the Goldstone mode $\pi$ (cf.\,\S\ref{subsubsec: EFTI}). For a spin-$s$ field, the simplest interactions are given by \citep[e.g.,][]{Lee:2016vti}
\beq\label{eq: massive-spinning-action}
    S_{\sigma,\pi} \supset \int d^4x\sqrt{-g}\frac{1}{a^{2s}}\bigg\{\rho_s\partial_{i_1\cdots i_s}\pi_c\hat{\sigma}_{i_1\cdots i_s}+\frac{1}{\Lambda_s^s}\dot{\pi}_c\partial_{i_1\cdots i_s}\pi_c\hat{\sigma}_{i_1\cdots i_s}+\lambda_s\dot{\pi}_c\hat{\sigma}^2_{i_1\cdots i_s}\bigg\}
\eeq
defining the canonically normalized field $\pi_c\equiv\pi\sqrt{2M_{\rm Pl}^2\dot{H}c_s}$ and $\hat{\sigma}\equiv\sigma-\mathrm{tr}\,\sigma$. Here, we find three main interactions: a linear conversion of $\pi$ from $\sigma$ (which generically involves a slow-roll factor of $\dot{\phi}$ for inflaton $\phi$), a quadratic mixing of two $\pi$ fields with $\sigma$ and a quadratic mixing of a two $\sigma$ fields with $\pi$; the amplitudes of each are independent (except for spin-$1$), though constrained by perturbativity.\footnote{For spin-zero, one instead uses $\dot{\pi}\sigma$, $(\partial_\mu\pi)^2\sigma$ and $\dot{\pi}\sigma^2$ to ensure shift-symmetry.} The lowest-order bispectra contributions involve the $\pi\pi\sigma$ and $\pi\sigma$ vertices (scaling as $\rho_s\Lambda_s^{-s}$), whilst the trispectra contributions involve two $\pi\pi\sigma$ couplings (scaling as $\Lambda_s^{-2s}$); as such, the trispectrum amplitude could be much larger than that of the bispectrum (strongly exceeding the Suyama-Yamaguchi bound). 

In the limit of very massive particles (with $m\gg H$), one may `integrate-out' $\sigma$ in the perturbative action, effectively performing an expansion in $m^{-1}$. At leading order, this corresponds to $\hat{\sigma}_{i_1\cdots i_s}\to \lambda/m^2\hat{\partial}_{i_1\cdots i_s}\pi_c$, sourcing, for example $\dot{\pi}(\hat{\partial}_{i_1\cdots i_s}\pi)^2$ \citep[e.g.,][]{MoradinezhadDizgah:2018ssw}. These manifest as $m$-independent (though $s$-dependent) self-interactions, with the $s=0$ term recovering the EFT of inflation templates discussed in \S\ref{subsubsec: EFTI}. Due to this, and the intrinsic Boltzmann suppression of high-mass particles, we primarily restrict our attention to particles with masses around the Hubble scale, which yield distinctive signatures. 

\vskip 4pt
\paragraph{Curvature Correlators} As discussed in \citep{Kehagias:2012td,Arkani-Hamed:2015bza}, the limiting forms of exchange interactions can be obtained by working in the `operator product expansion' limit. Utilizing the above interaction vertices and dropping derivatives for clarity, the squeezed bispectrum and collapsed trispectra take the schematic form: 
\beq
    \lim_{k_3\to0}\av{\zeta(\vk_1)\zeta(\vk_2)\zeta(\vk_3)}_c &\sim& \int_{\vK,\vK'}\av{\pi(\vk_1)\pi(\vk_2)\sigma^{(0)}(\vK)}\frac{1}{\av{\sigma^{(0)}(\vK)\sigma^{(0)}(\vK')}}\av{\pi(\vk_3)\sigma^{(0)}(\vK')}\\\nonumber
    \lim_{|\vk_1+\vk_2|\to0}\av{\zeta(\vk_1)\zeta(\vk_2)\zeta(\vk_3)\zeta(\vk_4)}_c &\sim& \sum_{\lambda\lambda'}\int_{\vK,\vK'}\av{\pi(\vk_1)\pi(\vk_2)\sigma^{(\lambda)}(\vK)}\frac{1}{\av{\sigma^{(\lambda)}(\vK)\sigma^{(\lambda')}(\vK')}}\av{\pi(\vk_3)\pi(\vk_4)\sigma^{(\lambda')}(\vK')},
\eeq
where $\zeta = -H\pi+\cdots$, as before. Notably, the bispectrum involves only the $\lambda=0$ helicity state, due to the transverse condition in \eqref{eq: polarization-tensors} (since $\av{\pi\sigma^{(\lambda)}}$ must contain $k^{i_1}\cdots k^{i_s}\varepsilon^{(\lambda)}_{i_1\cdots i_s}(\hk) \propto \delta^{\rm K}_{\lambda 0}$), whilst the trispectrum contains all helicity states with factors $\hat{k}_1^{i_1}\cdots \hat{k}_1^{i_s}\varepsilon_{i_1\cdots i_s}^{(\lambda)}(\hK) \propto e^{i\lambda\varphi}\mathcal{L}_s^\lambda(\cos\vartheta)$ (from \eqref{eq: polarization-tensors}, noting that $\vk_1\approx -\vk_2$). This again indicates that trispectrum non-Gaussianity can be much larger than bispectrum non-Gaussianity, particularly if the de Sitter symmetries relating helicity power spectra are broken. 

% By symmetry, the angular components of the $\pi\pi\sigma^{(\lambda)}$ three-point function involve the following contractions (from \eqref{eq: polarization-tensors}, noting that $\vk_1\approx -\vk_2$):
% \beq\label{eq: pol-contraction}
%     \hat{k}_1^{i_1}\cdots \hat{k}_1^{i_s}\varepsilon_{i_1\cdots i_s}^{(\lambda)}(\hK) \propto e^{i\lambda\varphi}\mathcal{L}_s^\lambda(\cos\vartheta)
%     %&=& \hat{Y}_s^\lambda(\vartheta,\varphi) \equiv \frac{1}{(2|\lambda|-1)!!}e^{i\lambda\varphi}\mathcal{L}_s^{|\lambda|}(\cos\vartheta),
% \eeq
% where $\hat{Y}^\lambda_s$ is a spherical harmonic with altered normalization (see \citep{Baumann:2017jvh} for a detailed description) and $(\vartheta,\varphi)$ are the polar and azimuthal angles of $\hk$ about $\hK$. For $\lambda=0$, this is just a Legendre polynomial, leading to a characteristic $\mathcal{L}_s(\hk_1\cdot\hk_3)$ dependence in the bispectrum, as in \eqref{eq: direction-dependent-Bk}. For the trispectrum, the angular dependence is more complex, and depends on $\hk_1$, $\hk_3$ and $\hK$ (cf.\,\S\ref{subsubsec: direction-templates}), as we see below.

Computing the full correlators and their amplitudes requires a fair amount of algebra, and is usually performed via the Schwinger-Keldysch formalism, bootstrap approaches or numerical techniques \citep[e.g.][]{Lee:2016vti,Arkani-Hamed:2018kmz,Pimentel:2022fsc,Wang:2022eop,Jazayeri:2023xcj,Bordin:2018pca,Pinol:2023oux,Werth:2023pfl}. However, the overall shapes are constrained by symmetries and the angular structure of the polarization  tensors, which yield a relatively simple squeezed limit. As shown in \citep{Pimentel:2022fsc,MoradinezhadDizgah:2017szk,Lee:2016vti,Assassi:2012zq,Noumi:2012vr}, bispectra generated by the $\dot\pi\partial_{i_1\cdots i_s}\pi\sigma_{i_1\cdots i_s}$ interaction take the form\footnote{Strictly this applies only for $s>0$ due to the non-conformal scalar couplings \citep{Arkani-Hamed:2015bza}. The difference is a factor of $|3/2-i\mu_0|^2 = m^2$ which can be absorbed into the amplitude.}
\beq
    \lim_{k_3\to0}\av{\zeta(\vk_1)\zeta(\vk_2)\zeta(\vk_3)}'_c &\propto& \mathcal{L}_s(\hk_1\cdot\hk_3)P_\zeta(k_1)P_\zeta(k_3)\left(\frac{k_3}{4k_1}\right)^{3/2}\\\nonumber
    &&\,\times\,\left[(1+i\sinh\pi\mu_s)\frac{5/2+s+i\mu_s}{3/2-s-i\mu_s}\frac{\Gamma(-i\mu_s)}{\Gamma(1/2-i\mu_s)}\left(\frac{k_3}{4k_1}\right)^{i\mu_s}+\text{c.c.}\right]\\\nonumber
    &\propto& \mathcal{L}_s(\hk_1\cdot\hk_3)P_\zeta(k_1)P_\zeta(k_3)\left(\frac{k_3}{k_1}\right)^{3/2}\cos\left[\mu_s\log\frac{k_3}{k_1}+\delta(\mu_s)\right],
\eeq
with the Legendre polynomial appearing due to the contractions of the $\lambda=0$ polarization tensor. For heavy particles with real $\mu_s$, the bispectrum displays oscillations in the logarithm of $k_3/k_1$ with some $\mu$-dependent (and model-dependent) phase; this is sourced by particle production, with $\log k_3/k_1$ tracking the number of $e$-folds of inflation. This shape applies also for intermediate and light particles (with the preferred foliation changing only the amplitude); thence, the bispectrum scales as $(k_1/k_3)^{3/2-\nu_s}$ in the squeezed limit, breaking the single-field consistency condition of \citep{Maldacena:2002vr} in a mass-dependent manner. If $s$ is odd, the exact squeezed limit vanishes exactly; in general, bispectra with odd $s$ are suppressed by $k_3/k_1$ relative to those with even $s$, due to exchange symmetries.% We additionally caution that the level of non-Gaussianity may be small unless we invoke additional assumptions such as broken conformal invariance \citep{Lee:2016vti}. 

The collapsed limit of the trispectrum takes a similar form \citep{Arkani-Hamed:2015bza,Lee:2016vti,Bordin:2018pca,Arkani-Hamed:2018kmz,Jazayeri:2023xcj,Assassi:2012zq}
\beq\label{eq: collapsed-trispectrum}
    &&\lim_{|\vk_1+\vk_2|\to0}\av{\zeta(\vk_1)\zeta(\vk_2)\zeta(\vk_3)\zeta(\vk_4)}_c \propto P_\zeta(k_1)P_\zeta(k_3)P_\zeta(K)\left(\frac{K^2}{16k_1k_3}\right)^{3/2}\\\nonumber
    &&\qquad\qquad\,\times\,\left[\left(\frac{K^2}{16k_1k_3}\right)^{i\mu_s}(1+i\sinh\pi\mu_s)\frac{(5/2+s+i\mu_s)^2}{(3/2+s-i\mu_s)^2}\Gamma(-i\mu_s)^2\Gamma(1/2+s+i\mu_s)^2\Theta(\hk_1,\hk_3;\hK,\mu_s)+\text{c.c.}\right];
\eeq
this once again displays oscillations for real $\mu_s$ and scales as $(K^2/k_1k_3)^{3/2+\nu_s}$ for real $\nu_s$. Here, the angular piece is encoded in the $\Theta$ function, which takes the following form assuming de Sitter symmetries \citep{Arkani-Hamed:2015bza}:
\beq\label{eq: theta-function}
    \Theta(\hk_1,\hk_3;\hK,\mu_s) &\equiv& \sum_{\lambda=-s}^se^{i\lambda(\varphi-\varphi')}\mathcal{L}_s^\lambda(\cos\vartheta)\mathcal{L}_s^\lambda(\cos\vartheta')\frac{(s-\lambda)!}{(s+\lambda)!}W_\lambda(s,\mu_s),
\eeq
\resub{where $(\vartheta, \vartheta')$ are the angles of $\hat{\bf k}_1$ and $\hat{\bf k}_3$ to $\hat{\bf K}$ and $\varphi - \varphi'$ is the angle between the projections of $\hat{\bf k}_1$ and $\hat{\bf k}_3$ onto the plane perpendicular to $\hat{\bf K}$.}
%where $(\vartheta,\varphi)$ are the angles of $\hk_1$ and $\hk_3$ to $\hK$.
\resub{Additionally,} the angular dependence arises from the contractions of the polarization tensor with $\hk_1,\hk_3$, the factor of $(s-\lambda)!/(s+\lambda)!$ enters due to switching $\mathcal{L}_s^{-\lambda}$ to $\mathcal{L}_s^{\lambda}$, \resub{and} $W_\lambda(s,\mu_s)$ comes from the power spectrum of $\sigma^{(\lambda)}$ in \eqref{eq: sigma-2pt}. This form is fairly generic; a number of modifications to the de Sitter case including broken boost symmetries, higher-spins, partially massless states, and preferred foliation states also lead to trispectra of this form (and its imaginary $\mu_s$ equivalent), albeit with modified $W_\lambda$ \citep[e.g.,][]{Arkani-Hamed:2018kmz,Lee:2016vti,Pimentel:2022fsc,Alexander:2019vtb,Baumann:2017jvh,Wang:2022eop,Jazayeri:2023xcj}. Notably, the phenomenology is specified by only two factors: its scaling and/or oscillation frequency (set by the mass) and its angular dependence (set by the particle spin), making it an excellent diagnostic for new physics.\footnote{Note that couplings of the exchange field to the inflaton can alter the oscillation frequency \citep{Werth:2023pfl}. We here ignore this effect, assuming weak mixing.}

In \S\ref{sec: estimators-exchange}, we will require an explicitly separable form for the trispectrum. To this end, we can rewrite the angular dependence in the following form, which is derived in Appendix \ref{app: angle-simplifications}:
\beq\label{eq: theta-function-separable}
    \Theta(\hk_1,\hk_3;\hK,\mu_s) &=& \frac{(4\pi)^{3/2}}{2s+1}\sum_{S=0}^{2s}\sqrt{2S+1}\sum_{\lambda}(-1)^{S+\lambda}W_\lambda(s,\mu_s)\tj{s}{s}{S}{\lambda}{-\lambda}{0}\\\nonumber
    &&\,\times\,\sum_{\lambda_1\lambda_3\Lambda}\tj{s}{s}{S}{\lambda_1}{\lambda_3}{\Lambda}Y_{s\lambda_1}(\hk_1)Y_{s\lambda_3}(\hk_3)Y_{S\Lambda}(\hK)\\\nonumber
    &\equiv&\sum_{S=0}^{2s}\mathcal{C}_s(S,\mu_s)\sum_{\lambda_1\lambda_3\Lambda}\tj{s}{s}{S}{\lambda_1}{\lambda_3}{\Lambda}Y_{s\lambda_1}(\hk_1)Y_{s\lambda_3}(\hk_3)Y_{S\Lambda}(\hK).
\eeq
This is a sum over isotropic combinations of $\hk_1,\hk_3,\hK$ (akin to the phenomenological direction-dependent templates of \S\ref{subsubsec: direction-templates}), with the weighting function, $\mathcal{C}_s$, set by $W_\lambda$. Notably, only terms with even $S$ contribute; this occurs since the correlators are parity even with $W_{\lambda}=W_{-\lambda}$. For $\mu=0$, $W_\lambda=1$, and we find the simpler expression:
\beq\label{eq: Theta-conformal}
    \Theta(\hk_1,\hk_3;\hK,\mu_s) &=& \frac{4\pi}{2s+1}\sum_{\lambda}Y_{s\lambda}(\hk_1)Y_{s\lambda}^*(\hk_3) \equiv \mathcal{L}_s(\hk_1\cdot\hk_3);
\eeq
this matches the result of \citep{Arkani-Hamed:2015bza}; in this limit, there is no dependence on $\hK$. If the system does not satisfy special conformal transform symmetries (e.g., in the light spin-$s\geq 2$ case), $W_\lambda$ will take a different (model-dependent) form, encoding the sound speeds for each helicity state (with $W_\lambda \propto c_\lambda^{-2\nu}$ in the notation of \citep{Bordin:2019tyb}). 

\vskip 4pt
\paragraph{Templates} To search for inflationary massive particle interactions, we must construct trispectrum templates that capture the main phenomenology whilst being practically computable. Since the full (non-collapsed) trispectra are difficult to compute and are not explicitly separable, we will focus on the collapsed limit given in \eqref{eq: collapsed-trispectrum}, and construct approximate templates both for the heavy and intermediate/light mass regimes.\footnote{An alternative approach would be to compute approximate templates using the bootstrap formalism \citep{Pimentel:2022fsc}, as discussed in \citep{Sohn:2024xzd} for bispectra. Whilst the resulting templates would be valid everywhere, they take a far more complex form, and are difficult to estimate with the tools discussed in \S\ref{sec: estimators}. Moreover, the non-squeezed parts of the template are degenerate with single-field models, thus restricting to collapsed limits probes the `smoking gun' of cosmological collider physics.} To ensure that we remain within the regime of validity, one can apply the following factor to all trispectra:
\beq\label{eq: coll-restriction}
    \Theta_{\rm H}(k_1-\alpha_{\rm coll} K)\Theta_{\rm H}(k_3-\alpha_{\rm coll} K)
\eeq
for Heaviside function $\Theta_{\rm H}$ and constant $\alpha_{\rm coll}\geq 1$, which practically restricts to $K\leq k_{1,3}/\alpha_{\rm coll}$ (and thus $k_1\approx k_2$, $k_3\approx k_4$ for large $\alpha_{\rm coll}$, given the triangle conditions). This matches the approach of \citep{Floss:2022grj,Meerburg:2016zdz}, with \citep{Floss:2022grj} assuming $\alpha_{\rm coll}=2$ (albeit with a symmetrized template), which restricted to the collapsed regime whilst limiting overlap with contact diagrams. This choice does not impact templates with large $\nu_s$ (which asymptote to the local shape), but is important for more massive particles, whence the collapsed behavior is suppressed. In practice, we use a separable approximation to \eqref{eq: coll-restriction} (see \S\ref{subsec: collider-estimator}) and typically marginalize over the equilateral EFT of inflation shapes, to reduce from contributions outside the collapsed limits. As shown in Appendix \ref{app: forecasts}, the resulting templates are well correlated to the full forms, particularly for large $\nu_s$.

For heavy particles, we utilize the following template
\begin{empheq}[box=\fbox]{align}\label{eq: heavy-exchange-template}
    \av{\zeta(\vk_1)\zeta(\vk_2)\zeta(\vk_3)\zeta(\vk_4)}'_c &\supset \tau_{\rm NL}^{\rm heavy}(s,\mu_s)\sum_{S=0}^{2s}\left|\mathcal{C}_s(S,\mu_s)\right|\sum_{\lambda_1\lambda_3\Lambda}\tj{s}{s}{S}{\lambda_1}{\lambda_3}{\Lambda}Y_{s\lambda_1}(\hk_1)Y_{s\lambda_3}(\hk_3)Y_{S\Lambda}(\hK)\\\nonumber
    &\,\times\,\frac{1}{2}\left[\left(\frac{K^2}{k_1k_3}\right)^{3/2+i\mu_s}e^{i\omega_s(S,\mu_s)}+\left(\frac{K^2}{k_1k_3}\right)^{3/2-i\mu_s}e^{-i\omega_s(S,\mu_s)}\right]\\\nonumber
    &\,\times\,\Theta_{\rm H}(k_1-\alpha_{\rm coll} K)\Theta_{\rm H}(k_3-\alpha_{\rm coll} K)P_\zeta(k_1)P_\zeta(k_3)P_\zeta(K)+\text{11 perms.},
\end{empheq}
where $\mu_s>0$ is real and 
\beq\label{eq: theta-function-phase}
    \omega_s(S,\mu_s) = \mathrm{arg}\left[2^{-4i\mu_s}(1+i\sinh\pi\mu_s)\frac{(5/2+s+i\mu_s)^2}{(3/2+s-i\mu_s)^2}\Gamma(-i\mu_s)^2\Gamma(1/2+s+i\mu_s)^2\right]+\mathrm{arg}\left[\mathcal{C}_s(S,\mu_s)\right]+\pi
\eeq
encodes the phases of each term (including the contribution from $W_\lambda$).\footnote{We add $\pi$ to this expression wlog to ensure that the heavy and light particle templates coincide at $\mu_s=\nu_s=0$. We stress that this phase depends on model assumptions such as the inflaton sound-speed, $c_\pi$ \citep[cf.][]{Lee:2016vti}.} For intermediate and light mediators (hereafter labelled `light' for brevity), we assume
\begin{empheq}[box=\fbox]{align}\label{eq: light-exchange-template}
    \av{\zeta(\vk_1)\zeta(\vk_2)\zeta(\vk_3)\zeta(\vk_4)}'_c &\supset \tau_{\rm NL}^{\rm light}(s,\nu_s)\sum_{S=0}^{2s}\mathcal{C}_s(S,i\nu_s)\sum_{\lambda_1\lambda_3\Lambda}\tj{s}{s}{S}{\lambda_1}{\lambda_3}{\Lambda}Y_{s\lambda_1}(\hk_1)Y_{s\lambda_3}(\hk_3)Y_{S\Lambda}(\hK)\\\nonumber
    &\,\times\,\Theta_{\rm H}(k_1-\alpha_{\rm coll} K)\Theta_{\rm H}(k_3-\alpha_{\rm coll} K)\left(\frac{K^2}{k_1k_3}\right)^{3/2-\nu_s}P_\zeta(k_1)P_\zeta(k_3)P_\zeta(K)+\text{11 perms.}
\end{empheq}
where $\mathcal{C}_s(S,i\nu_s)$ is real. In this work, we will assume the de Sitter solutions for phases and mode amplitudes, respecting the Higuchi bound; as emphasized above, our templates are considerably more general. We note that templates with similar $\nu_s$ are expected to be correlated given the limited dynamic range of CMB data; this is quantified in Appendix \ref{app: forecasts} (and demonstrated explicitly in \papertwo) and implies that there is little utility in testing pairs of templates with $|\nu_s-\nu_s'|\ll 1/2$.

Whilst the above templates correctly capture the non-local behavior of the trispectrum in the collapsed limit, they do not attempt to describe the equilateral components (often known as the `analytic' pieces \citep{MoradinezhadDizgah:2018ssw}). These are, in general, model-dependent and require in-in calculation or bootstrap methods to calculate. Furthermore, they are often degenerate with the EFT inflationary shapes \citep{Meerburg:2016zdz,Floss:2022grj,Cabass:2024wob}, making their presence less discriminative. The amplitude of our templates are encoded by $\tau_{\rm NL}^{\rm heavy/light}$, which, by construction, have matching amplitudes for $\mu_s\to0,\nu_s\to0$. For scalar exchange, the light particle template matches the local shape of \eqref{eq: tauNL-shape}, with $\tau_{\rm NL}^{\rm light}(0,3/2)=\taunl$ (noting that the collapsed limit dominates, \citep[cf.][]{Regan:2010cn}). For conformally coupled particles, $\mu_s=\nu_s=0$, whence $\mathcal{C}_s(S,0) = (4\pi)^{3/2}(-1)^s/\sqrt{2s+1}\times\delta_{S0}^{\rm K}$ and the templates take the limiting form
\beq\label{eq: light-exchange-template-spin0}
    \av{\zeta(\vk_1)\zeta(\vk_2)\zeta(\vk_3)\zeta(\vk_4)}'_c &\supset& \tau_{\rm NL}^{\rm light}(s,0)\mathcal{L}_{s}(\hk_1\cdot\hk_3)\left(\frac{K^2}{k_1k_3}\right)^{3/2}\\\nonumber
    &&\,\times\,\Theta_{\rm H}(k_1-\alpha_{\rm coll} K)\Theta_{\rm H}(k_3-\alpha_{\rm coll} K)P_\zeta(k_1)P_\zeta(k_3)P_\zeta(K)+\text{11 perms.}
\eeq
using \eqref{eq: Theta-conformal}, with $\tau_{\rm NL}^{\rm light}(s,0)=\tau_{\rm NL}^{\rm heavy}(s,0)$. For spin-$0$ particles, the templates again simplify \citep[cf.][]{Assassi:2012zq}; 
\beq\label{eq: massive-spin0}
    \av{\zeta(\vk_1)\zeta(\vk_2)\zeta(\vk_3)\zeta(\vk_4)}'_c &\supset& \tau_{\rm NL}^{\rm light}(0,\nu_0)\left(\frac{K^2}{k_1k_3}\right)^{3/2-\nu_0}\\\nonumber
    &&\,\times\,\Theta_{\rm H}(k_1-\alpha_{\rm coll} K)\Theta_{\rm H}(k_3-\alpha_{\rm coll} K)P_\zeta(k_1)P_\zeta(k_3)P_\zeta(K)+\text{11 perms.}\\\nonumber    \av{\zeta(\vk_1)\zeta(\vk_2)\zeta(\vk_3)\zeta(\vk_4)}'_c &\supset& \tau_{\rm NL}^{\rm heavy}(0,\mu_0)\left(\frac{K^2}{k_1k_3}\right)^{3/2}\cos\left[\mu_0\log\frac{K^2}{k_1k_3}+\omega_0(0,\mu_0)\right]\\\nonumber
    &&\,\times\,\Theta_{\rm H}(k_1-\alpha_{\rm coll} K)\Theta_{\rm H}(k_3-\alpha_{\rm coll} K)P_\zeta(k_1)P_\zeta(k_3)P_\zeta(K)+\text{11 perms.};
\eeq
the light template is the well-known quasi-single-field regime. Finally, for massless spin $2$ particles (violating the Higuchi bound), we can set $W_\lambda=1$, finding
\beq\label{eq: light-exchange-template-spin2}
    \av{\zeta(\vk_1)\zeta(\vk_2)\zeta(\vk_3)\zeta(\vk_4)}'_c &\supset& \tau_{\rm NL}^{\rm light}(2,3/2)\mathcal{L}_2(\hk_1\cdot\hk_3)\\\nonumber
    &&\,\times\,\Theta_{\rm H}(k_1-\alpha_{\rm coll} K)\Theta_{\rm H}(k_3-\alpha_{\rm coll} K)P_\zeta(k_1)P_\zeta(k_3)P_\zeta(K)+\text{11 perms.}
\eeq
without a squeezed limit divergence; this is akin to the $\tau_{\rm NL}^{220}$ direction-dependent template of \S\ref{subsubsec: direction-templates}, and takes the same form as that sourced by graviton exchange \citep{Seery:2008ax} (though this has a negligible amplitude $\tau_{\rm NL}\sim r$).\footnote{Setting $\nu_s = 3/2$, $\alpha_{\rm coll}=0$ and $\mathcal{C}_s(S,i\nu_s) = \delta^{\rm K}_{S\sigma}$ for integer $\sigma\in\{0,\cdots,2s\}$, the light spin-$s$ template reduces exactly to that of $\tau_{\rm NL}^{nn\sigma}$. Whilst this violates the Higuchi bound, it is a useful consistency test.}

\subsection{Other Templates}
\noindent In the above sections, we have introduced a wealth of trispectrum templates, both phenomenological and theory-inspired, whose amplitudes we will constrain below. This list is far from exhaustive however. A (still inexhaustive) set of templates we have not considered include:
\begin{itemize}
    \item Folded templates arising from non Bunch-Davies initial conditions, including thermal initial states and dissipative effects \citep[e.g.,][]{Meerburg:2009ys,Salcedo:2024smn,Mylova:2021eld,Chen:2006nt}. Phenomenologically similar effects can also arise from higher-derivative interactions \citep{Bartolo:2010di}. The associated bispectrum templates feature a signature enhancement at $k_1\approx k_2+k_3$.
    \item Oscillatory non-Gaussianity sourced by sharp features in the inflaton potential breaking approximate scale-invariance \citep{Wang:1999vf,Chen:2008wn}, or resonant features in axion models of inflation \citep{Flauger:2009ab,Flauger:2010ja}. These lead to linear and logarithmic oscillations in $k$, possibly with additional scale-dependence \citep{Achucarro:2012fd,Adshead:2011jq,Chen:2008wn,Planck:2015zfm,DuasoPueyo:2023kyh}.
    \item Isocurvature non-Gaussianity arising from multi-field inflation models \citep[e.g.,][]{Langlois:2008wt,Bartolo:2001cw}. Depending on the interactions in play, one could produce both auto- and cross-spectra of the isocurvature perturbation with adiabatic fluctuations. These are usually assumed to have a local-type spectrum, but differ from the $\gnl$ and $\taunl$ templates due to the isocurvature transfer functions.
    \item Scale-dependent non-Gaussianity, including that exhibited by multi-field models where each component has a different power spectrum \citep{Byrnes:2010ft,Wang:2022eop}. This can lead to running of the local non-Gaussianity parameters, \textit{i.e.}\ $\tau_{\rm NL}\to \tau_{\rm NL}(k)$. A similar effect can arise in equilateral models such as DBI inflation \citep{Bartolo:2010im}. 
    \item Parity-violating scalar trispectra, such as those sourced by interactions in ghost inflation, violations of scale-invariance, and massive spinning field exchange \citep[e.g.,][]{Cabass:2022rhr}. These can peak in either equilateral or collapsed regimes and feature a peculiar angular dependence.
    \item Equilateral collider signatures arising from collider models with broken boost symmetries, featuring oscillations outside the collapsed limit \citep[e.g.,][]{Wang:2022eop,Jazayeri:2022kjy}. This is generated when the sound-speed of the exchange particle is much less than the inflaton: $c_\sigma\ll c_s$. In the opposite limit, one forms a low-speed collider, with a distinctive resonance for mildly-squeezed configurations \citep{Jazayeri:2023xcj}.
    \item Tachyonic collider signatures sourced by unstable negative-mass fields present during inflation (which could arise in phase transitions or non-locality) \citep{McCulloch:2024hiz}. These give correlators analogous to those of light spin-zero fields, but with enhanced scalings: the mass parameter becomes $\sqrt{9/4+\tilde{m}^2/H^2}\geq 3/2$ for mass $-\tilde{m}$, leading to a stronger-than-local enhancement in the collapsed limit.
    \item Non-Gaussianity from partially massless fields in inflation \citep{Baumann:2017jvh,Franciolini:2018eno}. In de Sitter space, spinning particles with discrete masses below the Higuchi limit (\textit{i.e.}\ with $m^2/H^2<s(s-1)$) are possible -- these do not generate a scalar bispectrum, but can source a scalar four-point function. This requires a template similar to \eqref{eq: light-exchange-template} but with a different angular dependence and a scaling violating the Higuchi bound. An additional signal of interest comes from fermion exchange diagrams, for example those predicted in supersymmetric models \citep{Alexander:2019vtb}. 
    \item Recombination-era features from compensated isocurvature perturbations (CIPs) \citep{Grin:2013uya}. If present, CIPs would spatially modulate acoustic physics at reionization, sourcing a collapsed trispectrum proportional to the CIP power spectrum (analogous to the gravitational lensing trispectrum). 
\end{itemize}    
The bispectrum equivalents of many such models have been constrainted using \textit{Planck} data \citep{2014A&A...571A..24P,Planck:2015zfm,Planck:2019kim,Sohn:2024xzd}. In most cases, corresponding trispectrum analyses are both feasible and interesting, though they will require theoretical computation of the relevant four-point functions and the development of separable templates.

\section{Estimation: General Forms}\label{sec: estimators}
\noindent Next, we construct optimal estimators for the trispectra considered above. In each case, our goal is to estimate the template amplitude, $A_{\alpha}$ (e.g., $A_\alpha = \gnl$), from CMB temperature and polarization anisotropies. These derivations draw on the binned trispectrum estimators of \citep{Philcox:2023uwe,Philcox:2023psd} and the contact non-Gaussianity estimators of \citep{2015arXiv150200635S}, but specialize to templates rather than $\ell$-bins, fully incorporate mask and polarization effects, make use of some optimizations presented in \citep{Philcox:2024rqr}, and, most importantly, extend to a wide variety of templates including the (collapsed) cosmological collider. We will first discuss the general properties of the estimators before presenting the specific forms for contact, exchange, and late-time trispectra in \S\ref{sec: estimators-contact},\,\ref{sec: estimators-exchange},\,\&\,\ref{sec: estimators-late-time}.

\subsection{Quartic Estimators}

\noindent Given a dataset $d$ with indices $i$, we may form correlators such as the two- and four-point functions $\mathsf{C}^{i_1i_2} \equiv \av{d^{i_1}d^{i_2*}}$ and $\mathsf{T}^{i_1i_2i_3i_4} \equiv \av{d^{i_1}d^{i_2}d^{i_3}d^{i_4}}$. Here and in the remainder of this work, our convention is that $i,j,\ldots$ indices refer either to `spin/pixel'-space (\textit{i.e.}\ $d^i = {}_sd(\hn)$ for spin $s$ and angular position $\hn$) or `polarization/harmonic'-space (\textit{i.e.}\ $d^i = d_{\ell m}^X$ for $X\in\{T,E,B\}$ and angular momentum indices $\ell,m$). In this section, we will use both definitions interchangeably.

Assuming that the likelihood for the data can be approximated as an Edgeworth series (\textit{i.e.}\ it is a small perturbation around a Gaussian), the optimal estimator, $\widehat{A}_\alpha$, for some parameter $A_\alpha$ appearing only in the trispectrum is given by
\beq\label{eq: optimal-estimators}
    \widehat{A}_\alpha &=& \sum_{\beta}\F^{-1}_{\alpha\beta}\widehat{N}_\beta\\\nonumber
    \widehat{N}_\alpha &=& \frac{1}{4!}\frac{\partial\T^{i_1i_2i_3i_4}}{\partial A_\alpha}\mathcal{H}^*_{i_1i_2i_3i_4}[\Ci d]\\\nonumber
    \mathcal{F}_{\alpha\beta} &=& \frac{1}{4!}\left[\left(\frac{\partial\T^{i_1i_2i_3i_4}}{\partial A_\alpha}\right)^*\Ci_{i_1j_1}\Ci_{i_2j_2}\Ci_{i_3j_3}\Ci_{i_4j_4}\frac{\partial\T^{j_1j_2j_3j_4}}{\partial A_\beta}\right]^*,
\eeq
where the Hermite tensor is defined as
\beq\label{eq: hermite}
    \mathcal{H}_{i_1i_2i_3i_4}[h] = h_{i_1}h_{i_2}h_{i_3}h_{i_4}-\left(h_{i_1}h_{i_2}\av{h_{i_3}h_{i_4}}+\text{5 perms.}\right)+\left(\av{h_{i_1}h_{i_2}}\av{h_{i_3}h_{i_4}}+\text{2 perms.}\right).
\eeq
This is simply obtained by maximizing the likelihood with respect to $A_\alpha$, assuming a fiducial value of $A_\alpha=0$ \citep{Philcox:2023psd,2017arXiv170903452S,Hamilton:2005ma,Oh:1998sr,2011MNRAS.417....2S}.\footnote{An alternative approach is to first compress the trispectrum to a set of bins or modal coefficients, and then compare the data to a similarly processed model. This is discussed in \citep{Fergusson:2010gn,Regan:2010cn,2011arXiv1105.2791F} and \citep{Philcox:2023uwe,Philcox:2023psd} respectively, and can be close-to-optimal in practice, depending on the model in question.}

The above estimator involves the inverse covariance matrix $\Ci$, which is used to weight the data and form the normalization matrix. As discussed in \citep{2015arXiv150200635S,Philcox:2023uwe,Philcox:2023psd}, this is often prohibitively expensive to compute; as such it may be desirable to use an approximate filtering $\tCi$, whence the estimators become
\beq\label{eq: general-estimators-1}
    \widehat{N}_\alpha &=& \frac{1}{4!}\frac{\partial\T^{i_1i_2i_3i_4}}{\partial A_\alpha}\mathcal{H}^*_{i_1i_2i_3i_4}[\tCi d]\\\nonumber
    \mathcal{F}_{\alpha\beta} &=& \frac{1}{4!}\left[\left(\frac{\partial\T^{i_1i_2i_3i_4}}{\partial A_\alpha}\right)^*\tCi_{i_1j_1}\tCi_{i_2j_2}\tCi_{i_3j_3}\tCi_{i_4j_4}\frac{\partial\T^{j_1j_2j_3j_4}}{\partial A_\beta}\right]^*.
\eeq
Multiple features are of note: (1) the Hermite tensor subtracts off the disconnected (Gaussian) pieces of the four-point function, ensuring \textbf{no contamination} from the two-point function; (2) the estimator is \textbf{unbiased} for any choice of weighting $\tCi$ (regardless of masking, inpainting, and leakage effects), such that $\mathbb{E}[\widehat{A}_\alpha] = A_\alpha$; (3) the estimator accounts for \textbf{correlations} between all templates included in the analysis; (4) in the limit of $\tCi\to\Ci$ and Gaussian statistics ($A_\alpha\to0$), the estimator achieves \textbf{minimum variance}, with $\mathrm{cov}(\widehat{A}_\alpha,\widehat{A}_\beta) = \F^{-1}_{\alpha\beta}$. As such, $\F$ is usually known as the `Fisher matrix'. 

\subsection{Data Model \& Optimal Weights}\label{subsec: data-model}
\noindent To build the estimators, we require an explicit form for the four-point function $\mathsf{T}$, which requires relating the observed data ($d_i$, which naturally exists in spin/pixel-space) to the underlying CMB field of interest ($a_i$, usually defined in polarization/harmonic-space). Here, we employ the standard linear definition \citep[e.g.,][]{1999ApJ...510..551O,Smith:2007rg}
\beq
    d_i = [\mathsf{P}a]_i + n_i
\eeq
where $\mathsf{P}$ is the `pointing matrix' and $n$ is a spin/pixel-space instrumental noise component with zero mean. Typically, one assumes that $\mathsf{P}$ can be represented by the following sequence of linear operations:
\begin{itemize}
    \item $\mathsf{L}$: lensing via some lensing potential $\phi$ (inducing non-Gaussianity);
    \item $\mathsf{B}$: convolution with the polarization-dependent instrumental beam, as well as any pixel window function;
    \item $\mathsf{Y}$: spherical harmonic synthesis from polarization/harmonic-space to spin/pixel-space;
    \item $\mathsf{W}$: multiplication by a (possibly non-invertible and possibly spin-dependent) mask,
\end{itemize}
such that $\mathsf{P} \equiv \mathsf{W}\mathsf{Y}\mathsf{B}\mathsf{L}$. In this series, we will ignore the effects of lensing in $\mathsf{P}$ (except through its modification to the two-point function), though we note that some form of `delensing' will likely be necessary to achieve tight non-Gaussianity constraints from future high-resolution experiments \citep[e.g.,][]{Green:2016cjr,Trendafilova:2023xtq}. Typically, $\mathsf{B}$ is diagonal in harmonic-space and $\mathsf{W}$ is diagonal in map-space, though they can correlate different spins and polarizations. For an ideal experiment, $\mathsf{W} = \mathsf{1}$, though in practice we usually wish to mask out poorly reconstructed regions of the survey, unobserved patches and point sources, such that the noise in the unmasked region is roughly translation-invariant and we can estimate $a_{\ell m}^X$ without bias. 

Denoting the signal and noise covariances by $\mathbb{C} \equiv \av{aa^\dag}$ and $\mathsf{N} \equiv \av{nn^\dag}$ respectively, we can write the full two- and four-point functions as:
\beq
    \mathsf{C} \equiv \av{dd^\dag} = \mathsf{P}\mathbb{C}\mathsf{P}^\dag + \mathsf{N}, \qquad \mathsf{T}^{i_1i_2i_3i_4} = \mathsf{P}^{i_1j_1}\cdots\mathsf{P}^{i_4j_4}\av{a_{j_1}\cdots a_{j_4}}_c + \text{noise},
\eeq
where $\mathsf{X}^\dag$ is the Hermitian conjugate of $\mathsf{X}$. The non-Gaussianity estimators become
\beq\label{eq: general-estimators}
    \widehat{N}_\alpha &=& \frac{1}{4!}\frac{\partial\av{a^{i_1}a^{i_2}a^{i_3}a^{i_4}}_c}{\partial A_\alpha}\mathcal{H}^*_{i_1i_2i_3i_4}[\Si d]\\\nonumber
    \mathcal{F}_{\alpha\beta} &=& \frac{1}{4!}\left[\left(\frac{\partial\av{a^{i_1}a^{i_2}a^{i_3}a^{i_4}}_c}{\partial A_\alpha}\right)^*[\Si\mathsf{P}]_{i_1j_1}[\Si\mathsf{P}]_{i_2j_2}[\Si\mathsf{P}]_{i_3j_3}[\Si\mathsf{P}]_{i_4j_4}\frac{\partial\av{a^{j_1}a^{j_2}a^{j_3}a^{j_4}}_c}{\partial A_\beta}\right]^*,
\eeq
defining the new weight $\Si \equiv \mathsf{P}^\dag \tCi$ \citep[cf.][]{Philcox:2024rqr} and ignoring noise contributions to $\mathsf{T}$ (which are not of cosmological interest).\footnote{This differs from the notation of \citep{Philcox:2023uwe,Philcox:2023psd}, whose $\Si$ was equal to our $\tCi$. Our redefinition simplifies the resulting estimators.} Here, $\Si$ represents the total weight applied to the data (including, for example, inpainting and harmonic-space filtering), and the $\Si\mathsf{P}$ factors ensure that the estimator is correctly normalized (assuming that $\mathsf{P}$ is precisely known).

Following the discussion above, the optimal estimator is obtained by identifying $\F^{-1}_{\alpha\beta}$ with the Gaussian covariance of $\widehat{A}_\alpha$, which requires
\beq\label{eq: weight-optimality}
    \Si\mathsf{C}\mathsf{S}^{-\dag} = \Si\mathsf{P} \quad \Leftrightarrow\quad \mathsf{P}^\dag\left[\tCi\mathsf{C}\tilde{\mathsf{C}}^{-\dag}-\tCi\right]\mathsf{P} = 0,
\eeq
(\textit{i.e.}\ $\tilde{\mathsf{C}}^{-1}\mathsf{C} = \mathsf{1}$ over pixels which are not killed by $\mathsf{P}$). This has the general solution
\beq\label{eq: opt-weight-def}
    \Si_{\rm opt} = \mathsf{P}^\dag\mathsf{C}^{-1} = \mathbb{C}^{-1}\left[\mathbb{C}^{-1}+\mathsf{P}^\dag\mathsf{N}^{-1}\mathsf{P}\right]^{-1}\mathsf{P}^\dag\mathsf{N}^{-1},
\eeq
using the Woodbury matrix identity. Up to the $\mathbb{C}^{-1}$ factor, this is simply a beam-deconvolved Wiener filter, and can be implemented by (a) multiplying by the full inverse covariance, and (b) multiplying by the pointing matrix.\footnote{Note that we require the inverse noise covariance, $\mathsf{N}^{-1}$, only in pixels for which the mask is non-vanishing, since $\mathsf{P}$ contains a factor of $\mathsf{W}$. This allows the estimator to be applied to cut-sky datasets, where $\mathsf{N}$ is formally infinite in masked pixels.} In the ideal limit of a unit mask and translation-invariant noise $N_\ell$, the application of $\Si$ on a polarization/harmonic-space map $d_{\ell m}^X$ is given by
\beq\label{eq: ideal-weight}
    [\Si_{\rm ideal}d]_{\ell m}^X = B_\ell^X\sum_Y\left[B_\ell^{\,}\mathbb{C}_\ell B^\dag_\ell+N_\ell\right]^{-1,XY}d_{\ell m}^Y,
\eeq
\textit{i.e.}\ we divide by the fiducial signal-plus-noise and remove the beam.\footnote{In this limit, the factors of $\Si\mathsf{P}$ in \eqref{eq: general-estimators} simplify to $B_\ell^X[B_\ell\mathbb{C}_\ell B_\ell^\dagger +N_\ell]^{-1,XY}B_\ell^Y$, which is simply the inverse beam-deconvolved power spectrum.} In realistic settings, one can (a) implement the optimal filtering of \eqref{eq: opt-weight-def}, using numerical methods such as conjugate gradient descent or machine learning to invert $\left[\mathbb{C}^{-1}+\mathsf{P}^\dag\mathsf{N}^{-1}\mathsf{P}\right]^{-1}$ (e.g., by solving $\left[\mathbb{C}^{-1}+\mathsf{P}^\dag\mathsf{N}^{-1}\mathsf{P}\right]d_{\rm WF} = \mathsf{P}^\dag \mathsf{N}^{-1}d$ for Wiener-filtered $d_{\rm WF}$) \citep[e.g.,][]{1999ApJ...510..551O,Munchmeyer:2019kng,Costanza:2023cgt,Costanza:2024rut}, or (b) use an approximate form for $\Si$ instead of the optimal solution. For suitably chosen approximations, this leads to minimal loss of signal to noise \citep[cf.][]{Philcox4pt3}, noting that the estimator is unbiased for any $\Si$. A typical choice is to first inpaint any small holes then apply the translation-invariant weighting of \eqref{eq: ideal-weight} -- this form was assumed in \citep{2015arXiv150200635S,Planck:2015zfm,Planck:2019kim} and will be used in \paperthree.

\subsection{Monte Carlo Summation}\label{subsec: monte-carlo}
\noindent In order for the estimators to be implemented efficiently, they must be rewritten in a separable form. As we will see below, the trispectrum term $\av{a^{i_1}a^{i_2}a^{i_3}a^{i_4}}_c$ can be explicitly split into a sum of terms (sum-)separable in $i_1,i_2,i_3,i_4$. We similarly require a separable form for the Hermite tensor \eqref{eq: hermite} in the numerator of \eqref{eq: general-estimators}. This can be achieved by replacing the averages $\av{h[d]_ih[d]_j}_d$ (where $h[d] = \Si d$ in our case) by Monte Carlo sums:
\beq
    \av{h[d]_{i_1}h[d]_{i_2}}_d \to \frac{1}{N_{\rm disc}}\sum_{n=1}^{N_{\rm disc}}h[\delta^{(n)}]_{i_1}h[\delta^{(n)}]_{i_2}
\eeq
where $\{\delta^{(n)}\}$ are  independent and identically distributed (iid) random fields with covariance $\mathsf{C}_{\rm disc}=\mathsf{C}$, \textit{i.e.}\ simulations that whose two-point statistics match those of the observational data. For the quadratic term $\av{h[d]_{i_1}h[d]_{i_2}}_d\av{h[d]_{i_3}h[d]_{i_4}}_d$, we require two sets of iid random fields to avoid correlations between the two Monte Carlo averages. If the covariance of the simulations does not match that of the data, the estimator incurs a bias $\mathcal{O}[(\mathsf{C}_{\rm disc}-\mathsf{C})^2]$.\footnote{An alternative trispectrum estiamtor is given schematically by $h^4-3\av{h^2}\av{h^2}$: this incurs a larger $\mathcal{O}[(\mathsf{C}_{\rm disc}-\mathsf{C})]$ bias.}

The normalization matrices, $\mathcal{F}_{\alpha\beta}$, are slightly more difficult to compute, though they are independent of the dataset $d$. % (but depend non-trivially on the pointing matrix $\mathsf{P}$ and weighting function $\Si$). 
For efficient computation, we adopt the approach of \citep{Oh:1998sr,2015arXiv150200635S} (which is commonly used in the applied mathematics community for computing the trace of high-dimensional matrices \citep{girard89,hutchinson90,Epperly_2024,meyer2021hutch}), first rewriting the product of $\Si$ filters appearing in \eqref{eq: general-estimators}:
\beq
    [\Si\mathsf{P}]_{i_2j_2}[\Si\mathsf{P}]_{i_3j_3}[\Si\mathsf{P}]_{i_4j_4}+\text{5 perms.} &=& [\Si\mathsf{P}]_{i_2i_2'}[\Si\mathsf{P}]_{i_3i_3'}[\Si\mathsf{P}]_{i_4i_4'}\\\nonumber
    &&\,\times\,\left[\A^{i_2'j_2'}\A^{i_3'j_3'}\A^{i_4'j_4'}+\text{5 perms.}\right]\Ai_{j_2'j_2}\Ai_{j_3'j_3}\Ai_{j_4'j_4},
\eeq
for a general invertible and symmetric matrix $\mathsf{A}$. Defining a set of iid maps $\{a^{(n)}\}$ with covariance $\av{a^{(n)}_ia^{(m)*}_j}_{a} = \delta_{\rm K}^{nm} \mathsf{A}_{ij}$,\footnote{For fast convergence, we require $\mathsf{A}^{-1} \approx \Si\mathsf{P}$, which is approximately the inverse of the beam-deconvolved power spectrum.} this can be decoupled into two pieces:
\beq
    [\Si\mathsf{P}]_{i_2j_2}[\Si\mathsf{P}]_{i_3j_3}[\Si\mathsf{P}]_{i_4j_4}+\text{5 perms.} &=& \left\langle[\Si\mathsf{P} a^{(1)}]_{i_2}[\Si\mathsf{P} a^{(2)}]_{i_3}[\Si\mathsf{P} a^{(3)}]_{i_4}\right.\\\nonumber
    &&\,\times\,\left.[\Ai a^{(1)}]_{j_2}[\Ai a^{(2)}]_{j_3}[\Ai a^{(3)}]_{j_4}\right\rangle_a+\text {5 perms.},
\eeq
where the expectation can be evaluated as a Monte Carlo summation over random fields $a$. In practice, we can utilize an alternative form that makes optimal use of only two pairs of Monte Carlo simulations \citep{2015arXiv150200635S}: this leads to the Fisher matrix:
\begin{empheq}[box=\dbox]{align}\label{eq: fisher-MC}
    \mathcal{F}_{\alpha\beta} = \frac{1}{48}\bigg[\left(F^{111,111}_{\alpha\beta}+F^{222,222}_{\alpha\beta}\right)\,+\,9\left(F^{112,112}_{\alpha\beta}+F^{122,122}_{\alpha\beta}\right)\,-\,3\left(F^{111,122}_{\alpha\beta}+F^{222,112}_{\alpha\beta}+F^{122,111}_{\alpha\beta}+F^{112,222}_{\alpha\beta}\right)\bigg]
\end{empheq}
subject to the definition
\begin{empheq}[box=\dbox]{align}\label{eq: fisher-MC2}
    F^{abc,def}_{\alpha\beta} &\equiv \frac{1}{4!}\bigg\langle\left(\frac{\partial\av{a^{i_1}a^{i_2}a^{i_3}a^{i_4}}_c}{\partial A_\alpha}\right)^*[\Si\mathsf{P} a^{(a)}]_{i_2}[\Si\mathsf{P} a^{(b)}]_{i_3}[\Si\mathsf{P} a^{(c)}]_{i_4}\,\times\,[\Si\mathsf{P}]_{i_1j_1}\\\nonumber
    &\qquad\,\times\,[\mathsf{A}^{-1}a^{(d)}]^*_{j_2}[\mathsf{A}^{-1}a^{(e)}]^*_{j_3}[\mathsf{A}^{-1}a^{(f)}]^*_{j_4}\frac{\partial\av{a^{j_1}a^{j_2}a^{j_3}a^{j_4}}_c}{\partial A_\beta}\bigg\rangle^*_a\nonumber\\
    &\equiv \frac{1}{4!}\bigg\langle\left(Q_{\alpha}[\Si \mathsf{P}a^{(a)},\Si\mathsf{P} a^{(b)},\Si \mathsf{P}a^{(c)}]\right)^* \cdot[\mathsf{S}^{-1}\mathsf{P}]\cdot\left(Q_\beta[\mathsf{A}^{-1} a^{(d)},\mathsf{A}^{-1} a^{(e)},\mathsf{A}^{-1} a^{(f)}]\right)\bigg\rangle^*_a.
\end{empheq}
%where we perform an additional symmetrization with respect to \citep{2015arXiv150200635S}. 
This defines the derivative maps
\beq\label{eq: Q-def}
    Q^i_\alpha[x,y,z] = \frac{\partial\av{a^ia^ja^ka^l}_c}{\partial A_\alpha}x^*_jy^*_kz^*_l;
\eeq
once these are computed, the Fisher matrix is obtained as a summation over the $i_1,j_1$ indices (usually in polarization/harmonic-space), which is straightforward to compute. The Fisher matrix can be asymmetric; symmetry is obtained only if $\Si\mathsf{P}=\mathsf{P}^\dag\mathsf{S}^{-\dag}$ (which is true for optimal $\Si$).

\subsection{Harmonic-Space Forms \& Summary}
\noindent The final ingredient in the estimators is the relation between the CMB correlators $\av{a_{i_1}a_{i_2}a_{i_3}a_{i_4}}_c$ and the amplitudes of interest $A_\alpha$ (representing $\gnl$, $\taunl$, \textit{et cetera}.). This requires the linear relation between CMB fluctuations and comoving curvature perturbations, $\zeta$, which is given by the standard polarization/harmonic-space form:
\beq\label{eq: transfer-def}
    a_{\ell m}^X = 4\pi i^\ell \int_{\vk}\mathcal{T}^X_\ell(k)Y_{\ell m}^*(\hk)\zeta(\vk).
\eeq
Here $\mathcal{T}_\ell^{X}(k)$ is the associated transfer function (defined by this equation), $Y_{\ell m}$ is a spherical harmonic and $X\in\{T,E,B\}$ indexes the field of interest.
%\footnote{For consistency, we will work only with the curvature perturbation, $\zeta$, in this work. This removes all $3/5$ factors relating curvature and potential.} 
Using \eqref{eq: transfer-def}, we can relate $\av{a_{i_1}a_{i_2}a_{i_3}a_{i_4}}_c$ to the trispectrum models discussed in \S\ref{sec: templates}:
\beq\label{eq: Tlm-general}
    T^{\ell_1\ell_2\ell_3\ell_4,X_1X_2X_3X_4}_{m_1m_2m_3m_4}\equiv \av{a_{\ell_1m_1}^{X_1}a_{\ell_2m_2}^{X_2}a_{\ell_3m_3}^{X_3}a_{\ell_4m_4}^{X_4}}_c = \prod_{i=1}^4\left[4\pi i^{\ell_i}\int_{\vk_i}\mathcal{T}_{\ell_i}^{X_i}(k_i)Y_{\ell_im_i}^*(\hk_i)\right]\av{\zeta(\vk_1)\zeta(\vk_2)\zeta(\vk_3)\zeta(\vk_4)}_c,
\eeq
keeping the indices explicit for clarity.
% To obtain the four-point derivative of interest, we first use the above expressions to relate $\av{a_{i_1}a_{i_2}a_{i_3}a_{i_4}}_c$ to the harmonic-space trispectrum $T^{\ell_1\ell_2\ell_3\ell_4,X_1X_2X_3X_4}_{m_1m_2m_3m_4}$:
% \beq\label{eq: Tijkl-from-harmonics}
%     \av{a^{i_1}a^{i_2}a^{i_3}a^{i_4}}_c \equiv \av{\prod_{i=1}^4{}_{s_i}a(\hn_i)}_c &=& \prod_{i=1}^4\left[\sum_{\ell_im_iX_i}{}_{s_i}\mathcal{R}_{X_i}{}_{s_i}Y_{\ell_im_i}(\hn)\right]\av{a_{\ell_1m_1}^{X_1}a_{\ell_2m_2}^{X_2}a_{\ell_3m_3}^{X_3}a_{\ell_4m_4}^{X_4}}_c\\\nonumber
%     &\equiv& \prod_{i=1}^4\left[\sum_{\ell_im_iX_i}{}_{s_i}\mathcal{R}_{X_i}{}_{s_i}Y_{\ell_im_i}(\hn)\right]T^{\ell_1\ell_2\ell_3\ell_4,X_1X_2X_3X_4}_{m_1m_2m_3m_4},
% \eeq
% which can itself be recast in terms of the four-point function of $\zeta$
% \beq\label{eq: Tlm-general}
%     T^{\ell_1\ell_2\ell_3\ell_4,X_1X_2X_3X_4}_{m_1m_2m_3m_4} &=& \prod_{i=1}^4\left[4\pi i^{\ell_i}\int_{\vk_i}\mathcal{T}_{\ell_i}^{X_i}(k_i)Y_{\ell_im_i}^*(\hk_i)\right]\av{\zeta(\vk_1)\zeta(\vk_2)\zeta(\vk_3)\zeta(\vk_4)}_c.
% \eeq
Combining expressions, the numerator of the general trispectrum estimator \eqref{eq: general-estimators} can be written in polarization/harmonic-space as
\begin{empheq}[box=\dbox]{align}\label{eq: general-estimators-harmonic}
    \widehat{N}_\alpha &= \widehat{\mathcal{N}}_\alpha[d,d,d,d]-\bigg(\av{\widehat{\mathcal{N}}_\alpha[d,d,\delta,\delta]}_{\delta}+\text{5 perms.}\bigg)+\bigg(\av{\widehat{\mathcal{N}}_\alpha[\delta^{(1)},\delta^{(1)},\delta^{(2)},\delta^{(2)}]}_\delta+\text{2 perms.}\bigg)\\\nonumber
    \widehat{\mathcal{N}}_\alpha[\alpha,\beta,\gamma,\delta] &\equiv \frac{1}{24}\frac{\partial\av{a^{i_1}a^{i_2}a^{i_3}a^{i_4}}_c}{\partial A_\alpha}[\Si \alpha]^*_{i_1}[\Si\beta]^*_{i_2}[\Si\gamma]^*_{i_3}[\Si\delta]^*_{i_4}\\\nonumber
    &=\frac{1}{24}\sum_{\ell_im_iX_i}\frac{T^{\ell_1\ell_2\ell_3\ell_4,X_1X_2X_3X_4}_{m_1m_2m_3m_4}}{\partial A_\alpha}[\Si\alpha]^{X_1*}_{\ell_1m_1}[\Si\beta]^{X_2*}_{\ell_2m_2}[\Si\gamma]^{X_3*}_{\ell_3m_3}[\Si\delta]^{X_4*}_{\ell_4m_4},
\end{empheq}
where the covariance of $\{\delta^{(1,2)}\}$ are expected to match that of the data, as before. We may similarly express the Fisher matrix given in \eqref{eq: fisher-MC} in polarization/harmonic-space; this requires a straightforward transform of the $Q$ filters \eqref{eq: Q-def}:
\begin{empheq}[box=\dbox]{align}\label{eq: Q-def-harmonic}
    Q_{\ell m,\alpha}^{X}[x,y,z] = \sum_{\ell_im_iX_i}\frac{\partial T^{\ell\ell_2\ell_3\ell_4,XX_2X_3X_4}_{mm_2m_3m_4}}{\partial A_\alpha}x^{X_2*}_{\ell_2m_2}y^{X_3*}_{\ell_3m_3}z^{X_4*}_{\ell_4m_4}.
\end{empheq}

To build the Fisher matrix and optimal $\Si$ weights, we require an explicit form for the pointing matrix, $\mathsf{P}$ (see \S\ref{subsec: data-model}). This is defined by its action on some polarization/harmonic-space map $x$:
\beq
    {}_s[\mathsf{P}x](\hn) \equiv {}_s[\mathsf{W}\mathsf{Y}\mathsf{B}x](\hn) = {}_sW(\hn)\sum_{\ell m}{}_sY_{\ell m}(\hn)\sum_X{}_s\mathcal{R}^XB_\ell^X x_{\ell m}^X
\eeq
where ${}_s\mathcal{R}_X$ % = \begin{pmatrix} 1 & 0 &0 \\ 0 & -1 & -i\\0 & -1 & +i\end{pmatrix}$ (which we will hereafter keep implicit) 
transforms from polarization- to spin-space, ${}_sW(\hn)$ is a spin-dependent mask, and $B_\ell^X$ is a beam, containing both experimental and pixel-window contributions. Here, we have neglected lensing, and assumed a translation-invariant beam with no temperature-to-polarization leakage, though these are not limitations of the method. 

Often, one wishes to `band-limit' the analysis, \textit{i.e.}\ to only include information only from a particular range of multipoles. To practically implement this, one can adopt two different methods. Firstly, one could apply an $\ell$-space cut (denoted by the $\mathsf{\Theta}$ operator) to the data, practically redefining the pointing matrix as $\mathsf{WYB}\to \mathsf{WY\Theta B}$. This is expensive to implement, since it applies only to the $\Si$-weighted leg of \eqref{eq: fisher-MC}, thus we must keep all $\ell$-modes in $\mathsf{A}^{-1}a$, and could be affected by aliasing effects. A second option is to assume that the theory only has support for a certain range of scales, practically setting $T_{\ell_i}$ to zero outside these regimes. This is cheaper; we can discard $(\ell,m)$ modes immediately after filtering the normalization maps by $\Si\mathsf{P}$ and $\mathsf{A}^{-1}$,\footnote{For exchange templates, we can additionally restrict the internal leg to lie in $[L_{\rm min},L_{\rm max}]$. Where there are multiple internal legs (for example for directional templates), we enforce these limits on all internal $L$-modes, \textit{i.e.}\ $L_{\rm min}\leq L,L'\leq L_{\rm max}$.} and will be assumed where relevant below.

Whilst our estimators are similar to the `pure MC' form proposed in \citep{2015arXiv150200635S} (and used in \citep{Planck:2015zfm,Planck:2019kim}), there are several important differences: (1) we include both temperature and polarization data, rather than just temperature; (2) we do not require non-Gaussian simulations to form the numerator; (3) we construct the Fisher matrix using arbitrary $\mathsf{A}$ (removing restrictive simulation constraints, and allowing unbiased estimation); (4) we make slightly more efficient use of simulations in \eqref{eq: fisher-MC2} via symmetrization; (5) unlike the form work, we do not take the difference of $\widehat{A}$ computed from the data and a suite of simulations. The latter choice is made to allow for verification for the estimators and to reduce computational costs (noting that we can remove lensing bias explicitly). Comparison to other types of four-point estimators can be found in \S\ref{sec: estimator-comparison}.

\subsection{Computation Strategy}\label{subsec: strategy}
\noindent By combining \eqref{eq: general-estimators-harmonic} with \eqref{eq: fisher-MC}\,\&\,\eqref{eq: Q-def-harmonic}, we obtain an efficient primordial template estimator. Below, we will discuss the practicalities of computing $\widehat{\mathcal{N}}_\alpha$ and $Q_\alpha$ for various (separable) primordial shapes; before doing so, however, we outline our basic computation strategy. For the numerators: 
\begin{enumerate}
    \item Define the data, $d$, and a pair of input simulations ($\delta^{(1)}, \delta^{(2)}$) whose two-point function matches that of the data (e.g., simulations). The random fields are used to compute the disconnected terms in the estimator numerator.
    \item Filter each map by the linear operator $\Si$, transforming to polarization/harmonic-space.
    \item Define the relevant transfer-function-weighted maps (see below, e.g., $P[x]$ and $Q[x]$ maps for $\gnl$ and $\taunl$) for each field of interest and radial component.
    \item Compute the estimator numerators, $\widehat{\mathcal{N}}_\alpha$, via spin/pixel-space summation and chained harmonic transforms.
    \item Iterate over the $N_{\rm disc}$ input simulations, and combine the data and simulations to form the combined estimator numerator via \eqref{eq: general-estimators-harmonic}.
\end{enumerate}
For the Fisher matrix, we follow a similar strategy:
\begin{enumerate}
    \item Define a pair of Gaussian random fields $a^{(1)},a^{(2)}$ with known covariance $\mathsf{A}$ (usually set to the fiducial beam-deconvolved power spectrum in polarization/harmonic-space).
    \item Filter these maps by the $\Ai$ and $\Si\mathsf{P}$ linear operators, returning the outputs in polarization/harmonic-space.
    \item Compute the $Q_\alpha$ maps for each template of interest via chained harmonic transforms and numerical quadrature.
    \item Compute the Fisher contributions $F_{\alpha\beta}$ as the inner product (in polarization/harmonic-space) of the pairs of $Q_\alpha$ derivatives.
    \item Iterate over pairs of Monte Carlo realizations to form the expectation $\mathcal{F}_{\alpha\beta}$.
\end{enumerate}
Finally, we form the complete estimators by combining the numerator with the inverse Fisher matrix $\mathcal{F}^{-1}$. The result is an unbiased and quasi-optimal estimator for $\{A_\alpha\}$, which accounts for correlations between the various templates, sky cuts, experimental beams, polarization, and beyond.

\section{Estimation: Contact Trispectra}\label{sec: estimators-contact}
\noindent We now derive estimators for the contact trispectra discussed in \S\ref{subsec: contact-templates}. Our starting point is the polarization/harmonic-space trispectrum \eqref{eq: Tlm-general}, inserting the contact definition \eqref{eq: contact-exchange-tzeta}:
\beq\label{eq: Tlm-contact}
    \left.T^{\ell_1\ell_2\ell_3\ell_4,X_1X_2X_3X_4}_{m_1m_2m_3m_4}\right|_{\rm contact} &=& \int d\vr\,\prod_{i=1}^4\left[(-1)^{\ell_i}Y_{\ell_im_i}^*(\hr)\frac{2}{\pi}\int_0^\infty k_i^2dk_i\mathcal{T}_{\ell_i}^{X_i}(k_i)j_{\ell_i}(k_ir)\right]T_\zeta(k_1,k_2,k_3,k_4),
\eeq
where we have rewritten the Dirac delta as an exponential and integrated over $\hk_i$. This could be further simplified by computing the $\hr_i$ integral, yielding $3j$ symbols and reduced trispectrum. We will not need such a form in this work.

\subsection{\texorpdfstring{$\gnl$}{Local Shape}}\label{subsec: loc-estimator}
The $\gnl$ trispectrum can be written in the above form and explicitly separated in $k_i$, yielding 
\beq\label{eq: tlm-gloc}
    \partial_{\gnl}T^{\ell_1\ell_2\ell_3\ell_4,X_1X_2X_3X_4}_{m_1m_2m_3m_4} &=& \frac{54}{25}\int d\hr\,Y_{\ell_1m_1}^*(\hr)Y_{\ell_2m_2}^*(\hr)Y_{\ell_3m_3}^*(\hr)Y_{\ell_4m_4}^*(\hr)\\\nonumber
    &&\,\times\,\int_0^\infty r^2dr\,p_{\ell_1}^{X_1}(r)p_{\ell_2}^{X_2}(r)p_{\ell_3}^{X_3}(r)q_{\ell_4}^{X_4}(r)+\text{3 perms.},
\eeq
with\footnote{These are analogs to the functions appearing in KSW-type bispectrum estimators, with the relations $p^X_\ell(r)=(5/3)(-1)^{\ell}\beta^X_\ell(r), q^X_\ell(r) = (3/5)(-1)^{\ell}\alpha^X_\ell(r)$ \citep[e.g.,][]{Komatsu:2003iq}.}
\beq\label{eq: p-q-def}
    p_\ell^X(r) &\equiv& (-1)^\ell \frac{2}{\pi}\int_0^\infty k^2dk\,\mathcal{T}^X_\ell(k)j_\ell(kr)P_\zeta(k), \qquad q_\ell^X(r) \equiv (-1)^\ell \frac{2}{\pi}\int_0^\infty k^2dk\,\mathcal{T}^X_\ell(k)j_\ell(kr).
\eeq
This trispectrum can be evaluated with a single radial integral. Inserting this into the unsymmetrized estimator numerator $\tau$ \eqref{eq: general-estimators-harmonic}, we find
\begin{empheq}[box=\dbox]{align}\label{eq: gNL-estimator}
    \widehat{\mathcal{N}}_{g^{\rm loc}_{\rm NL}}[\alpha,\beta,\gamma,\delta] &= \frac{9}{100}\int_0^\infty r^2dr\,\int d\hr\,P[\Si\alpha](\hr,r)P[\Si\beta](\hr,r)P[\Si\gamma](\hr,r)Q[\Si\delta](\hr,r)+\text{3 perms.},
\end{empheq}
defining the real scalar maps
\beq\label{eq: P-Q-def}
    P[x](\hr,r) &\equiv& \sum_{\ell m X}p_{\ell}^X(r)Y_{\ell m}(\hr)x^{X}_{\ell m}, \qquad Q[x](\hr,r) \equiv \sum_{\ell m X}q_{\ell}^X(r)Y_{\ell m}(\hr)x^{X}_{\ell m}.
\eeq
The full estimator numerator can thus be estimated using only linear operations, first computing the $\Si x$ maps in polarization/harmonic-space (for $x\in\{\alpha,\beta,\gamma,\delta\}$), then multiplying by the (precomputed) cosmology-dependent functions $p_\ell^X(r)$ and $q_\ell^X(r)$ to form $P$ and $Q$ maps, and finally computing $\widehat{\mathcal{N}}_{\gnl}$ as a pixel-space summation (for $\hr$), and a numerical integral over $r$. In practice, we can use a coarse sampling for the integral, as discussed in \S\ref{sec: optim}. Notably, the addition of polarization is essentially trivial; we simply sum over the polarization axis, weighted by the relevant $p_\ell^X$ factor.

The $\gnl$ contribution to the Fisher matrix can be obtained similarly. As discussed above, we simply require the $Q$-derivative, which can be written
\begin{empheq}[box=\dbox]{align}
    Q^X_{\ell m,\gnl}[x,y,z] &= \frac{9}{25}\int_0^\infty r^2dr\,p_{\ell}^{X}(r)\int d\hr\,Y_{\ell m}^*(\hr)P[x](\hr,r)P[y](\hr,r)Q[z](\hr,r)+\text{23 perms.},
\end{empheq}
following an analogous procedure to the above. This is straightforward to estimate by transforming $[P^2Q](\hr,r)$ to polarization/harmonic-space, then multiplying by $p_\ell^X(r)$ and performing a numerical integral (and adding permutations).

\subsection{\texorpdfstring{$\gcon$}{Constant Shape}}\label{subsec: con-estimator}
\noindent The constant-shape model defined in \eqref{eq: con-shape} is straightforward to compute following the above techniques. From the template definition, we can immediately write
\beq
    \partial_{g_{\rm NL}^{\rm con}}T^{\ell_1\ell_2\ell_3\ell_4,X_1X_2X_3X_4}_{m_1m_2m_3m_4} &=& \frac{216}{25}\int d\hr\,\int_0^\infty r^2dr\,\prod_{i=1}^4\left[r_{\ell_i}^{X_i}(r)Y_{\ell_im_i}^*(\hr)\right],
\eeq
with
\beq
    r_{\ell}^{X}(r) \equiv (-1)^\ell\frac{2}{\pi}\int_0^\infty k^2dk\,\mathcal{T}_\ell^X(k)j_\ell(kr)P^{3/4}_\zeta(k).
\eeq
This yields the full estimator and the $Q$ derivative map:
\begin{empheq}[box=\dbox]{align}
    \widehat{\mathcal{N}}_{g^{\rm con}_{\rm NL}}[\alpha,\beta,\gamma,\delta] &= \frac{9}{25}\int_0^\infty r^2dr\,\int d\hr\,R[\Si\alpha](\hr,r)R[\Si\beta](\hr,r)R[\Si\gamma](\hr,r)R[\Si\delta](\hr,r)\\\nonumber
    Q^X_{\ell m,g_{\rm NL}^{\rm con}}[x,y,z] &= \frac{216}{25}\int_0^\infty r^2dr\,r_{\ell}^{X}(r)\int d\hr\,Y_{\ell m}^*(\hr)R[x](\hr,r)R[y](\hr,r)R[z](\hr,r),
\end{empheq}
subject to the definition
\beq
    R[x](\hr,r) &\equiv& \sum_{\ell m X}r_{\ell}^X(r)Y_{\ell m}(\hr)x^{X}_{\ell m}.
\eeq
This can be computed as for the $\gnl$ local shape, and requires only one radial integral.

\subsection{EFT of Inflation Templates}\label{subsec: efti-estimators}
\subsubsection{\texorpdfstring{$\gnldotdot$}{Shape 1}}\label{subsec: efti1-estimator}
\noindent As discussed in \citep{2015arXiv150200635S}, the EFT of inflation templates can be computed using similar methods to the local shape. Starting from \eqref{eq: Tlm-contact} and the shape definition \eqref{eq: EFTI-schwinger}, we can write
\beq
    \partial_{\gnldotdot}T^{\ell_1\ell_2\ell_3\ell_4,X_1X_2X_3X_4}_{m_1m_2m_3m_4} &=& \frac{9216}{25}\int d\vr\,\int_{-\infty}^0d\tau\,\tau^4\left[\prod_ia_{\ell_i}^{X_i}(r,\tau)Y_{\ell_im_i}^*(\hr)\right]
\eeq
defining the functions
\beq
    a_{\ell}^{X}(r,\tau) \equiv (-1)^\ell\frac{2}{\pi}\int_0^\infty k^2dk\,\mathcal{T}_\ell^X(k)j_\ell(kr)e^{k\tau}k^{5/4}P^{3/4}_\zeta(k),
\eeq
for a suitably discretized set of points in $r$ and $\tau$. Note that we have symmetrically replaced the scale-invariant power spectrum, $A_\zeta$, with $k^3P_\zeta(k)$, to ensure the correct overall scaling for $n_s\neq 1$. This yields the numerator (via \ref{eq: general-estimators-harmonic})
\begin{empheq}[box=\dbox]{align}
    \widehat{\mathcal{N}}_{\gnldotdot}[\alpha,\beta,\gamma,\delta] = \frac{384}{25}\int_0^\infty r^2dr\,\int_{-\infty}^0d\tau\,\tau^4\,\int d\hr\,A[\Si\alpha](\hr,r,\tau)A[\Si\beta](\hr,r,\tau)A[\Si\gamma](\hr,r,\tau)A[\Si\delta](\hr,r,\tau)
\end{empheq}
where 
\beq
    A[x](\hr,r,\tau) \equiv \sum_{\ell m X}a_{\ell}^{X}(r,\tau)Y_{\ell m}(\hr)x^X_{\ell m}.
\eeq
This can be computed similarly to $\gnl$, but now involves a double integral over $\tau,r$ instead of a single integral over $r$, due to the conformal time integral present in the template definition. As before, polarization is trivially included by summing over $X$.

The contribution to the Fisher matrix is analogous to the above, and yields
\begin{empheq}[box=\dbox]{align}
    Q_{\ell m,\gnldotdot}^{X}[x,y,z] = \frac{9216}{25}\int_0^\infty r^2dr\,\int_{-\infty}^0d\tau\,\tau^4a_{\ell}^{X}(r,\tau)\int d\hr\,Y_{\ell m}^*(\hr)A[x](\hr,r,\tau)A[y](\hr,r,\tau)A[z](\hr,r,\tau),
\end{empheq}
which can be computed via harmonic transforms, as before.

\subsubsection{\texorpdfstring{$\gnldotdel$}{Shape 2}}\label{subsec: efti2-estimator}
\noindent To form the trispectrum corresponding to the $\gnldotdel$ shape, we must take special care of the $\vk_3\cdot\vk_4$ factor (following \citep{2015arXiv150200635S}). Starting from \eqref{eq: Tlm-general}, the polarization/harmonic-space trispectrum involves integrals over $\hk_i$ of the form
\beq
    \int_{\vk}e^{i\vk\cdot\vr}\,Y_{\ell m}^*(\hk)f(k)\vk \equiv -i\nabla_{\vr}\left[\int_{\vk}e^{i\vk\cdot\vr}\,Y_{\ell m}^*(\hk)f(k)\right] = i^{\ell-1}\frac{2}{\pi}\int_0^\infty k^2dk\,f(k)\nabla_{\vr}\bigg(j_\ell(kr)Y_{\ell m}^*(\hr)\bigg),
\eeq
for some scalar valued function $f(k)$. The inner product of two derivatives can be expanded in radial and angular components as
\beq
    \nabla\,f(\vr)\cdot\nabla\,g(\vr) = \partial_rf(\vr)\partial_rg(\vr)+\frac{1}{2r^2}\left[\edth f(\vr)\bar\edth g(\vr)+\bar\edth f(\vr)\edth g(\vr)\right],
\eeq
for spin-raising and spin-lowering operators $\edth$ and $\bar\edth$ \citep[e.g.,][]{Castro:2005bg}.\footnote{This is straightforwardly proven by writing $\nabla \equiv (\partial_r,\partial_\theta/r, \csc\theta\partial_\phi/r)$ in spherical polars, and inserting the definition of $\edth$ and $\bar\edth$, noting that $f$ and $g$ are spin-$0$.} Noting that $\edth Y^*_{\ell m}(\hr) = -\sqrt{\ell(\ell+1)}{}_{-1}Y_{\ell m}^*(\hr)$ and $\bar\edth Y^*_{\ell m}(\hr) = \sqrt{\ell(\ell+1)}{}_{+1}Y^*_{\ell m}(\hr)$, the derivatives corresponding to $\vk_3\cdot\vk_4$ can be written
\beq
    &&\nabla_{\vr}\bigg(j_\ell(k_3r)Y_{\ell_3 m_3}^*(\hr)\bigg)\cdot\nabla_{\vr}\bigg(j_{\ell_4}(k_4r)Y_{\ell_4 m_4}^*(\hr)\bigg) = k_3k_4j_{\ell_3}'(k_3r)j_{\ell_4}'(k_4r)Y_{\ell_3 m_3}^*(\hr)Y_{\ell_4m_4}^*(\hr)\\\nonumber
    &&\qquad\qquad\qquad\qquad\qquad\qquad\qquad\qquad\,-\,\sum_{\mu=\pm1}\frac{j_{\ell_3}(k_3r)j_{\ell_4}(k_4r)}{2r^2}\sqrt{\ell_3(\ell_3+1)\ell_4(\ell_4+1)}{}_{+\mu}Y^*_{\ell_3m_3}(\hr){}_{-\mu}Y_{\ell_4m_4}^*(\hr).
\eeq
This is a sum of three terms, each explicitly separable in $k_3$ and $k_4$.

The $\gnldotdel$ trispectrum can thus be written
\beq\label{eq: dot-del-Tlm}
    \partial_{\gnldotdel}T^{\ell_1\ell_2\ell_3\ell_4,X_1X_2X_3X_4}_{m_1m_2m_3m_4} &=& \frac{13824}{325}\int d\vr\,Y_{\ell_1m_1}^*(\hr)Y_{\ell_2m_2}^*(\hr)\int_{-\infty}^0d\tau\,\tau^2\,a_{\ell_1}^{X_1}(r,\tau)a_{\ell_2}^{X_2}(r,\tau)\\\nonumber
    &&\,\times\,\bigg[b_{\ell_3}^{X_3}(r,\tau)b_{\ell_4}^{X_4}(r,\tau)Y_{\ell_3m_3}^*(\hr)Y_{\ell_4m_4}^*(\hr)\\\nonumber
    &&\qquad\quad\,-\,\frac{1}{2}c_{\ell_3}^{X_3}(r,\tau)c_{\ell_4}^{X_4}(r,\tau)\sum_{\mu=\pm 1}{}_{+\mu}Y^*_{\ell_3m_3}(\hr){}_{-\mu}Y^*_{\ell_4m_4}(\hr)\bigg]\,+\,\text{5 perms.}
\eeq
subject to the definitions
\beq
    b_\ell^X(r,\tau) &\equiv& (-1)^\ell\frac{2}{\pi}\int_0^\infty k^2dk\,\mathcal{T}_{\ell}^{X}(k)j'_{\ell}(kr)(1-k\tau)k^{1/4}P^{3/4}_\zeta(k)e^{k\tau}\\\nonumber
    c_\ell^X(r,\tau) &\equiv& (-1)^\ell\frac{\sqrt{\ell(\ell+1)}}{r}\frac{2}{\pi}\int_0^\infty k^2dk\,\mathcal{T}_{\ell}^{X}(k)j_{\ell}(kr)(1-k\tau)k^{-3/4}P^{3/4}_\zeta(k)e^{k\tau},
\eeq
where the Bessel function derivatives can be evaluated using the identity $j_\ell'(x) = [\ell j_{\ell-1}(x)-(\ell-1)j_{\ell+1}(x)]/(2\ell+1)$.

As before, the estimator numerator is formed by contracting this with the data, yielding
\begin{empheq}[box=\dbox]{align}
    \widehat{\mathcal{N}}_{\gnldotdel}[\alpha,\beta,\gamma,\delta] &= \frac{288}{325}\int_0^\infty r^2dr\,\int_{-\infty}^0d\tau\,\tau^2\,\int d\hr\,A[\Si\alpha](\hr,r,\tau)A[\Si\beta](\hr,r,\tau)\\\nonumber
    &\,\times\,\bigg\{B[\Si\gamma](\hr,r,\tau)B[\Si\delta](\hr,r,\tau)-\mathrm{Re}\left[{}_{-1}C[\Si\gamma](\hr,r,\tau){}_{+1}C[\Si\delta](\hr,r,\tau)\right]\bigg\}\\\nonumber
    &\,+\,\text{11 perms.},
\end{empheq}
with
\beq
    B[x](\hr,r,\tau) &\equiv& \sum_{\ell m X}b_\ell^X(r,\tau)Y_{\ell m}(\hr)x^{X}_{\ell m}, \qquad     {}_sC[x](\hr,r,\tau) \equiv \sum_{\ell m X}c_\ell^X(r,\tau){}_{s}Y_{\ell m}(\hr)x^{X}_{\ell m}.
\eeq
Note that ${}_{-s}C[x](\hr,r,\tau)=(-1)^s{}_s\left(C[x](\hr,r,\tau)\right)^*$; this implies that $\{{}_{+s}C,(-1)^s{}_{-s}C\}$ is a spin-$s$ pair of maps. This numerator is slightly harder to compute than the $\gnldotdot$ numerator due to the spin-$\pm1$ transforms and the more complex permutation structures.

For the Fisher matrix derivatives, we proceed analogously, finding
\begin{empheq}[box=\dbox]{align}
    Q_{\ell m,\gnldotdel}^X[x,y,z] &= \frac{6912}{325}\int_0^\infty r^2dr\,\int_{-\infty}^0d\tau\,\tau^2\bigg\{a_{\ell}^{X}(r,\tau)\int d\hr\,A[x](\hr,r,\tau)Y_{\ell m}^*(\hr)\\\nonumber
    &\,\times\,\,\bigg(B[y](\hr,r,\tau)B[z](\hr,r,\tau)-\mathrm{Re}\left[{}_{-1}C[y](\hr,r,\tau){}_{+1}C[z](\hr,r,\tau)\right]\bigg)\\\nonumber
    &\,+\,\int d\hr\,A[y](\hr,r,\tau)A[z](\hr,r,\tau)\\\nonumber
    &\,\times\,\bigg[b_{\ell}^{X}(r,\tau)B[x](\hr,r,\tau)Y_{\ell m}^*(\hr)+\frac{1}{2}c_{\ell}^{X}(r,\tau)\sum_{\mu=\pm 1}{}_{+\mu}C[x](\hr,r,\tau){}_{+\mu}Y^*_{\ell m}(\hr)\bigg]\bigg\}\,+\,\text{5 perms.},
\end{empheq}
separating out two classes of permutations. This can be computed via spin-$0$ and spin-$\pm1$ harmonic transforms as well as a discretized sum over $r$ and $\tau$.

\subsubsection{\texorpdfstring{$\gnldeldel$}{Shape 3}}\label{subsec: efti3-estimator}
\noindent The $\gnldeldel$ trispectrum is similar to that of $\gnldotdel$ but involves two $\vk_i\cdot\vk_j$ scalar products. Following the above logic (see also \citep{2015arXiv150200635S}), we find the polarization/harmonic-space trispectrum
% \beq
%     \partial_{\gnldeldel} T^{\ell_1\ell_2\ell_3\ell_4,X_1X_2X_3X_4}_{m_1m_2m_3m_4} &=& \frac{82944}{2575}\int d\vr\,\int_{-\infty}^0d\tau\,\\\nonumber
%     &&\times\,\left[b_{\ell_1}^{X_1}(r,\tau)b_{\ell_2}^{X_2}(r,\tau)Y^*_{\ell_1m_1}(\hr)Y^*_{\ell_2m_2}(\hr)-\frac{1}{2}c_{\ell_1}^{X_1}(r,\tau)c_{\ell_2}^{X_2}(r,\tau)\sum_{\mu=\pm1}{}_{+\mu}Y^*_{\ell_1m_1}(\hr){}_{-\mu}Y^*_{\ell_2m_2}(\hr)\right]\\\nonumber
%     &&\times\,\left[b_{\ell_3}^{X_3}(r,\tau)b_{\ell_4}^{X_4}(r,\tau)Y^*_{\ell_3m_3}(\hr)Y^*_{\ell_4m_4}(\hr)-\frac{1}{2}c_{\ell_3}^{X_3}(r,\tau)c_{\ell_4}^{X_4}(r,\tau)\sum_{\nu=\pm1}\,{}_{+\nu}Y^*_{\ell_3m_3}(\hr){}_{-\nu}Y^*_{\ell_4m_4}(\hr)\right]\\\nonumber
%     &&\,+\,\text{2 perms.}
% \eeq
\beq
    \partial_{\gnldeldel} T^{\ell_1\ell_2\ell_3\ell_4,X_1X_2X_3X_4}_{m_1m_2m_3m_4} &=& \frac{82944}{2575}\int d\vr\,\int_{-\infty}^0d\tau\,\bigg[b_{\ell_1}^{X_1}(r,\tau)b_{\ell_2}^{X_2}(r,\tau)Y^*_{\ell_1m_1}(\hr)Y^*_{\ell_2m_2}(\hr)\\\nonumber
    &&\qquad\qquad\qquad\qquad\qquad\,-\frac{1}{2}c_{\ell_1}^{X_1}(r,\tau)c_{\ell_2}^{X_2}(r,\tau)\sum_{\mu=\pm1}{}_{+\mu}Y^*_{\ell_1m_1}(\hr){}_{-\mu}Y^*_{\ell_2m_2}(\hr)\bigg]\\\nonumber
    &&\qquad\qquad\qquad\qquad\,\times\,\bigg[\{\ell_1,\ell_2,X_1,X_2\}\leftrightarrow\{\ell_3,\ell_4,X_3,X_4\}\bigg]+\text{2 perms.},
\eeq
which now involves two pairs of spin-$\pm1$ spherical harmonics.
%This bears strong similarities to \eqref{eq: dot-del-Tlm}, but with two derivative terms, appearing from the pair of $(\partial\sigma)^2$ factors. 
The estimator numerator follows straightforwardly:
\begin{empheq}[box=\dbox]{align}
    \widehat{\mathcal{N}}_{\gnldeldel}[\alpha,\beta,\gamma,\delta] &=\frac{1728}{2575}\int_0^\infty r^2dr\,\int_{-\infty}^0d\tau\,\int d\hr\,\\\nonumber
    &\,\times\,\bigg(B[\Si\alpha](\hr,r,\tau)B[\Si\beta](\hr,r,\tau)-\mathrm{Re}\left[{}_{-1}C[\Si\alpha](\hr,r,\tau){}_{+1}C[\Si\beta](\hr,r,\tau)\right]\bigg)\\\nonumber
    &\,\times\,\bigg(B[\Si\gamma\,](\hr,r,\tau)B[\Si\delta](\hr,r,\tau)-\mathrm{Re}\left[{}_{-1}C[\Si\gamma\,](\hr,r,\tau){}_{+1}C[\Si\delta](\hr,r,\tau)\right]\bigg)\\\nonumber
    &\,+\,\text{5 perms.};
\end{empheq}
this is easily evaluated via map-space summation, given the $B$ and ${}_{\pm 1}C$ maps discussed above.

The Fisher matrix is formed similarly, and involves the derivative
\begin{empheq}[box=\dbox]{align}
    Q_{\ell m,\gnldeldel}^X[x,y,z] &= \frac{41472}{2575}\int_0^\infty r^2dr\,\int_{-\infty}^0d\tau\,\int d\hr\,\\\nonumber
    &\,\times\,\left(b_{\ell}^{X}(r,\tau)B[x](\hr,r,\tau)Y^*_{\ell m}(\hr)+\frac{1}{2}c_{\ell}^{X}(r,\tau)\sum_{\mu=\pm1}{}_{+\mu}C[x](\hr,r,\tau){}_{+\mu}Y^*_{\ell m}(\hr)\right)\\\nonumber
    &\,\times\,\bigg(B[y](\hr,r,\tau)B[z](\hr,r,\tau)-\mathrm{Re}\left[{}_{-1}C[y](\hr,r,\tau){}_{+1}C[z](\hr,r,\tau)\right]\bigg)\\\nonumber
    &\,+\,\text{5 perms.}.
\end{empheq}
This can be computed by forming the pixel-space maps $B(B^2-\mathrm{Re}[{}_{-1}C{}_{+1}C])$ and ${}_{\pm1}C(B^2-\mathrm{Re}[{}_{-1}C{}_{+1}C])$ and performing spin-$0$ and spin-$\pm1$ weighted inverse harmonic transforms.

\section{Estimation: Exchange Trispectra}\label{sec: estimators-exchange}
\noindent Next, we build estimators for exchange-factorizable templates. First, we rewrite the trispectrum definition in a more convenient form following \eqref{eq: contact-exchange-tzeta}:
\beq\label{eq: exchange-Tk}
    \left.T^{\ell_1\ell_2\ell_3\ell_4,X_1X_2X_3X_4}_{m_1m_2m_3m_4}\right|_{\rm exchange} &=&
    \int_{\vK}\,T_\zeta(
    \vk_1,\vk_2,\vk_3,\vk_4;\vK)\\\nonumber
    &&\,\times\,\left(\int d\vr\,e^{-i\vK\cdot\vr}\prod_{i=1,2}\left[\int_{\vk_i}4\pi i^{\ell_i}\mathcal{T}_{\ell_i}^{X_i}(k_i)Y_{\ell_im_i}^*(\hk_i)e^{i\vk_i\cdot\vr}\right]\right)\\\nonumber
    &&\,\times\,\left(\int d\vr'\,e^{i\vK\cdot\vr'}\prod_{i=3,4}\left[\int_{\vk_i}4\pi i^{\ell_i}\mathcal{T}_{\ell_i}^{X_i}(k_i)Y_{\ell_im_i}^*(\hk_i)e^{i\vk_i\cdot\vr'}\right]\right).
\eeq
As before, we have rewritten the Dirac deltas as exponentials, which now involve a pair of $\vr,\vr'$ dummy integrals. Further simplification of this expression depends on the form of $T_\zeta$; as we see below, the expressions are considerably simpler if the trispectrum has no explicit dependence on $\hk_i$ and $\hK$.

\subsection{\texorpdfstring{Local Trispectra: $\taunl$}{Local Trispectra}}\label{subsec: loctau-estimator}
\noindent From \eqref{eq: tauNL-shape}, the local $\taunl$ trispectrum is explicitly separable in $k_i$ and does not contain additional $\hk_i$ and $\hK$ factors. In this case, we can simplify the general trispectrum form \eqref{eq: exchange-Tk} by expanding the exponentials and performing the angular momentum integrals by spherical harmonic orthgonality; this leads to
\beq\label{eq: exchange-Tlm}
    \left.T^{\ell_1\ell_2\ell_3\ell_4,X_1X_2X_3X_4}_{m_1m_2m_3m_4}\right|_{\rm exchange} &=&
    \frac{2}{\pi}\int_0^\infty K^2dK\,T_\zeta(k_1,k_2,k_3,k_4,K)\sum_{LM}\\\nonumber
    &&\,\times\,\left(\int d\vr\,Y^*_{LM}(\hr)\prod_{i=1,2}\left[(-1)^{\ell_i}Y_{\ell_im_i}^*(\hr)\frac{2}{\pi}\int_0^\infty k_i^2dk_i\,\mathcal{T}_{\ell_i}^{X_i}(k_i)j_{\ell_i}(k_ir)\right]j_L(Kr)\right)\\\nonumber
    &&\,\times\,\left(\int d\vr'\,Y_{LM}(\hr')\prod_{i=3,4}\left[(-1)^{\ell_i}Y_{\ell_im_i}^*(\hr')\frac{2}{\pi}\int_0^\infty k_i^2dk_i\,\mathcal{T}_{\ell_i}^{X_i}(k_i)j_{\ell_i}(k_ir')\right]j_L(Kr')\right),
\eeq
for $L\geq 1$. The $\taunl$ trispectrum is given by 
\beq\label{eq: Tlm-tauNL}
    \partial_{\taunl}T^{\ell_1\ell_2\ell_3\ell_4,X_1X_2X_3X_4}_{m_1m_2m_3m_4} &=& \sum_{LM}\int_0^\infty r^2dr\,\int_0^\infty r'^2dr'\,F_L(r,r')\,p_{\ell_1}^{X_1}(r)q_{\ell_3}^{X_3}(r)\,p_{\ell_2}^{X_2}(r')q_{\ell_4}^{X_4}(r')\\\nonumber
    &&\,\times\,\left(\int d\hr\,Y_{\ell_1m_1}^*(\hr)Y_{\ell_3m_3}^*(\hr)Y^*_{LM}(\hr)\right)\left(\int d\hr'\,Y_{\ell_2m_2}^*(\hr')Y_{\ell_4m_4}^*(\hr')Y_{LM}(\hr')\right)+\text{11 perms.},
\eeq
where $p_\ell^X$ and $q_\ell^X$ were defined in \eqref{eq: p-q-def}, and we introduce the coupling \citep[cf.][]{2015arXiv150200635S}:
\beq\label{eq: FL-def}
    F_L(r,r') \equiv \frac{2}{\pi}\int_0^\infty K^2dK\,j_L(Kr)j_L(Kr')P_\zeta(K).
\eeq
For a power-law primordial cosmology with $P_\zeta(k) = 2\pi^2/k^3\times A_\zeta(k/k_*)^{n_s-1}$), the $F_L$ function can be simplified analytically \citep[6.574.1]{nist_dlmf}:
\beq
    F_{L}(r,r') &=& \frac{2\pi^2 A_\zeta}{(k_\ast r')^{n_s-1}}\left(\frac{r}{r'}\right)^L\frac{2^{n_s-3}\Gamma\left(L-\frac{1}{2}+\frac{n_s}{2}\right)}{\Gamma\left(2-\frac{n_s}{2}\right)\Gamma\left(L+\frac{3}{2}\right)}{}_2F_1\left(L-\frac{1}{2}+\frac{n_s}{2},\frac{n_s}{2}-1,L+\frac{3}{2},\left(\frac{r}{r'}\right)^2\right),
\eeq
fixing $r<r'$ wlog. This uses the Gamma function $\Gamma$ and the confluent hypergeometric function ${}_2F_1$. \eqref{eq: Tlm-tauNL} could be further simplified by analytic integration over $\hr,\hr'$, though this does not aid us in our quest for separable estimators.

Inserting the $\taunl$ trispectrum into the general estimator \eqref{eq: general-estimators-harmonic}, we find
\begin{empheq}[box=\dbox]{align}\label{eq: tauNL-loc-num}
    \widehat{\mathcal{N}}_{\taunl}[\alpha,\beta,\gamma,\delta] &=\frac{1}{24}\sum_{LM}F_L(r,r')\left(\int_0^\infty r^2dr\,\int d\hr\,Y^*_{LM}(\hr)P[\Si\alpha](\hr,r)Q[\Si\gamma](\hr,r)\right)\\\nonumber
    &\qquad\,\times\,\left(\int_0^\infty r'^2dr'\,\int d\hr'\,Y_{LM}^*(\hr')P[\Si\beta](\hr',r')Q[\Si\delta](\hr',r')\right)^*+\text{11 perms.};
\end{empheq}
roughly speaking, this is the angular power spectrum of the locally measured power spectrum \citep[cf.][]{Hanson:2009gu,Feng:2015pva,Marzouk:2022utf}.\footnote{This analogy will be discussed in detail in \S\ref{sec: estimator-comparison}, allowing for comparison with standard estimators.} As for contact trispectra, this can computed with linear operations, in particular:
\begin{enumerate}
    \item Filter the $\alpha,\beta,\gamma,\delta$ data or random-field maps by the weighting scheme $\Si$.
    \item Compute the (real) $P,Q$ pixel-space maps for each $r$ of interest using inverse spherical harmonic transforms.
    \item Compute the quadratic combination $[PQ]_{LM}(r) \equiv \int d\hr\,Y_{LM}^*(\hr)P(\hr,r)Q(\hr,r)$ via a harmonic transform and sum over $r$, weighting by $F_L(r,r')$.
    \item Combine $[PQ]^*_{LM}(r')$ and $(F_L [PQ])_{LM}(r')$ via numerical integration and a polarization/harmonic-space sum.
\end{enumerate}
The contribution to the Fisher matrix can be obtained similarly and involves the derivative operator:
\begin{empheq}[box=\dbox]{align}\label{eq: Q-tauNL}
    Q^X_{\ell m,\taunl}[x,y,z] &= \frac{1}{2}\int_0^\infty r^2dr\,\left[p_{\ell}^{X}(r)\int d\hr\,Y_{\ell m}^*(\hr)Q[x](\hr,r)+q_{\ell}^{X}(r)\int d\hr\,Y_{\ell m}^*(\hr)P[x](\hr,r)\right]\\\nonumber
    &\qquad\,\times\,\left(\sum_{LM}Y_{LM}(\hr)\left(\int_0^\infty r'^2dr' F_L(r,r')\int d\hr'\,P[y](\hr',r')Q[z](\hr',r')Y_{LM}^*(\hr')\right)\right)\\\nonumber
    &+\text{11 perms.},
\end{empheq}
separating out the two permutations of interest. This can be computed using chained harmonic transforms and integration over $r,r'$ as for $\widehat{\mathcal{N}}_{\taunl}$.

\subsection{\texorpdfstring{Direction-Dependent Trispectra: $\tau_{\rm NL}^{n_1n_3n}$}{Direction-Dependent Trispectra}}\label{subsec: direction-estimator}\label{subsec: direc-estimator}

\noindent Due to the additional angular dependence, the trispectrum templates given in \S\ref{subsubsec: direction-templates} require a slightly more nuanced computation strategy. Starting from the generalized $\tau_{\rm NL}^{n_1n_3n}$ template \eqref{eq: dnnn-def}, we can write the polarization/harmonic-space trispectrum as
\beq
    \partial_{\tau_{\rm NL}^{n_1n_3n}}T^{\ell_1\ell_2\ell_3\ell_4,X_1X_2X_3X_4}_{m_1m_2m_3m_4} &=&
    \frac{1}{2}\sum_{\mu_1\mu_3\mu}\tj{n_1}{n_3}{n}{\mu_1}{\mu_3}{\mu}\int_{\vK}P_\zeta(K)e^{i\vK\cdot(\vr'-\vr)}Y_{n\mu}(\hK)\\\nonumber
    &&\,\times\,\left(\int d\vr\,\,q^{X_2}_{\ell_2}(r)\,\,Y_{\ell_2m_2}^*(\hr)\,\,\left[\int_{\vk_1}4\pi i^{\ell_1}\mathcal{T}_{\ell_1}^{X_1}(k_1)P_\zeta(k_1)Y_{\ell_1m_1}^*(\hk_1)Y_{n_1\mu_1}(\hk_1)e^{i\vk_1\cdot\vr}\,\right]\right)\\\nonumber
    &&\,\times\,\left(\int d\vr'\,q_{\ell_4}^{X_4}(r')Y^*_{\ell_4m_4}(\hr')\left[\int_{\vk_3}4\pi i^{\ell_3}\mathcal{T}_{\ell_3}^{X_3}(k_3)P_\zeta(k_3)Y_{\ell_3m_3}^*(\hk_3)Y_{n_3\mu_3}(\hk_3)e^{i\vk_3\cdot\vr'}\right]\right)\\\nonumber
    &&\,+\,\text{23 perms.}
\eeq
using \eqref{eq: exchange-Tk} and inserting the $q_\ell^X$ factors from \eqref{eq: p-q-def}. The $\vK$-integral can be simplified analogous via
\beq\label{eq: partial-K-int}
	\int_{\vK}P_\zeta(K)e^{i\vK\cdot(\vr'-\vr)}Y_{n\mu}(\hK) &=& \sum_{LL'MM'}Y^*_{LM}(\hr)Y^*_{L'M'}(\hr')i^{L'-L}\left(\frac{2}{\pi}\int_0^\infty K^2dK\,j_L(Kr)j_{L'}(Kr')P_\zeta(K)\right)\G^{LL'n}_{MM'\mu}\nonumber\\
	&\equiv&	\sum_{LL'MM'}Y^*_{LM}(\hr)Y^*_{L'M'}(\hr')i^{L'-L}F_{LL'}(r,r')\G^{LL'n}_{MM'\mu}.
\eeq
Here, we have expanded exponential factors and defined a coupling matrix $F_{LL'}(r,r')$ (which reduces to the $F_L(r,r')$ function of \ref{eq: FL-def} for $L=L'$); this can again be expressed analytically for a power-law cosmology \citep[6.574.1]{nist_dlmf}:
\beq\label{eq: FLL-analyt} 
    F_{LL'}(r,r') &=& \frac{2\pi^2 A_\zeta}{(k_\ast r')^{n_s-1}}\left(\frac{r}{r'}\right)^{L}\frac{2^{n_s-3}\Gamma\left(\frac{L+L'}{2}-\frac{1}{2}+\frac{n_s}{2}\right)}{\Gamma\left(\frac{L'-L}{2}+2-\frac{n_s}{2}\right)\Gamma\left(L+\frac{3}{2}\right)}\\\nonumber
    &&\,\times\,{}_2F_1\left(\frac{L+L'}{2}-\frac{1}{2}+\frac{n_s}{2},\frac{L-L'}{2}+\frac{n_s}{2}-1,L+\frac{3}{2},\left(\frac{r}{r'}\right)^2\right)
\eeq
assuming $r<r'$ wlog. In \eqref{eq: partial-K-int}, we have also introduced the Gaunt factor, $\G$, as the angular integral over three spherical harmonics:
\beq
    \mathcal{G}^{\ell_1\ell_2\ell_3}_{m_1m_2m_3} \equiv \sqrt{\frac{(2\ell_1+1)(2\ell_2+1)(2\ell_3+1)}{4\pi}}\tj{\ell_1}{\ell_2}{\ell_3}{m_1}{m_2}{m_3}\tjo{\ell_1}{\ell_2}{\ell_3}.
\eeq
For small $n,\mu$, this is cheap to compute since $L'$ is restricted to $|L-n|\leq L'\leq L+n$, with $n=\mu=0$ enforcing $L=L'$, $M+M'=0$, recovering the $\taunl$ form. We can additionally simplify the $\hk_{1,3}$ integrals via 
\beq:
	\int_{\vk}4\pi i^{\ell}\mathcal{T}_{\ell}^{X}(k)P_\zeta(k)Y_{\ell m}^*(\hk)Y_{n\mu}(\hk)e^{i\vk\cdot\vr} &=& \sum_{\ell'm'}i^{\ell'-\ell}\left((-1)^{\ell}\frac{2}{\pi}\int_0^\infty k^2dk\,\mathcal{T}_{\ell}^{X}(k)j_{\ell'}(xr)P_\zeta(k)\right)Y_{\ell'm'}(\hr)(-1)^{\mu}\G^{\ell \ell'n}_{mm'(-\mu)}\nonumber\\
	&\equiv& \sum_{\ell'm'}i^{\ell'-\ell}p^X_{\ell \ell'}(r)Y_{\ell'm'}(\hr)(-1)^\mu\G^{\ell \ell'n}_{mm'(-\mu)}
\eeq
defining $p^X_{\ell \ell'}(r)$ as the generalization of $p_\ell^X$ \eqref{eq: p-q-def}, with $p_{\ell\ell}^X=p_\ell^X$. 

Combining the above ingredients, we obtain the trispectrum numerator for $\tau_{\rm NL}^{n_1n_3n}$:
\begin{empheq}[box=\dbox]{align}
    \widehat{\mathcal{N}}_{\tau_{\rm NL}^{n_1n_3n}}[\alpha,\beta,\gamma,\delta] &= \frac{1}{48}\sum_{\mu_1\mu_3\mu}\tj{n_1}{n_3}{n}{\mu_1}{\mu_3}{\mu}\sum_{LL'MM'}\int_0^\infty r^2dr\,\int_0^\infty r'^2dr'\,i^{L'-L}F_{LL'}(r,r')\G^{LL'n}_{MM'\mu}\\\nonumber
    &\,\times\,\int d\hr\,Y^*_{LM}(\hr)P_{n_1\mu_1}[\Si\alpha](\hr,r)Q[\Si\beta](\hr,r)\\\nonumber
    &\,\times\,\int d\hr'\,Y^*_{L'M'}(\hr')P_{n_3\mu_3}[\Si\gamma](\hr',r')Q[\Si\delta](\hr',r')\,+\,\text{23 perms.},
\end{empheq}
defining the maps
\beq
    P_{n\mu}[x](\hr,r) &\equiv& \sum_{\ell'm'}\left(\sum_{\ell m X}i^{\ell'-\ell}(-1)^\mu \G^{\ell \ell'n}_{mm'(-\mu)}p_{\ell \ell'}^{X}(r)x^{X*}_{\ell m}\right)Y_{\ell'm'}(\hr),
\eeq
which can be computed via a spherical harmonic transform. These have the conjugate relation $P_{n\mu}^*[x](\hr,r) = (-1)^{n+\mu}P_{n(-\mu)}[x](\hr,r)$, and satisfy $\sqrt{4\pi}\,P_{00}[x](\hr,r)=P[x](\hr,r)$ \eqref{eq: P-Q-def}. As for the $\hK$-integral, the presence of the Gaunt symbol does not greatly increase computation time, since triangle conditions restrict the range of allowed momenta to $|\ell-n|\leq \ell'\leq \ell+n$ \citep[cf.][]{Duivenvoorden:2019ses}. To compute the overall estimator, we first obtain $P_{n\mu}$ and $Q$ via harmonic transforms of the data or random fields, then perform an inverse harmonic transform of $P_{n_3\mu_3}Q$, a matrix product with $\G$ and $F_{LL'}$, a forward harmonic transform for $Y_{LM}$, and a summation over $r$ and $\mu_i$. This requires $\sim (2n+1)^3$ times more harmonic transforms than the $\taunl$ estimator, but can still be computed with only linear operations.

Gradient maps $Q_{\tau_{\rm NL}^{n_1n_3n}}$ are computed analogously to the above. Skipping a laborious computation, we find the lengthy result
% \beq
%     Q^X_{\ell m,\tau_{\rm NL}^{n_1n_3n}}[x,y,z] &=& \frac{1}{4}\int_0^\infty r^2dr\,\sum_{\mu_1}\\\nonumber
%     &&\,\times\,\left[\sum_{L_1M_1}i^{L_1-\ell}(-1)^{\mu_1}\G^{\ell L_1n_1}_{mM_1(-\mu_1)}p_{\ell L_1}^X(r)\int d\hr\,Y_{L_1M_1}(\hr)Q[x](\hr,r)+q_{\ell}^X(r)\int d\hr\,Y_{\ell m}^*(\hr)P_{n_1\mu_1}[x](\hr,r)\right]\\\nonumber
%     &&\,\times\,
%     \sum_{\mu_3\mu}\tj{n_1}{n_3}{n}{\mu_1}{\mu_3}{\mu}\left(\sum_{LM}Y_{LM}^*(\hr)\left(\sum_{L'M'}\int_0^\infty r'^2dr'\,i^{L'-L}F_{LL'}(r,r')\G^{LL'n}_{MM'\mu}\right.\right.\\\nonumber
%     &&\qquad\,\times\,\left.\left.\int d\hr'\,P_{n_3\mu_3}[y](\hr',r')Q[z](\hr',r')Y^*_{L'M'}(\hr')\right)\right)\,+\,\text{23 perms.},
% \eeq
\begin{empheq}[box=\dbox]{align}
    Q^X_{\ell m,\tau_{\rm NL}^{n_1n_3n}}[x,y,z] &= q^X_{\ell m,\tau_{\rm NL}^{n_1n_3n}}[x,y,z]+(-1)^{n_1+n_3}q^X_{\ell m,\tau_{\rm NL}^{n_3n_1n}}[x,y,z]\\\nonumber
    q^X_{\ell m,\tau_{\rm NL}^{n_1n_3n}}[x,y,z]   
    &= \frac{1}{4}\int_0^\infty r^2dr\,\sum_{\mu_1}\left[\sum_{\ell'm'}i^{\ell'-\ell}(-1)^{m}\G^{\ell \ell'n_1}_{m(-m')(-\mu_1)}p_{\ell \ell'}^X(r)\int d\hr\,Y^*_{\ell'm'}(\hr)Q[x](\hr,r)\right.\\\nonumber
    &\qquad\qquad\qquad\qquad\qquad\,\left.+\,q_{\ell}^X(r)\int d\hr\,Y_{\ell m}^*(\hr)P_{n_1\mu_1}[x](\hr,r)\right]\\\nonumber
    &\times\,
    \sum_{\mu_3\mu}\tj{n_1}{n_3}{n}{\mu_1}{\mu_3}{\mu}\left(\sum_{LM}(-1)^MY_{LM}(\hr)\left(\sum_{L'M'}\int_0^\infty r'^2dr'\,i^{L'-L}F_{LL'}(r,r')\G^{LL'n}_{(-M)M'\mu}\right.\right.\\\nonumber
    &\qquad\,\times\,\left.\left.\int d\hr'\,P_{n_3\mu_3}[y](\hr',r')Q[z](\hr',r')Y^*_{L'M'}(\hr')\right)\right)\,+\,\text{11 perms.},
\end{empheq}
which can be implemented sequentially, as before. We note that the permutations include interchange of $n_1$ and $n_3$; this is necessary to ensure that the trispectrum definition is symmetric and imply that the Fisher matrix is singular if both $\tau_{\rm NL}^{n_1n_3n}$ and $\tau_{\rm NL}^{n_3n_1n}$ are included in the analysis. The gradient satisfies the conjugation relation
\beq
    \left[Q^X_{\ell m,\tau_{\rm NL}^{n_1n_3n}}[x,y,z]\right]^* &=& (-1)^mQ^X_{\ell(-m),\tau_{\rm NL}^{n_1n_3n}}[x,y,z],
\eeq
implying that its harmonic transform is a real spin-zero map. 

Although the above derivation focused on the generalized $\tau_{\rm NL}^{n_1n_3n}$ forms, the results can be recast as estimators for $\tau_{\rm NL}^{n, \rm even}$ and $\tau_{\rm NL}^{n, \rm odd}$ (as defined in \eqref{eq: dn-even-def}\,\&\,\eqref{eq: dn-odd-def}) using relations \eqref{eq: dn-even-dnnn}\,\&\,\eqref{eq: dn-odd-dnnn}. By linearity, we find
\begin{empheq}[box=\dbox]{align}
    \widehat{\mathcal{N}}_{\tau_{\rm NL}^{n,\rm even}} &= \frac{(-1)^n}{3}\frac{(4\pi)^{3/2}}{\sqrt{2n+1}}\left[\widehat{\mathcal{N}}_{\tau_{\rm NL}^{nn0}}+2\widehat{\mathcal{N}}_{\tau_{\rm NL}^{0nn}}\right]\\\nonumber
    \widehat{\mathcal{N}}_{\tau_{\rm NL}^{n, \rm odd}}&=\frac{\sqrt{2}(-1)^n}{3}(4\pi)^{3/2}\sum_{NN'}\sqrt{(2N+1)(2N'+1)}\tjo{N}{1}{n}\tjo{N'}{1}{n}\begin{Bmatrix} N & 1 & N'\\ 1 & n & 1\end{Bmatrix}\\\nonumber
    &\,\times\,\left[\widehat{\mathcal{N}}_{\tau_{\rm NL}^{NN'1}}+(-1)^n\widehat{\mathcal{N}}_{\tau_{\rm NL}^{N1N'}}+\widehat{\mathcal{N}}_{\tau_{\rm NL}^{1NN'}}\right]
\end{empheq}
noting that $\widehat{\mathcal{N}}_{\tau_{\rm NL}^n} =\sum_{NNN''}(\partial \tau_{\rm NL}^n/\partial \tau_{\rm NL}^{NN'N''})\widehat{\mathcal{N}}_{\tau_{\rm NL}^{NN'N''}}$. The results for $Q_{\tau_{\rm NL}^{n,\rm even/odd}}$ are analogous. If $n=0$, we can simplify the Gaunt symbols further, finding
\beq
    \widehat{\mathcal{N}}_{\tau_{\rm NL}^{n_1n_30}}[\alpha,\beta,\gamma,\delta] &=& \delta_{\rm K}^{n_1n_3}\frac{1}{48}\frac{4\pi}{2n_1+1}(-1)^{n_1}\sum_{\mu_1}\sum_{LM}\int_0^\infty r^2dr\,\int_0^\infty r'^2dr'\,F_{L}(r,r')\\\nonumber
    &&\,\times\,\int d\hr\,Y^*_{LM}(\hr)P_{n_1\mu_1}[\Si\alpha](\hr,r)Q[\Si\beta](\hr,r)\\\nonumber
    &&\,\times\,\left(\int d\hr'\,Y^*_{LM}(\hr')P_{n_1\mu_1}[\Si\gamma](\hr',r')Q[\Si\delta](\hr',r')\right)^*\,+\,\text{23 perms.};
\eeq
modulo factors of $\sqrt{4\pi}$, this is simply the $\taunl$ estimator with $P$ replaced with $P_{n_1\mu_1}$, summing over $\mu_1$.

\subsection{\texorpdfstring{Massive Spinning Particles: $\tau_{\rm NL}^{\rm heavy}(s,\mu_s), 
\tau_{\rm NL}^{\rm light}(s,\nu_s)$}{Massive Spinning Particles}}\label{subsec: collider-estimator}
\noindent Finally, we construct estimators for the cosmological collider templates discussed in \S\ref{subsubsec: collider}. These are similar to the direction-dependent templates (due to the spherical harmonics appearing in the polarization tensors), but have more complex radial parts resulting from the additional $K^2/k_1k_3$ factors.

For efficient computation, we require a separable form of the Heaviside functions $\Theta_{\rm H}(k_i-\alpha_{\rm coll} K)$, which restrict the estimator to the quasi-collapsed limit. A number of options are possible. Firstly, one could explicitly compute the $k_i$ parts of the estimators for all $k_i\geq \alpha_{\rm coll} K$, then integrate over $K$ in a final step. Whilst this is exact, it requires many more harmonic transforms and significantly increases computation time. Alternatively, one could rewrite the Heaviside function as an integral or infinite sum of Bessel functions, as in \citep{MoradinezhadDizgah:2018ssw,Slepian:2018vds,Philcox:2021bul}. If the sum can be approximated by relatively few terms, this approach could be much faster; however, it is still difficult to implement since it adds oscillatory functions to the $k_i$ and $K$ integrals.
%(particularly the modified version of $P_{n\mu}(\hr,r)$) and two more in the $K$ integrals (modifying $F_{LL'}(r,r')$). 
A simpler approximation can be obtained by replacing
\beq
    \Theta_{\rm H}(k_i-\alpha_{\rm coll} K) \to \Theta_{\rm H}(k_i-\alpha_{\rm coll} K_{\rm coll})\Theta_{\rm H}(K_{\rm coll}-K),
\eeq
\textit{i.e.}\ separately truncating the $k_i$ and $K$ integrals at scales $\alpha_{\rm coll},K_{\rm coll}$. This does not require additional harmonic transforms or integration. Given that the signal-to-noise is dominated by large $k_i$, and, if the signal peaks in the collapsed limit, low $K$, we do not expect this to lead to much loss of signal-to-noise if $K_{\rm coll}$ is set appropriately.\footnote{In practice, we additionally marginalize over the equilateral EFT of inflation templates which further reduces correlations. As discussed in \papertwo, we typically use $K_{\rm coll}\chi_{\rm rec}\lesssim \ell_{\rm max}/4$, where $\chi_{\rm rec}$ is the distance to last scattering.} We test this approach in Appendix\,\ref{app: forecasts} (through primordial Fisher forecasts) and \papertwo (with numerical estimates) and find that the simplification yields good correlations with the full template, though loses some information at low $\nu_s$. Finally, one could simply restrict the analysis to large $\ell$ and small $L$; assuming a rough correspondence between $\ell$ and $L$, this is analogous to the above approach.

To derive the separable estimators, we first define the polarization/harmonic-space trispectrum for heavy and light particle exchange, using the primordial definitions \eqref{eq: heavy-exchange-template}\,\&\,\eqref{eq: light-exchange-template}: 
\beq
    \partial_{\tau_{\rm NL}^{\rm heavy}(s,\mu_s)}T^{\ell_1\ell_2\ell_3\ell_4,X_1X_2X_3X_4}_{m_1m_2m_3m_4} &=& \frac{1}{2}\sum_{S=0}^{2s}\left|\mathcal{C}_s(S,\mu_s)\right|\sum_{\lambda_1\lambda_3\Lambda}\tj{s}{s}{S}{\lambda_1}{\lambda_3}{\Lambda}\sum_{\chi=\pm1}e^{i\chi\omega_s(S,\mu_s)}\int_{\vK;K\leq K_{\rm coll}}K^{3+2i\chi\mu_s}\nonumber\\
    &&\,\times\,\int d\vr\,Y_{\ell_2m_2}^*(\hr)q_{\ell_2}^{X_2}(r)\int d\vr'\,Y_{\ell_4m_4}^*(\hr')q_{\ell_4}^{X_4}(r')P_\zeta(K)e^{i\vK\cdot(\vr'-\vr)}Y_{S\Lambda}(\hK)\nonumber\\
    &&\,\times\,\nonumber\\
    &&\,\times\,\left[\int_{\vk_1;k_1\geq \alpha_{\rm coll} K_{\rm coll}}4\pi i^{\ell_1}\mathcal{T}_{\ell_1}^{X_1}(k_1)k_1^{-3/2-i\chi\mu_s}P_\zeta(k_1)Y_{\ell_1m_1}^*(\hk_1)Y_{s\lambda_1}(\hk_1)e^{i\vk_1\cdot\vr}\right]\nonumber\\
    &&\,\times\,\left[\int_{\vk_3;k_3\geq \alpha_{\rm coll} K_{\rm coll}}4\pi i^{\ell_3}\mathcal{T}_{\ell_3}^{X_3}(k_3)k_3^{-3/2-i\chi\mu_s}P_\zeta(k_3)Y_{\ell_3m_3}^*(\hk_3)Y_{s\lambda_3}(\hk_3)e^{i\vk_3\cdot\vr'}\right]\nonumber\\
    &&\,+\,\text{11 perms.}
\eeq
and
\beq
    \partial_{\tau_{\rm NL}^{\rm light}(s,\nu_s)}T^{\ell_1\ell_2\ell_3\ell_4,X_1X_2X_3X_4}_{m_1m_2m_3m_4} &=& \sum_{S=0}^{2s}\mathcal{C}_s(S,i\nu_s)\sum_{\lambda_1\lambda_3\Lambda}\tj{s}{s}{S}{\lambda_1}{\lambda_3}{\Lambda}\int_{\vK;K\leq K_{\rm coll}}K^{3-2\nu_s}\nonumber\\
    &&\,\times\,\int d\vr\,Y_{\ell_2m_2}^*(\hr)q_{\ell_2}^{X_2}(r)\int d\vr'\,Y_{\ell_4m_4}^*(\hr')q_{\ell_4}^{X_4}(r')P_\zeta(K)e^{i\vK\cdot(\vr'-\vr)}Y_{S\Lambda}(\hK)\nonumber\\
    &&\,\times\,\left[\int_{\vk_1;k_1\geq \alpha_{\rm coll} K_{\rm coll}}4\pi i^{\ell_1}\mathcal{T}_{\ell_1}^{X_1}(k_1)k_1^{-3/2+\nu_s}P_\zeta(k_1)Y_{\ell_1m_1}^*(\hk_1)Y_{s\lambda_1}(\hk_1)e^{i\vk_1\cdot\vr}\right]\nonumber\\
    &&\,\times\,\left[\int_{\vk_3;k_3\geq \alpha_{\rm coll} K_{\rm coll}}4\pi i^{\ell_3}\mathcal{T}_{\ell_3}^{X_3}(k_3)k_3^{-3/2+\nu_s}P_\zeta(k_3)Y_{\ell_3m_3}^*(\hk_3)Y_{s\lambda_3}(\hk_3)e^{i\vk_3\cdot\vr'}\right]\nonumber\\
    &&\,+\,\text{11 perms.},
\eeq
in terms of the helicity amplitude and phase defined in \eqref{eq: theta-function-separable}\,\&\,\eqref{eq: theta-function-phase}. This differs from the direction-dependent templates only by the sum over $S$, the sum over $\chi$ (encoding the mode function complex conjugate pair for heavy templates), and the additional powers of $k_i$ and $K$ in the relevant integrals (coming from the asymptotic scalings of massive particles in de Sitter space). Analogous to before, we define
%(leaving the $K_{\rm coll}$ and $\alpha_{\rm coll}$ arguments implicit)
\beq
    F^{(\beta)}_{LL'}(r,r') &\equiv& \frac{2}{\pi}\int_{0}^{K_{\rm coll}} K^{2+\beta}dK\,j_L(Kr)j_{L'}(Kr')P_\zeta(K)\\\nonumber
    p_{\ell\ell'}^{(\beta),X}(r) &\equiv& \frac{2}{\pi}\int_{\alpha_{\rm coll} K_{\rm coll}}^\infty k^{2+\beta}dk\,\mathcal{T}_{\ell}^{X}(k)j_{\ell'}(xr)P_\zeta(k),
\eeq
with $F^{(0)}_{LL'}(r,r')\approx F_{LL'}(r,r')$ and $p_{\ell\ell'}^{(0),X}(r)\approx p^X_{\ell\ell'}(r)$ (noting that the massless scalar regime is insensitive to $K_{\rm coll}$). For $K_{\rm coll}\to \infty$, the first function has an analytic form for power-law power spectra similar to \eqref{eq: FLL-analyt} \citep[6.574.1]{nist_dlmf}:
\beq
    F^{(\beta)}_{LL'}(r,r') &=& \frac{2\pi^2 A_\zeta }{(k_\ast r')^{n_s-1}r'^\beta}\left(\frac{r}{r'}\right)^{L}\frac{2^{n_s+\beta-3}\Gamma\left(\frac{L+L'}{2}-\frac{1}{2}+\frac{n_s+\beta}{2}\right)}{\Gamma\left(\frac{L'-L}{2}+2-\frac{n_s+\beta}{2}\right)\Gamma\left(L+\frac{3}{2}\right)}\\\nonumber
    &&\,\times\,{}_2F_1\left(\frac{L+L'}{2}-\frac{1}{2}+\frac{n_s+\beta}{2},\frac{L-L'}{2}+\frac{n_s+\beta}{2}-1,L+\frac{3}{2},\left(\frac{r}{r'}\right)^2\right)
\eeq
for $r<r'$ wlog, assuming $\mathrm{Re}(\beta)+L+L'+n_s>1$ and $\mathrm{Re}(\beta)<4-n_s$, which is always satisfied for $n_s=0.96$ and our mass ranges.\footnote{This solution is not particularly useful, since it diverges as $r\to r'$ limit for $\mathrm{Re}(\beta)\geq 3-n_s$ (corresponding to $\nu_s\leq n_s/2$). This highlights the importance of truncating the integral $K_{\rm coll}$.} In the realistic case of finite $K_{\rm coll}$, the integral must be computed numerically.

Armed with these definitions, we can write the trispectrum numerators as
\begin{empheq}[box=\dbox]{align}
    \widehat{\mathcal{N}}_{\tau_{\rm NL}^{\rm heavy}(s,\mu_s)}[\alpha,\beta,\gamma,\delta] &= \frac{1}{24}\mathrm{Re}\sum_{S=0}^{2s}|\mathcal{C}_s(S,\mu_s)|e^{i\omega_s(S,\mu_s)}\sum_{\lambda_1\lambda_3\Lambda}\tj{s}{s}{S}{\lambda_1}{\lambda_3}{\Lambda}\\\nonumber
    &\,\times\,\sum_{LL'MM'}\int_0^\infty r^2dr\int_0^\infty r'^2dr'\,i^{L'-L}F_{LL'}^{(3+2i\mu_s)}(r,r')\G^{LL'S}_{MM'\Lambda}\\\nonumber
    &\,\times\,\int d\hr\,Y^*_{LM}(\hr)P_{s\lambda_1}^{(-3/2-i\mu_s)}[\Si\alpha](\hr,r)Q[\Si\beta](\hr,r)\nonumber\\
    &\,\times\,\int d\hr'\,Y^*_{L'M'}(\hr')P_{s\lambda_3}^{(-3/2-i\mu_s)}[\Si\gamma](\hr',r')Q[\Si\delta](\hr',r')\,+\,\text{11 perms.}\nonumber
\end{empheq}
(noting that the two $\chi$ terms are complex conjugates) and
\begin{empheq}[box=\dbox]{align}
    \widehat{\mathcal{N}}_{\tau_{\rm NL}^{\rm light}(s,\nu_s)}[\alpha,\beta,\gamma,\delta] &= \frac{1}{24}\sum_{S=0}^{2s}\mathcal{C}_s(S,i\nu_s)\sum_{\lambda_1\lambda_3\Lambda}\tj{s}{s}{S}{\lambda_1}{\lambda_3}{\Lambda}\\\nonumber
    &\,\times\,\sum_{LL'MM'}\int_0^\infty r^2dr\int_0^\infty r'^2dr'\,i^{L'-L}F_{LL'}^{(3-2\nu_s)}(r,r')\G^{LL'S}_{MM'\Lambda}\\\nonumber
    &\,\times\,\int d\hr\,Y^*_{LM}(\hr)P_{s\lambda_1}^{(-3/2+\nu_s)}[\Si\alpha](\hr,r)Q[\Si\beta](\hr,r)\\\nonumber
    &\,\times\,\int d\hr'\,Y^*_{L'M'}(\hr')P_{s\lambda_3}^{(-3/2+\nu_s)}[\Si\gamma](\hr',r')Q[\Si\delta](\hr',r')\,+\,\text{11 perms.}
\end{empheq}
for
\beq
    P_{s\lambda}^{(\beta)}[x](\hr,r) &\equiv& \sum_{\ell'm'}\left(\sum_{\ell m X}i^{\ell'-\ell}(-1)^\lambda \G^{\ell\ell's}_{mm'(-\lambda)}p^{(\beta),X}_{\ell\ell'}(r)x^{X*}_{\ell m}\right)Y_{\ell'm'}(\hr),
\eeq
which satisfies $\left(P_{s\lambda}^{(\beta)}[x](\hr,r)\right)^* = (-1)^{s+\lambda}P_{s(-\lambda)}^{(\beta^*)}[x](\hr,r)$ and $P_{00}^{(\beta)}[x](\hr,r) =\sum_{\ell m X}p^{(\beta),X}_{\ell\ell}(r)x^{X}_{\ell m}Y_{\ell m}(\hr)/\sqrt{4\pi}$. 

Despite their ungainly length, the above expressions are not parametrically harder to implement than those in the direction-dependent template of \S\ref{subsec: direction-estimator}, though the heavy particle estimators require complex maps. In the limit $s=0$, $\nu_s=3/2$, we recover the $\taunl$ estimator, noting that the $K$ and $k$ integrals are dominated by $K\to 0$ and $k\to \infty$, as before. In the conformal-coupling limit ($\nu_s\to 0$ or $\mu_s\to 0$, with $\tau_{\rm NL}^{\rm light}(s,0)=\tau_{\rm NL}^{\rm heavy}(s,0)$), the estimators simplify considerably:
\beq
    \widehat{\mathcal{N}}_{\tau_{\rm NL}^{\rm light}(s,0)}[\alpha,\beta,\gamma,\delta] &=& \frac{1}{24}\frac{4\pi}{2s+1}(-1)^s\sum_{\lambda}\sum_{LM}\int_0^\infty r^2dr\int_0^\infty r'^2dr'\,F_{LL}^{(3)}(r,r')\\\nonumber
    &&\,\times\,\left[P_{s\lambda}^{(-3/2)}[\Si\alpha](\hr,r)Q[\Si\beta](\hr,r)\right]_{LM}\left[P_{s\lambda}^{(-3/2)}[\Si\gamma](\hr',r')Q[\Si\delta](\hr',r')\right]^*_{LM}\\\nonumber
    &&\,+\,\text{11 perms.},
\eeq
(denoting the harmonic transforms by $[x]_{LM} \equiv \int d\hn\,Y_{LM}(\hn)x(\hn)$). This occurs since the angular dependence reduces to a Legendre polynomial, as in \eqref{eq: Theta-conformal}. Furthermore, for spin-zero particles
\beq
    \widehat{\mathcal{N}}_{\tau_{\rm NL}^{\rm heavy}(0,\mu_0)}[\alpha,\beta,\gamma,\delta] &=& \frac{4\pi}{24}\mathrm{Re}\sum_{LM}e^{i\omega_0(0,\mu_0)}\int_0^\infty r^2dr\int_0^\infty r'^2dr'\,F_{LL}^{(3+2i\mu_0)}(r,r')\\\nonumber
    &&\,\times\,\left[P_{00}^{(-3/2-i\mu_0)}[\Si\alpha](\hr,r)Q[\Si\beta](\hr,r)\right]_{LM}\left[P_{00}^{(-3/2+i\mu_0)}[\Si\gamma](\hr',r')Q[\Si\delta](\hr',r')\right]^*_{LM}\\\nonumber
    &&\,+\,\text{11 perms.}\\\nonumber
    \widehat{\mathcal{N}}_{\tau_{\rm NL}^{\rm light}(0,\nu_0)}[\alpha,\beta,\gamma,\delta] &=& \frac{4\pi}{24}
    \sum_{LM}\int_0^\infty r^2dr\int_0^\infty r'^2dr'\,F_{LL}^{(3-2\nu_0)}(r,r')\\\nonumber
    &&\,\times\,\left[P_{00}^{(-3/2+\nu_0)}[\Si\alpha](\hr,r)Q[\Si\beta](\hr,r)\right]_{LM}\left[P_{00}^{(-3/2+\nu_0)}[\Si\gamma](\hr',r')Q[\Si\delta](\hr',r')\right]^*_{LM}\\\nonumber
    &&\,+\,\text{11 perms.}
\eeq
Finally, if we consider $\nu_s\to 3/2, K_{\rm coll}\to\infty$ and assert the coupling $\mathcal{C}_s(S,3i/2) = \delta^{\rm K}_{SS'}$ (noting that this is outside the Higuchi limit, thus the usual $\mathcal{C}_s$ is undefined), we recover the direction-dependent estimator for $\tau_{\rm NL}^{ssS'}$.

The $Q$ derivatives used in the normalization of the above estimators are computed by a now familiar strategy. These are given by
\begin{empheq}[box=\dbox]{align}
    Q^X_{\ell m,\tau_{\rm NL}^{\rm heavy}(s,\mu_s)}[x,y,z] &= \frac{1}{4}\sum_{\chi=\pm1}\sum_{S=0}^{2s}\left|\mathcal{C}_s(S,\mu_s)\right|e^{i\chi\omega_s(S,\mu_s)}\int_0^\infty r^2dr\,\sum_{\lambda_1\lambda_3\Lambda}\tj{s}{s}{S}{\lambda_1}{\lambda_3}{\Lambda}\\\nonumber
    &\,\times\,\left[\sum_{\ell'm'}i^{\ell'-\ell}(-1)^{m}\G^{\ell \ell's}_{m(-m')(-\lambda_1)}p^{(-3/2-i\chi\mu_s),X}_{\ell \ell'}(r)\int d\hr\,Y^*_{\ell'm'}(\hr)Q[x](\hr,r)\right.\\\nonumber
    &\qquad\qquad\left.\,+\,q_{\ell}^X(r)\int d\hr\,Y_{\ell m}^*(\hr)P^{(-3/2-i\chi\mu_s)}_{s\lambda_1}[x](\hr,r)\right]\\\nonumber
    &\,\times\,
    \left(\sum_{LM}(-1)^MY_{LM}(\hr)\left(\sum_{L'M'}\int_0^\infty r'^2dr'\,i^{L'-L}F^{(3+2i\chi\mu_s)}_{LL'}(r,r')\G^{LL'S}_{(-M)M'\Lambda}\right.\right.\\\nonumber
    &\qquad\,\times\,\left.\left.\int d\hr'\,P^{(-3/2-i\chi\mu_s)}_{s\lambda_3}[y](\hr',r')Q[z](\hr',r')Y^*_{L'M'}(\hr')\right)\right)\,+\,\text{11 perms.}
\end{empheq}
Before $\chi$ summation, we have the symmetry $Q_{\ell m}^*(\chi) = (-1)^mQ_{\ell(-m)}(-\chi)$, which implies that the total $Q_{\ell m}$ are the conjugate components of a scalar field. For light fields, we find
\begin{empheq}[box=\dbox]{align}
    Q^X_{\ell m,\tau_{\rm NL}^{\rm light}(s,\nu_s)}[x,y,z] &= \frac{1}{2}\sum_{S=0}^{2s}\mathcal{C}_s(S,i\nu_s)\int_0^\infty r^2dr\,\sum_{\lambda_1\lambda_3\Lambda}\tj{s}{s}{S}{\lambda_1}{\lambda_3}{\Lambda}\\\nonumber
    &\,\times\,\left[\sum_{\ell'm'}i^{\ell'-\ell}(-1)^{m}\G^{\ell \ell's}_{m(-m')(-\lambda_1)}p^{(-3/2+\nu_s),X}_{\ell\ell'}(r)\int d\hr\,Y^*_{\ell'm'}(\hr)Q[x](\hr,r)\right.\\\nonumber
    &\qquad\qquad\left.\,+\,q_{\ell}^X(r)\int d\hr\,Y_{\ell m}^*(\hr)P^{(-3/2+\nu_s)}_{s\lambda_1}[x](\hr,r)\right]\\\nonumber
    &\,\times\,
    \left(\sum_{LM}(-1)^MY_{LM}(\hr)\left(\sum_{L'M'}\int_0^\infty r'^2dr'\,i^{L'-L}F^{(3-2\nu_s)}_{LL'}(r,r')\G^{LL'S}_{(-M)M'\Lambda}\right.\right.\\\nonumber
    &\qquad\,\times\,\left.\left.\int d\hr'\,P^{(-3/2+\nu_s)}_{s\lambda_3}[y](\hr',r')Q[z](\hr',r')Y^*_{L'M'}(\hr')\right)\right)\,+\,\text{11 perms.}
\end{empheq}
With these gargantuan expressions derived, we complete our study of primordial exchange templates.

\section{Estimation: Non-Primordial Trispectra}\label{sec: estimators-late-time}
\noindent Finally, we consider two non-primordial contributions to the CMB trispectrum: weak gravitational lensing and unclustered point sources. Whilst other secondary effects exist (including clustering of the Cosmic Infrared Background source galaxies, dipole effects and beyond \citep[e.g,][]{Penin:2013zya,Challinor:2002zh,Amendola:2010ty}), these two are expected to be the main contaminants relevant to four-point primordial non-Gaussianity analyses.

\subsection{\texorpdfstring{Point Sources: $t_{\rm ps}$}{Point Sources}}\label{subsec: point-source-estimator}
\noindent A collection of point source objects emitting in CMB frequencies (such as radio galaxies) naturally generates a connected four-point function \citep[e.g.][]{Toffolatti:1997dk,Hobson:1998sc}. To model this, we adopt a similar prescription to the the WMAP and \textit{Planck} three-point function analyses \citep[e.g.,][]{Komatsu:2001rj,Planck:2015zfm}, assuming a collection of $N$ point sources at locations $\{\hn_n\}$ each sourcing a temperature perturbation $\delta T_n$ in the CMB, \textit{i.e.}\ 
\beq\label{eq: ps-model}
    {}_sa(\hn) \supset \delta^{\rm K}_{s0}\sum_{n=1}^N\delta T_n\delta_{\rm D}(\hn-\hn_n), \qquad a_{\ell m}^X \supset \delta_{\rm K}^{XT}\sum_{n=1}^N\delta T_nY^*_{\ell m}(\hn_n).
\eeq
Here, we have assumed that the point source positions and intensities are uncorrelated (\textit{i.e.}\ they are Poissonian) and ignored any polarization contributions, since neither of these effects were detected in (suitably masked) \textit{Planck} three-point PNG analyses \citep{Planck:2015zfm}. Correlations could be sourced by dusty star-forming galaxies (which form the Cosmic Infrared Background) and would source additional scale- (and model-)dependence in the estimators below.

Model \eqref{eq: ps-model} leads to the following (temperature-only) three- and four-point functions:
\beq\label{eq: point-source-tspec}
    B^{\ell_1\ell_2\ell_3,TTT}_{m_1m_2m_3} &\supset& b_{\rm ps}\int d\hn\,Y_{\ell_1m_1}^*(\hn)Y_{\ell_2m_2}^*(\hn)Y_{\ell_3m_3}^*(\hn) \equiv b_{\rm ps}\G^{\ell_1\ell_2\ell_3}_{m_1m_2m_3}\\\nonumber
    T^{\ell_1\ell_2\ell_3\ell_4,TTTT}_{m_1m_2m_3m_4} &\supset & t_{\rm ps}\int d\hn\,Y_{\ell_1m_1}^*(\hn)Y_{\ell_2m_2}^*(\hn)Y_{\ell_3m_3}^*(\hn)Y_{\ell_4m_4}^*(\hn) \equiv \sum_{LM}t_{\rm ps}(-1)^M\G^{\ell_1\ell_2L}_{m_1m_2-M}\G^{\ell_3\ell_4L}_{m_3m_4M},
\eeq
where all scale-dependence is encoded in the Gaunt functions. These are contact-factorizable templates specified by the reduced bispectrum and trispectrum amplitudes $b_{\rm ps} = \av{\sum_{n=1}^N\delta T^3_n}/(4\pi)$ and $t_{\rm ps} = \av{\sum_{n=1}^N\delta T^4_n}/(4\pi)$.

Inserting \eqref{eq: point-source-tspec} into the harmonic-space trispectrum estimator of \eqref{eq: general-estimators-harmonic}, we find the following estimator numerator:
\begin{empheq}[box=\dbox]{align}\label{eq: ps-estimator}
    \widehat{\mathcal{N}}_{t_{\rm ps}}[\alpha,\beta,\gamma,\delta] &=\frac{1}{24}\sum_{\ell_im_iX_i}\int d\hn\,{}_0U^T[\Si\alpha](\hn){}_0U^T[\Si\beta](\hn){}_0U^T[\Si\gamma](\hn){}_0U^T[\Si\delta](\hn),
\end{empheq}
defining ${}_0U^T[x](\hn) \equiv \sum_{\ell m}Y_{\ell m}(\hn)x_{\ell m}^T$ as the (real) pixel-space filtered map (with notation matching \eqref{eq: U-V-def}). The Fisher derivative is obtained similarly:
\begin{empheq}[box=\dbox]{align}
    Q_{\ell m,t_{\rm ps}}^{X}[x,y,z] = \delta_{\rm K}^{XT}\int d\hn\,Y_{\ell m}^*(\hn){}_0U^T[x](\hn){}_0U^T[y](\hn){}_0U^T[z](\hn).
\end{empheq}
Since the estimators do not involve radial integrals, they are trivial to implement and can be used in joint analyses to assess any bias to a primordial estimator induced by point sources.

\subsection{\texorpdfstring{Gravitational Lensing: $A_{\rm lens}$}{Gravitational Lensing}}\label{subsec: lensing-estimator}

\noindent CMB lensing generates an exchange-factorizable trispectrum, whose amplitude, denoted $A_{\rm lens}$ (or $A^{\phi\phi}$ in \textit{Planck} analyses), can be estimated using similar methods to the $\tau_{\rm NL}$-type templates. %\footnote{An interesting extension is to build optimal estimators for the lensing \textit{bandpowers}, by introducing some vector of coefficients $\{A_{\rm lens}^i\}$, parametrizing the amplitude in each $L$-bin. Up to non-Gaussian bias contributions, this is a slight improvement on the standard quadratic estimator approach used in, for example, \citep{Planck:2018lbu}, featuring a mask and weighting dependent normalization matrix, $\F_{A_{\rm lens}^iA_{\rm lens}^j}$, as well as the (now-standard) realization-dependent debiasing contributions. It could be straightforwardly extended to include Gaussian lensing information present in the two-point function. These extensions, of course, are far beyond the scope of this series.}
As well as providing a robust four-point measurement of the lensing amplitude, including this in the analysis removes late-time lensing bias in the estimation of primordial shape coefficients. This was briefly considered in \citep{2015arXiv150200635S}, though only for the scalar case (with polarization being significantly more nuanced, as we see below). Unlike for primordial templates, the fiducial lensing amplitude is non-zero (with $A^{\rm fid}_{\rm lens}=1$); this requires a slight change to the formalism (as discussed in \citep{Hanson:2010rp}). In practice, we can obtain estimators similar to \eqref{eq: general-estimators} by Taylor expanding the likelihood around $A^{\rm fid}_{\rm lens}$; these are not quite optimal, however, due to (a) the non-Gaussian contributions to the estimator covariance and (b) the omission of higher-order estimators, such as six-point functions and cross-correlations with the integrated Sachs-Wolfe effect \citep{Hill:2018ypf} \resub{(which are considered in detail in \citep{Philcox:2025lxt})}.
% {we should probably mention that the estimator can be derived as a Taylor expansion around $A_{\rm lens}=1$, giving the NL corrections suggested in \citep{Hanson:2010rp}. It's also exact in NL corrections to $\phi\phi$, though misses bispectra etc.}

Following \citep{Okamoto:2003zw} (see also \citep{Lewis:2011fk}), a lensing potential $\phi$ causes the following transformation in the temperature and polarization fields:
\beq\label{eq: lensing-effect}
    a^X_{\ell m} \to a_{\ell m}^X + \sum_{LM\ell'm'}(-1)^{m}\tj{\ell}{\ell'}{L}{m}{-m'}{-M}\phi_{LM}\left[\epsilon_{\ell\ell'L}\,a_{\ell'm'}^{X}+\beta_{\ell\ell'L}\,a_{\ell'm'}^{\bar{X}}\right]{}_{s_X}F_{\ell L \ell'} \equiv a_{\ell m}^X + \delta a_{\ell m}^X[\phi],
\eeq
under the Born approximation at leading order. This uses the standard definitions
\beq
    &&\epsilon_{\ell\ell'L} \equiv \frac{1+(-1)^{\ell+\ell'+L}}{2}, \qquad \beta_{\ell\ell'L} \equiv \frac{1-(-1)^{\ell+\ell'+L}}{2i},\\\nonumber
    &&{}_{s}F_{\ell L \ell'} \equiv \frac{1}{2}\left[L(L+1)+\ell'(\ell'+1)-\ell(\ell+1)\right]\sqrt{\frac{(2\ell+1)(2L+1)(2\ell'+1)}{4\pi}}\tj{\ell}{L}{\ell'}{s}{0}{-s}
\eeq
with $a_{\ell m}^{\bar{T}}=0$, $a_{\ell m}^{\bar{E}} = -a_{\ell m}^{B}$ and $a_{\ell m}^{\bar{B}} = a_{\ell m}^{E}$, such that the $a_{\ell m}^{\bar{X}}$ terms will contribute only if $B$-modes are included in the data-vector. From \eqref{eq: lensing-effect}, we can form the trispectrum from two perturbed and two unperturbed fields, which takes the form
\beq\label{eq: lens-trispectrum-def}
    \left.T^{\ell_1\ell_2\ell_3\ell_4,X_1X_2X_3X_4}_{m_1m_2m_3m_4}\right|_{\rm lens} &=& \av{\delta a_{\ell_1m_1}^{X_1}[\phi]\delta a_{\ell_2m_2}^{X_2}[\phi]a^{X_3}_{\ell_3m_3}a_{\ell_4m_4}^{X_4}}+\text{5 perms.}
\eeq
This is clearly exchange-factorizable: the CMB fields in the perturbed legs will correlate with their unperturbed equivalents, whilst the two perturbed legs will `exchange' the $\phi$ field. 

Writing the power spectrum of $\phi$ as $\av{\phi_{LM}\phi^*_{LM}} \equiv  A_{\rm lens}C_L^{\phi\phi}$, for characteristic amplitude $A_{\rm lens}$, we can create a quartic estimator for $A_{\rm lens}$, following the above procedures (implicitly performing a Taylor expansion around $A_{\rm lens}^{\rm fid}$). After a somewhat lengthy calculation detailed in Appendix \ref{app: lensing}, this leads to the estimator numerator
\begin{empheq}[box=\dbox]{align}\label{eq: lensing-num}
    \widehat{\mathcal{N}}_{A_{\rm lens}}[\alpha,\beta,\gamma,\delta] &= \frac{1}{24}\sum_{LM}L(L+1)\Phi_{LM}[\Si\alpha,\Si\gamma]\Phi_{LM}^*[\Si\beta,\Si\delta]C^{\phi\phi}_L+\text{11 perms.},
\end{empheq}
involving the quadratic lensing estimator (which satisfies $\Phi_{LM}^* = (-1)^M\Phi_{L-M}$)
\beq\label{eq: PhiLM-simp}
    \Phi_{LM}[x,y] &\equiv&-\frac{1}{4}\sum_{X_1}(-1)^{s_{X_1}}\sum_{\lambda=\pm1}\\\nonumber
    &&\,\bigg\{\int d\hn\,\left[{}_{s_{X_1}}U^{X_1}[x](\hn){}_{s_{X_1}}V_\lambda^{X_1*}[y](\hn)-{}_{s_{X_1}}U^{X_1*}[x](\hn){}_{s_{X_1}}V^{X_1}_{-\lambda}[y](\hn)\right]{}_{+\lambda}Y^*_{LM}(\hn)\bigg\},
\eeq
where $s_X$ is the spin of component $X$, and we adopt the definitions
\beq\label{eq: U-V-def}
    {}_{s_X}U^X[x](\hn) &\equiv& \sum_{\ell m}{}_{s_X}Y_{\ell m}(\hn)x^X_{\ell m}\\\nonumber
    {}_{s_X}V^{X}_\lambda[x](\hn) &\equiv& \sum_{\ell m Z}{}_{s_X-\lambda}Y_{\ell m}(\hn)\sqrt{(\ell+\lambda s_X)(\ell-\lambda s_X+1)}\left(C_{\ell}^{XZ}-iC_{\ell}^{\bar{X}Z}\right)x^{Z}_{\ell m},
\eeq
each of which can be evaluated using spin-weighted spherical harmonic transforms (with ${}_0U^T[x] = {}_0x(\hn)$). Following \citep{Hanson:2010rp,Lewis:2011fk} we evaluate ${}_{s_X}V^{X}_\lambda$ using lensed CMB spectra, such that the estimator captures the excess power relative to $A_{\rm lens}^{\rm fid}$.

Estimator \eqref{eq: lensing-num} is closely related to the usual lensing numerator, with $\Phi_{LM}$ being an unnormalized quadratic estimator for $\phi_{LM}$. Notably, $\Phi_{LM}$ combines temperature and polarization information to form a single field. This is in contrast to many lensing estimators in the literature \citep[e.g.,][]{Hu:2001kj,Okamoto:2003zw}, which build a separate quadratic estimator $\Phi_{LM}^{(XY)}$ from each pair of fields ($X,Y$), then combine via $\Phi_{LM} = \sum_{X,Y}w_{XY}\Phi_{LM}^{(XY)}$ for minimum-variance weights $w_{XY}$. As shown in \citep{Maniyar:2021msb}, this leads to a slight loss of signal-to-noise and one should instead perform a single global optimization; our approach naturally realizes this solution.

Finally, we require the lensing contributions to the Fisher matrix. As before, these act both to normalize the estimator and to account for leakage between templates (which \paperthree will find to be greatly important). The corresponding $Q_{\ell m}^X$ map is derived in Appendix \ref{app: lensing} and reads
\begin{empheq}[box=\dbox]{align}\label{eq: Qlm-lens}
    Q_{\ell m,A_{\rm lens}}^{X}[x,y,z] &= \frac{1}{4}\sum_{\lambda}\bigg\{\int d\hn\,{}_{-s_X}Y_{\ell m}^*(\hn){}_{s_X}V_\lambda^{X*}[y](\hn){}_{-\lambda}W[x,z](\hn)\\\nonumber
    &\qquad\qquad\qquad+\int d\hn\,{}_{+s_X}Y_{\ell m}^*(\hn){}_{s_X}V_\lambda^{X}[y](\hn){}_{-\lambda}W^*[x,z](\hn)\bigg\}\nonumber\\
    &\,\,+\frac{1}{4}\sum_{\lambda X_3}\sqrt{(\ell+\lambda s_{X_3})(\ell-\lambda s_{X_3}+1)}\nonumber\\
    &\,\times\,\bigg\{\left(C_{\ell}^{X_3X}+iC_\ell^{\bar{X}_3X}\right)\int d\hn\,{}_{s_{X_3}}U^{X_3}[y](\hn){}_{-\lambda}W[x,z](\hn){}_{s_{X_3}-\lambda}Y_{\ell m}^*(\hn)\nonumber\\
    &\qquad\,-\,\left(C_{\ell}^{X_3X}-iC_\ell^{\bar{X}_3X}\right)\int d\hn\,{}_{s_{X_3}}U^{X_3*}[y](\hn){}_{-\lambda}W^*[x,z](\hn){}_{\lambda-s_{X_3}}Y_{\ell m}^*(\hn)\bigg\}\,+\,\text{5 perms.},
\end{empheq}
with
\beq
    {}_{\lambda}W[x,y](\hn) \equiv \sum_{LM}{}_{\lambda}Y_{LM}(\hn)L(L+1)\Phi_{LM}[x,y]C_L^{\phi\phi}
\eeq
which satisfies ${}_{-\lambda}W=(-1)^\lambda{}_\lambda W^*$. This can be computed using spin-weighted harmonic transforms as before. 
Similar estimators can be used to optimally constrain the lensing band-powers; these are formed by replacing $C_L^{\phi\phi}$ with its derivative with respect to a bin (which turns the Fisher normalization into an $N_{\rm bins}\times N_{\rm bins}$ matrix).

\section{Comparison to Standard Estimators}\label{sec: estimator-comparison}
\noindent The estimators derived in \S\ref{sec: estimators} through \S\ref{sec: estimators-late-time} appear fairly different to the standard forms used in the literature. As discussed in \S\ref{sec: estimators}, our $\gnl$, $\gnldotdot$, $\gnldotdel$ and $\gnldeldel$ estimators generalize those of \citep{2015arXiv150200635S}, which have been applied to both WMAP and \textit{Planck} temperature anisotropies \citep{2015arXiv150200635S,Planck:2015zfm,Planck:2019kim}. The key distinction is the inclusion of polarization, we make several additional technical upgrades. In contrast, our exchange trispectrum estimators appear significantly altered compared to those of previous works, particularly with respect to former $\taunl$ estimators \citep[e.g.,][]{2014A&A...571A..24P,Munshi:2009wy,Smidt:2010ra,Munshi:2010bh,Feng:2015pva,Marzouk:2022utf} and the canonical lensing estimators \citep[e.g.,][]{SPT:2014puc,Planck:2018lbu,ACT:2023dou,Carron:2022edh}. Below, we unpack this apparent difference in detail.

\subsection{Local Anisotropy Estimators}
\noindent As discussed in \citep{Hanson:2009gu}, exchange interactions such as $\taunl$ \resub{can be modeled as} a source of `local anisotropy', \textit{i.e.}\ a spatial variation of the local CMB power spectrum. As such, one can estimate $\taunl$ (and other trispectra peaking in the collapsed limit) using tools developed for CMB lensing analyses. Formally, any pair of (short-scale) CMB fields, $X,Y$, can be used to build an estimator, $\Phi$, for the modulation field $\phi(\hn)$, such that $\phi(\hn) \sim \Phi[X,Y](\hn)$ (symmetrizing over $X\leftrightarrow Y$). The square of this estimator encodes the characteristic amplitude, $A$ (\textit{i.e.}\ $A_{\rm lens}$ or $\tau_{\rm NL}$). Schematically,
\beq\label{eq: general-A-estimator}
    \widehat{\mathcal{N}} = \sum_{LM}w_L\Phi_{LM}[X,Y]\Phi^*_{LM}[X',Y'] \equiv \mathcal{A}[\Phi_{XY},\Phi_{X'Y'}]
\eeq
where $w_L$ is a minimum-variance weight and we introduce the short-hand $\mathcal{A}$.\footnote{In the case of CMB lensing, $\Phi[X,Y]$ estimates the lensing potential $\phi$, and the $\mathcal{A}$ operator computes the lensing power spectrum, averaged over some $L$-range and weighted by a fiducial $C_L^{\phi\phi}$.} Setting $X=Y=X'=Y' = d$ for some observed dataset $d$, we find the basic estimator $\widehat{N} = \mathcal{A}[\Phi_{dd},\Phi_{dd}]$. 

In practice, lensing and local anisotropy estimators must correct for two main sources of bias. Firstly $\Phi_{dd}$ is non-zero even in the absence of a non-Gaussian signal (\textit{i.e.}\ $\av{\Phi_{dd}}_d \neq \phi$) due to masking and inhomogeneous noise. This `mean-field' contribution is removed with simulations, such that $\widehat{N} = \mathcal{A}[\Phi_{dd}-\av{\Phi_{ss}}_s,\Phi_{dd}-\av{\Phi_{ss}}_s]$, where $\av{\cdots}_s$ represents the Monte Carlo average over some set of simulations $\{s\}$. Secondly, Gaussian noise biases the estimator, such that $\mathbb{E}[\widehat{A}]\neq A$ even after mean-field subtraction and normalization. This is conventionally removed with `realization-dependent noise' methods \citep{Namikawa:2012pe}, subtracting a term specified by:
\beq
    \mathrm{RDN}^{(0)} = \av{4\mathcal{A}[\Phi_{ds},\Phi_{ds}]-2\mathcal{A}[\Phi_{ss'},\Phi_{ss'}]}_{s,s'},
\eeq
where $\{s\}$ and $\{s'\}$ are two uncorrelated sets of simulations. The full estimator (under null hypotheses) is thus given by
\beq\label{eq: local-aniso-estimator}
    \widehat{N}_{\rm previous} = \mathcal{A}[\Phi_{dd}-\av{\Phi_{ss}}_s,\Phi_{dd}-\av{\Phi_{ss}}_s] - \mathrm{RDN}^{(0)}.
\eeq
This form (optionally with higher-order additions discussed below) has been used to estimate $\taunl$ \citep{Marzouk:2022utf} and the lensing power spectrum \citep[e.g.,][]{SPT:2014puc,Planck:2018lbu,ACT:2023dou} from data.

\subsection{Equivalence with Maximum-Likelihood Estimators}
\noindent Below, we demonstrate that the local anisotropy estimators are equivalent to the quartic estimators discussed in \S\ref{sec: estimators}. Due to the dominance of collapsed regimes in the underlying inflationary correlators, the primordial trispectrum estimators of \S\ref{sec: estimators-exchange} can be written as the large-scale power spectrum of two short-scale quadratic estimators. Explicitly, the quartic $\taunl$ numerator \eqref{eq: tauNL-loc-num} can be rewritten
\beq
    \widehat{\mathcal{N}}_{\taunl}[\alpha,\beta,\gamma,\delta] &=&\frac{1}{12}\mathcal{A}^{\taunl}[\Phi^{\taunl}_{\alpha\beta},\Phi^{\taunl}_{\gamma\delta}]+\text{11 perms.},
\eeq
subject to the power spectrum and quadratic estimator definitions:
\beq\label{eq: our-A-tau-def}
    \mathcal{A}^{\taunl}[\Phi_{\alpha\beta},\Phi_{\gamma\delta}] &\equiv& \frac{1}{2}\sum_{LM}\int_0^\infty\,r^2dr\int_0^\infty\,r'^2dr'\,F_L(r,r')\Phi_{LM}[\alpha,\beta](r)\Phi_{LM}^*[\gamma,\delta](r')\\\nonumber
    \Phi^{\taunl}_{LM}[\alpha,\beta](r) &=& \int d\hr\,Y^*_{LM}(\hr)P[\Si\alpha](\hr,r)Q[\Si\beta](\hr,r).
\eeq
This is clearly analogous to \eqref{eq: general-A-estimator}, but features additional radial integrals, which account for the finite width of the last-scattering surface (unlike, for example, \citep{Marzouk:2022utf}). These have negligible impact for $\taunl$, but become important for other templates, where the integrand is less sharply peaked at $r\sim r_\star$. 

The building block of every exchange estimator (trivially including the lensing estimator of \eqref{eq: lensing-num}) can be written in a similar form. From \eqref{eq: general-estimators-harmonic}, we find the full estimator numerator for a general exchange amplitude $A$ (keeping template labels implicit):
\beq\label{eq: our-estimator-perm}
    \widehat{N}_{\rm this\,\,work} = \mathcal{A}[\Phi_{dd},\Phi_{dd}] - \av{4\mathcal{A}[\Phi_{ds},\Phi_{ds}]+2\mathcal{A}[\Phi_{dd},\Phi_{ss}]}_s + \av{2\mathcal{A}[\Phi_{ss'},\Phi_{ss'}]+\mathcal{A}[\Phi_{ss},\Phi_{s's'}]}_{s,s'}
\eeq
in terms of a suitably defined power spectrum operator $\mathcal{A}$, quadratic estimator $\Phi$, and two uncorrelated sets of simulations, $\{s\}$ and $\{s'\}$. A simple reordering yields equivalence with the local anisotropy estimator (up to finite recombination effects):
\beq\label{eq: estimator-rewritten}
    \widehat{N}_{\rm this\,\,work} &=& \mathcal{A}[\Phi_{dd}-\av{\Phi_{ss}}_s,\Phi_{dd}-\av{\Phi_{ss}}_s] - \av{4\mathcal{A}[\Phi_{ds},\Phi_{ds}]-2\mathcal{A}[\Phi_{ss'},\Phi_{ss'}]}_{s,s} \equiv \widehat{N}_{\rm previous}.
\eeq
As such, our quasi-optimal estimators naturally include both mean-field and realization-dependent bias contributions.

\subsection{Discussion}
\noindent As derived above, our exchange estimators are analogous to the local anisotropy forms used in the literature for both $\taunl$ and $A_{\rm lens}$ (or $C_L^{\phi\phi}$). Whilst this is certainly no surprise (the optimal estimator is unique in the Gaussian limit), it is worth pointing out explicitly given the complex form of our direction-dependent and collider estimators, as well as the various non-trivial biases. Furthermore, the explicitly-quadratic form of \eqref{eq: estimator-rewritten} will be used in \paperthree to motivate a non-Gaussian sampling distribution for $\widehat{\tau}_{\rm NL}^{\rm loc}$. 
%The utility of our optimal estimator approach is clear; we gain all these theoretical developments essentially `for free'. 

In practice, local anisotropy estimators often contain a number of additional features to address specific biases and approximations:
\begin{itemize}
    \item \textbf{Normalization}: In the above discussion, we ignored the normalization term, $\F$. In lensing and local anisotropy estimators \citep[e.g.,][]{Marzouk:2022utf} this is usually computed under idealized conditions and then `corrected' by some multiplicative factor obtained using non-Gaussian simulations. In this work, we compute the normalization numerically (without non-Gaussian simulations), fully accounting for any masks and transfer functions. We do not require any empirical calibration factor and can use the estimators' performance on non-Gaussian simulations as a validation test.
    \item \textbf{$\text{N}^{(1)}$ term}: Lensing estimators often subtract an additional $\text{N}^{(1)}$ bias from \eqref{eq: local-aniso-estimator}, proportional to the fiducial lensing amplitude, $A^{\rm fid}_{\rm lens}$. This is sourced by from lensing contributions from the `wrong' legs of the quadratic estimator (noting that the $\Phi$ estimator is asymmetric). In our formalism, these contributsion are naturally accounted for in the normalization and do not need to be subtracted. Formally, this results in a slight boost in signal-to-noise (which is likely negligible in practice).  
    \item \textbf{$\text{N}^{(3/2)}$ term}: The lensing potential $\phi$ is a non-linear field, thus the quartic $A_{\rm lens}$ estimator can be biased by non-Gaussian correlators such as $\av{\phi^3}$. This can be ameliorated by subtracting an $\text{N}^{(3/2)}$ bias, which scales as $A_{\rm lens}^{3/2}$. Given that our focus is on primordial physics (which vanishes under null assumptions), we do not attempt to model such terms.
    \item \textbf{Monte Carlo term}: Finally, both $\taunl$ and lensing estimators sometimes include a Monte Carlo (MC) term to account for an overall bias. This is given by $\widehat{A}^{\rm MC} = \av{\widehat{A}^{\rm fid}}_s$, averaging over simulations $\{s\}$ with some fiducial non-Gaussianity amplitude $A^{\rm fid}$ (which could be zero). This could be used to remove lensing contributions to $\tau_{\rm NL}$ or to account for noise or foreground non-Gaussianity present in the simulations. Here, we instead remove lensing contamination by subtracting the (experiment-specific) bias using the Fisher matrix, $\F_{\tau_{\rm NL}A_{\rm lens}}$, or by performing a joint analysis of $\tau_{\rm NL}$ and $A_{\rm lens}$. Any residual sources of non-Gaussianity can be estimated by applying the pipeline to realistic simulations.
\end{itemize}

Finally, we note that local anisotropy estimators can be formulated in terms of cross-spectra between two distinct datasets \citep[e.g.,][]{Feng:2015pva,Marzouk:2022utf}, which functions as a useful consistency test. This can be realized in our optimal estimators via a small modification to \eqref{eq: our-estimator-perm}
\beq\label{eq: cross-estimator}
    \widehat{A}^{\rm cross}_{\rm this\,\,work} &=& \mathcal{A}[\Phi_{d_Ad_B},\Phi_{d_Ad_B}]\\\nonumber
    &&\,-\,\av{\mathcal{A}[\Phi_{d_As_B},\Phi_{d_As_B}]+    \mathcal{A}[\Phi_{s_Ad_B},\Phi_{d_As_B}]+    \mathcal{A}[\Phi_{d_As_B},\Phi_{s_Ad_B}]+    \mathcal{A}[\Phi_{s_Ad_B},\Phi_{s_Ad_B}]+    2\mathcal{A}[\Phi_{d_Ad_B},\Phi_{s_As_B}]}_s \\\nonumber
    &&\,+\,\av{\mathcal{A}[\Phi_{s_As'_B},\Phi_{s_As'_B}]+\mathcal{A}[\Phi_{s'_As_B},\Phi_{s_As'_B}]+\mathcal{A}[\Phi_{s_As_B},\Phi_{s'_As'_B}]}_{s,s'}
\eeq
where $d_A,d_B$ are two splits of the data (e.g., different frequency channels or half-mission splits). Assuming that $d_A$ and $d_B$ trace the same underlying CMB, the Fisher matrix is unchanged. \eqref{eq: cross-estimator} is an alternative estimator for $A$ which reduces bias from poorly understood noise at the expense of a slight increase in variance (since the optimal estimator would include auto-spectra, e.g., $\Phi_{d_Ad_A}$).

\section{Optimization}\label{sec: optim}
\noindent Primordial inflationary templates usually involve radial integration, either to implement momentum conservation and finite recombination effects (the $r$ and $r'$ integrals) or to integrate over the conformal coordinate $\tau$ (appearing in the EFT templates). A simple way to compute these is with na\"ive quadrature: however, this requires a large number of sampling points, $N_s$, to converge. This represents a significant computational challenge, given that the numerators and Fisher matrices involve $\mathcal{O}(N_s)$ harmonic transforms. Furthermore, when computing exchange trispectra, it is most efficient to hold an array of all $N_s$ maps in memory (of size $\mathcal{O}(N_{\rm pix}N_s)$), as well as the coupling matrix $F_L(r,r')$ (of size $\mathcal{O}(L_{\rm max}N_s^2)$), which quickly becomes limiting.

To ameliorate these problems, optimization schemes have been developed, such as that of \citep{2011MNRAS.417....2S} (and built upon in \citep{2015arXiv150200635S,Munchmeyer:2014nqa}). The basic idea is to replace the radial or time integrals with a finely-discretized sum, then approximate this sum with a small number of sampling points
\textit{i.e.}\ 
\begin{alignat*}{3}
    &\int_0^\infty r^2dr\,f(r) \quad && \approx \quad\sum_{I=1}^{N_s}r_I^2f(r_I)\delta r_I \quad &&\to\quad \sum_{i=1}^{N_{\rm opt}}w_i\,\times\,r_i^2f(r_i)\delta r_i\\\nonumber
    &\int_0^\infty r^2dr\,\int_{-\infty}^0 d\tau\,f(r,\tau) \quad &&\approx\quad \sum_{I=1}^{N_s}r_I^2f(r_I,\tau_I)\delta r_I\delta\tau_I \quad &&\to\quad \sum_{i=1}^{N_{\rm opt}}w_i\,\times\,r_i^2f(r_i)\delta r_i\delta\tau_i
\end{alignat*}
where the $N_{\rm opt}$ radial points $\{r_i\}$ (or $\{r_i,\tau_i\}$ in two dimensions) in the optimized representation are a subset of the $N_s$ points in the unoptimized representation $\{r_I\}$ (or $\{r_I,\tau_I\}$). This has two components: (1) a small set of sampling points; (2) associated weights $\{w_i\}$. To obtain the weights given some set of integration points, \citep{2011MNRAS.417....2S} defined a distance metric between the exact and optimized templates $\mathsf{T}$ and $\mathsf{T}_{\rm approx}$:
\beq\label{eq: score-func}
    \mathcal{S}[w] = \mathcal{F}[\mathsf{T}-\mathsf{T}_{\rm approx}[w],\mathsf{T}-\mathsf{T}_{\rm approx}[w]],
\eeq
where the `true' template is computed as a discrete sum with $N_s\gg 1$ points. This uses the Fisher distance:
\beq
    \mathcal{F}[\mathsf{T},\mathsf{T}'] = \frac{1}{4!}\left[\mathsf{T}^{abcd*}\Ci_{aa'}\Ci_{bb'}\Ci_{cc'}\Ci_{dd'}\mathsf{T}'^{a'b'c'd'}\right]^*,
\eeq
noting that $\F[\mathsf{T},\mathsf{T}]$ is the usual Fisher matrix. For the purposes of optimization, it is sufficient to compute $\mathcal{S}$ under ideal assumptions (\textit{i.e.}\ translation-invariant noise, Gaussian statistics and unit mask), which significantly reduces computation time. Note that $\mathcal{S}[w]$ is related to the `cosine', $\rho$, between shapes \citep{2011arXiv1105.2791F} via $\rho=\sqrt{(1-\mathcal{S})\F_{\rm true}/\F_{\rm approx}}$, where $\F_{\rm true}$ and $\F_{\rm approx}$ are the true and approximated Fisher matrices; as such, minimizing $\mathcal{S}$ minimizes the cosine.

Before discussing the specific optimization routines for contact and exchange trispectra (which will be validated numerically in \papertwo), we note some general properties of this approach. Let us assume that the Universe possesses a physical trispectrum described by template $\T$ and amplitude $g$. Performing an (idealized) analysis using the approximate template $\T_{\rm approx}$ will result in a (formally) biased estimate of $g$:
\beq\label{eq: optim-err}
    \mathbb{E}[\widehat{g}] = \left(1+\F[\T_{\rm approx},\T-\T_{\rm approx}]/\F[\T_{\rm approx},\T_{\rm approx}]\right)g,
\eeq
with an error scaling as $\sqrt{\mathcal{S}[w]/\F[T,T]}$. If the Universe is Gaussian ($g=0$), the bias vanishes -- this implies that our optimization procedure cannot induce spurious non-Gaussianity.\footnote{In practice, gravitational lensing always creates non-Gaussianity (with $A_{\rm lens}^{\rm fid}=1$). Provided that we always perform joint analyses of the lensing template with $\T_{\rm approx}$ or subtract off the expected bias using the Fisher matrix, this cannot source a spurious signal in the estimation of $g$ since we account for the correlation of $\T_{\rm approx}$ and $\T_{\rm lens}$.} In practice, replacing $\T$ with $\T_{\rm approx}$ corresponds to performing a CMB search for a (slightly) different template. As long as the two are well correlated (which is true in all the analyses of \papertwo and \paperthree), this is an efficient way in which to proceed.%, particularly given the various additional assumptions that go into template construction.

\subsection{Contact Trispectra}
\noindent For contact trispectra such as $\gnl$ and $\gnldotdot$, the templates are linear in the weights $w$, such that
\beq
    \mathsf{T}^{\rm approx}[w] = \sum_{i=1}^{N_{\rm opt}}w_i\mathsf{T}_i, \qquad \mathsf{T} = \sum_{i=1}^{N_{\rm opt}}\mathsf{T}_i+\sum_{I=N_{\rm opt}+1}^{N_s}\mathsf{T}_I,
\eeq
where we denote components present in the optimized representation by lower case indices and define $\mathsf{T}_i$ as the integrand at sampling point $i$. For $\gnl$, this takes the form
\beq\label{eq: tlm-gloc-opt}
    \mathsf{T}_i^{(\gnl)} = \frac{54}{25}r_i^2\delta r_i\,p_{\ell_1}^{X_1}(r_i)p_{\ell_2}^{X_2}(r_i)p_{\ell_3}^{X_3}(r_i)q_{\ell_4}^{X_4}(r_i)\int d\hr\,Y_{\ell_1m_1}^*(\hr)Y_{\ell_2m_2}^*(\hr)Y_{\ell_3m_3}^*(\hr)Y_{\ell_4m_4}^*(\hr)+\text{3 perms.},
\eeq
in polarization/harmonic-space, using \eqref{eq: tlm-gloc}. Defining $\mathcal{F}_{xy} \equiv \mathcal{F}[\mathsf{T}_x,\mathsf{T}_y]$ as the idealized Fisher distance between the template contributions at $x,y$, the score $\mathcal{S}[w]$ can be minimized exactly, yielding the optimal weights:
\beq\label{eq: optimal-weight}
    w_i^{\rm opt} = 1+\sum_{j=1}^{N_{\rm opt}}\sum_{K=N_{\rm opt}+1}^{N_s}\mathcal{F}^{-1}_{ij}\mathcal{F}_{jK}.
\eeq
This assumes an efficient method to compute $\F_{xy}$; as shown in Appendix \ref{app: analytic-fisher} (building on \citep{2015arXiv150200635S}), the matrix can be computed exactly for all the contact templates we consider in this work without harmonic transforms or Monte Carlo summation. Given some relatively large set of $N_s$ unoptimized sampling points, computation of $\F_{xy}$ has $\mathcal{O}(N_s^2\ell_{\rm max}^2)$ complexity, and only needs to be performed only once for a given beam, fiducial power spectrum and $\ell$-range. 

We must additionally choose the basis points themselves. As described in \citep{2011MNRAS.417....2S}, this can be done via a greedy algorithm, starting from an empty set ($N_{\rm opt}=0$) and, at each iteration, choosing the index which most improves the Fisher score, $\mathcal{S}[w^{\rm opt}]$. Given a previous set of basis indices $\{i\}$, the best choice is given by
\beq\label{eq: optimal-index} 
    \mathrm{arg\,max}_{I\notin\{i\}}\left[\frac{\left(\sum_{J\notin\{i\}}\left(\mathcal{F}_{IJ}-\sum_{ij}\mathcal{F}_{Ii}\mathcal{F}^{-1}_{ij}\mathcal{F}_{jJ}\right)\right)^2}{\mathcal{F}_{II}-\sum_{ij}\mathcal{F}_{Ii}\mathcal{F}^{-1}_{ij}\mathcal{F}_{jI}}\right].
\eeq
In practice, we continually add indices until a desired precision has been reached, for example
\beq\label{eq: convergence}
    \mathcal{S}[w_{\rm opt}]\leq f_{\rm thresh}\mathcal{S}_{\rm init} \equiv f_{\rm thresh}\mathcal{F}[\mathsf{T},\mathsf{T}],
\eeq
typically with $f_{\rm thresh}\sim 10^{-4}$, which should yield a trispectrum estimator with a multiplicative bias below $\sim 1\%$ \eqref{eq: optim-err}, which usually requires $\sim 10-100$ integration points. .In some settings (particularly for the EFT of inflation templates), computation of the ideal Fisher matrix is rate-limiting due to the large $\ell_{\rm max}$. In this case, one can adopt a two-step procedure \citep[cf.][]{2015arXiv150200635S}, splitting the unoptimized representation into $M>1$ subsets, each of which are separately optimized to find $M$ sets of sampling points, before performing a final optimization on the combination. 

\subsection{Exchange Trispectra}
\subsubsection{Algorithm}
\noindent For exchange-factorizable templates, computing an optimized representation is more difficult and has not been considered previously. However, the associated estimators are more expensive to implement due to the double $r,r'$ integration (unless one works in the local anisotropy limit discussed in \S\ref{sec: estimator-comparison}), thus the optimization is even more desirable. The complexity is two-fold: (a) the templates are now quadratic in the weights rather than linear (and thus the score is quartic, so linear optimization schemes fail); (b) the Fisher matrix cannot be efficiently computed analytically, even in idealized limits (as discussed in Appendix \ref{app: analytic-fisher}).

Here we sidestep these problems through small adjustments to the above algorithm (which \papertwo show to work well in practice). Our starting point is the quadratic template definition
\beq
    \T^{\rm approx}[w] \equiv \sum_{i=1}^{N_{\rm opt}}\sum_{j=1}^{N_{\rm opt}}w_iw_j\T_{ij}
\eeq
where $\T_{ij}$ denotes the template integrand at sampling points $i,j$, and the full template is obtained by setting $N_{\rm opt}=N_s$ and $w_i = 1$. A central part of the contact optimization algorithm was the Fisher distance between any two template contributions $\T_i$ and $\T_j$: here, we replace this by the distance between the derivatives of $\T_{\rm approx}[w]$ with respect to weights $w_i$ and $w_j$:\footnote{If $\T[w]$ is linear in $w$, this recovers the previous definition up to an unimportant factor of four.}
\beq
    \F_{ij} = \frac{1}{4}\F\left[\frac{\partial}{\partial w_i}\T_{\rm approx}[w],\frac{\partial}{\partial w_j}\T_{\rm approx}[w]\right],
\eeq
evaluated at $w=1$ and $N_s = N_{\rm opt}$ (\textit{i.e.}\ including all terms). As shown below, this can be efficiently computed using Monte Carlo methods.

Given the initial $N_s\times N_s$ matrix $\F_{ij}$, we optimize the templates as follows. Starting from some initial set of indices and weights (either empty, if we are performing optimization from scratch, else taken from a previous run):
\begin{enumerate}
    \item Pick a starting index based on \eqref{eq: optimal-index}, and guess the weights, $w^{\rm quad}$, using \eqref{eq: optimal-weight}. Here, we are approximating the quartic score function as quadratic around its minimum. If this is valid, then the $w^{\rm quad}$ weights are close to optimal.\footnote{Technically, this also assumes that $\F_{ij}$ defines the Hessian of $\mathcal{S}[w]$. This requires that $\F_{ij}$ dominates over the cross-term $\F[\partial_{w_iw_j}\T,\T]$, which is usually true.}
    \item Compute the true quartic score $\mathcal{S}[w^{\rm guess}]$ from the current set of indices and weights by evaluating \eqref{eq: score-func} at $w^{\rm guess}=w^{\rm quad}$. This must be performed numerically (as discussed below), and requires computation of $Q[w^{\rm guess}]$, as well as the full $Q$ arrays used to compute the unoptimized $\F_{ij}$, each of which is a sum over $N_{\rm it}$ (common) Monte Carlo realizations. To limit excess computation, we store the $N_{\rm it}(N_{\rm opt}-1)$ maps used in previous iterations.  
    \item If $\mathcal{S}[w^{\rm guess}]$ is lower than that computed with $N_{\rm opt}-1$ indices, accept the new index. If not, repeat steps (1) and (2) with the next best index, obtained from the quadratic prescription \eqref{eq: optimal-index}. 
    \item If $\mathcal{S}[w^{\rm guess}]$ satisfies our convergence criterion \eqref{eq: convergence} break, else repeat steps (1) through (3), adding a new index to the optimized representation.
\end{enumerate}
A further enhancement would be to perform numerical optimization to compute the weights at each step, \textit{i.e.}\ evaluate step (2) using $w^{\rm guess}=w^{\rm opt}$ rather than $w^{\rm quad}$. These could be computed using the BFGS optimization algorithm, given the score function $\mathcal{S}[w]$, its analytic derivative $\partial_{w_i}\mathcal{S}[w]$, and some initial guess of the weights $w^{\rm init}$ (for example, the value of $w^{\rm opt}$ at the previous step, supplemented by the difference between $w^{\rm quad}$ at the current and previous indices). Whilst this approach can improve convergence, it requires additional computation time (since evaluation of $\mathcal{S}$ is rate-limiting) and will not be applied in this series.

Whilst the procedure described above is not fully optimal, due to the approximate weights used in each step and the quadratic procedure used to select new indices, this is not a practical limitation. Assuming that our guessed weights are close to true solutions, $w^{\rm opt}$, the score can be well approximated as quadratic, thus the algorithm will converge quickly. Moreover, since we check for convergence using the true Fisher scores (instead of the quadratic approximations), we still expect to find a low-dimensional representation that closely represents the full finely-sampled trispectrum, even though our sampling points are not strictly optimal (provided criterion \eqref{eq: convergence} is satisfied). 
%In practice, we find that the guessed weights are close to the truth, in the sense that the final score, $\mathcal{S}[w^{\rm opt}]$ is close to $\mathcal{S}[w^{\rm guess}]$; as such, we perform the numerical optimization only every $5-10$ steps.{do we do this in practice?}

In many scenarios, we wish to compute multiple trispectra, accounting for their correlations. Whilst one could optimize the integration points and weights for each template separately, it is more efficient to define a global scheme, such that the relevant functions (e.g., $p_\ell^X(r)$, $B(\hr,r,\tau)$) can be reused. To this end, we first perform optimization for a single template (e.g., $\taunl$), computing the relevant weights and integration points. We then iterate over all remaining templates, starting from the previous sampling points in each case (which typically adds only a few indices). 
%The procedure is iterated across all templates, and ensures that (a) all representations are optimized to the desired tolerance, $f_{\rm thresh}$, and (b) we can optimize templates in order of efficiency. 
Strictly, this approach assumes all templates are uncorrelated, since we do not consider the joint Fisher matrices. % of, for example, $\gnl$ and $\gnldotdot$. 
Whilst one could instead optimize the joint Fisher matrix (which may lead to a slightly lower dimensional representation), it is both CPU- and memory-expensive, thus we adopt the above approach in practice. We finally note that one can split the optimization over $M$ partitions (as for the contact templates), which reduces the computation by almost a factor of $M$, possibly at the expense of a few additional integration points. 

\subsubsection{Computing the Fisher Matrix Derivative}
\noindent The above algorithm requires repeated computation of the ideal Fisher matrix between $\partial_{w_i}\T^{\rm approx}[w]$ and $\partial_{w_j}\T^{\rm approx}[w]$. Here, we discuss how this can be efficiently estimated. First, we rewrite the derivative matrix as a Monte Carlo sum as in \S\ref{subsec: monte-carlo}:
\beq
    F^{abc,def}_{ij} \equiv \frac{1}{4!}\bigg\langle\left(\frac{\partial}{\partial w_i}Q[\Ai a^{(a)},\Ai a^{(b)},\Ai a^{(c)}]\right)^* \cdot\Ai\cdot\left(\frac{\partial}{\partial w_j}Q[\Ai a^{(d)},\Ai a^{(e)},\Ai a^{(f)}]\right)\bigg\rangle^*_a,
\eeq
where $\av{}_a$ is evaluated as a sum over $N_{\rm  it}$ paired random fields as in \eqref{eq: fisher-MC2}. As for the contact optimization, we work under idealized conditions, setting $\Ai=\Si\mathsf{P}=\mathsf{P}^\dag\Ci\mathsf{P}$, with translation-invariant noise. Since we do not require an accurate estimation of the matrix amplitude (just its structure and principal components), we can use small $N_{\rm it}\sim1-\mathrm{few}$. In principle, noise in the Fisher matrix could hamper the optimization, since noisy $\F_{ij}$ could have `false minima'; in practice, we find this effect to be negligible \citep{Philcox4pt2}.

Working in polarization/harmonic-space, the derivative matrix can be written
\beq
    F^{abc,def}_{ij} \equiv \frac{1}{4!}\sum_{\ell m XY}\bigg\langle \frac{\partial}{\partial w_i}Q_{\ell m}^{X*}[\tilde{a}^{(a)},\tilde{a}^{(b)},\tilde{a}^{(c)}]B_\ell^XS_{\ell}^{-1,XY}B_\ell^Y\frac{\partial}{\partial w_j}Q_{\ell m}^Y[\tilde{a}^{(d)},\tilde{a}^{(e)}\tilde{a}^{(f)}]\bigg\rangle^*_a,
\eeq
for signal-plus-noise spectra $S_\ell$ and $\tilde{a}_{\ell m}^{X} \equiv B_\ell^X\sum_Y S_{\ell}^{-1,XY}B_\ell^Ya_{\ell m}^Y$. Due to our idealized assumptions, $\F_{ij}$ can be computed as a direct polarization/harmonic-space product, obviating the need to transform each bin of $\partial_{w_i}Q$ to map-space -- this significantly reduces computational cost. The $Q$ derivative is given by:
\beq
    \frac{\partial}{\partial w_i}Q_{\ell m}^X[x,y,z] = \sum_{\ell_im_iX_i}\frac{\partial}{\partial w_i}\frac{\partial }{\partial\tau}T^{\ell\ell_2\ell_3\ell_4,XX_2X_3X_4}_{mm_2m_3m_4}x_{\ell_2m_2}^{X_2*}y_{\ell_2m_2}^{X_2*}z_{\ell_3m_3}^{X_3*}
\eeq
for some template with amplitude $\tau$ (cf.\,\ref{eq: Q-def-harmonic}). As an example, the $\taunl$ derivative takes the form
\beq
    \frac{\partial}{\partial w_i}Q^X_{\ell m,\taunl}[x,y,z] &=& r_i^2\delta r_i\,\left[p_{\ell}^{X}(r_i)\int d\hr\,Y_{\ell m}^*(\hr)Q[x](\hr,r_i)+q_{\ell}^{X}(r_i)\int d\hr\,Y_{\ell m}^*(\hr)P[x](\hr,r_i)\right]\\\nonumber
    &&\,\times\,\left(\sum_{LM}Y_{LM}(\hr)\sum_{j=1}^{N_{\rm opt}} w_jr_j^2\delta r_j\,F_L(r_i,r_j)\int d\hr'\,P[y](\hr',r_j)Q[z](\hr',r_j)Y_{LM}^*(\hr')\right)+\text{11 perms.},
\eeq
using \eqref{eq: Q-tauNL}, noting that we sum only over $j$. Here, we have invoked symmetry of the Fisher matrix (upon realization averaging) to apply the derivative only to the first weight in $Q_{\ell m}^X$; this is trivial to compute alongside the full $Q_{\ell m}^X$ map. Using outer products, this can be used to compute both the unoptimized $N_{s}\times N_{s}$ matrix $\F_{ij}$ and any necessary derivatives. 
%is then computed as an outer product of the two derivatives (considering all sampling points, \textit{i.e.}\ setting $w=1$ and $N_s=N_{\rm opt}$). 
When optimizing a number of templates in turn (for a joint analysis of $\tau_{\rm NL}^{0,\rm even}$ and $\tau_{\rm NL}^{1,\rm even}$ for instance), we compute all $\F_{ij}$ matrices initially to avoid recomputation. We further note that the optimization can be further expedited by using only low-resolution maps (down to \textsc{HEALPix} $N_{\rm side}\approx \ell_{\rm max}$ \citep{Gorski:2004by}) since we do not include a mask.

\section{Summary \& Next Steps}\label{sec: discussion}
\noindent To date, most analyses of primordial non-Gaussianity have focused on the bispectrum. Are there interesting signatures hidden in higher-point functions? In this work, we have begun a detailed search for primordial four-point physics using the observed CMB trispectrum. In particular, we have scoured the cosmology and high-energy physics literature to find well-motivated inflationary models that could source detectable four-point signatures. A broad conclusion is that many sources of non-Gaussianity exists, such as single-field self-interactions, spinning massive particle exchange, solid inflation, gauge fields, and beyond. Building on many previous works, we have outlined a set of primordial trispectrum templates whose amplitudes can be constrained in order to probe many of the above models. Each is either a `contact' or an `exchange' estimator, with the latter depending on some exchange momentum. To derive efficient estimators, we require these templates to be \textit{separable}; in some cases (e.g., local-type non-Gaussianity) this is manifest, whilst others require are separable only with an additional time integral (the EFT of inflation templates), careful treatment of angular factors (the direction-dependent templates) or by restricting to collapsed limits (the collider templates). 

Most of this work has been devoted to deriving optimal direct estimators for the trispectrum amplitudes, building on earlier works including \citep{Sekiguchi:2013hza,2015arXiv150200635S}. By construction, these have a number of useful features:
\begin{itemize}
    \item \textbf{Unbiased}: Due to our application-dependent normalization matrix, $\F$, the estimators are not biased by masking, beams, inpainting or leakage effects, regardless of how the data is weighted. 
    %This is due to our application-dependent normalization matrix, $\F$, which we demonstrate can be efficiently computed using Monte Carlo summation, spherical harmonic transforms and simple optimization algorithms.
    \item \textbf{Non-Gaussian}: We subtract off the Gaussian contributions to the estimators using a `realization-dependent' approach similar to CMB lensing estimators \citep{Namikawa:2012pe}, using high-resolution simulations. This reduces any errors from an incorrectly assumed fiducial cosmology to second order in $C_\ell^{\rm true}-C_\ell^{\rm assumed}$. In addition, we demonstrate that our estimators match the standard `local anisotropy' forms in the relevant limits.
    \item \textbf{Decorrelated}: By performing a joint analysis of multiple templates, we can fully account for the correlations between them. This can also be used to remove contamination from late-time effects such as CMB lensing and unresolved point sources.
    \item \textbf{Optimal}: An important input to the estimators is the choice of weighting scheme, $\Si$, applied to the data (which could include inpainting, $\ell$-space filtering and beyond). As $\Si$ approaches the optimal solution $\mathsf{P}^\dag \mathsf{C}^{-1}$ (for pointing matrix $\mathsf{P}$ and pixel covariance $\mathsf{C}$), the estimators become minimum-variance, with Gaussian covariance given by $\F^{-1}$. By the Cram\'er-Rao theorem, they are thus optimal.
    \item \textbf{Efficient}: Using spherical harmonic transforms and Monte Carlo summation, our estimators can be implemented in $\mathcal{O}(N_{\rm pix}\log N_{\rm pix})$ time, which is a \textit{huge} improvement over the na\"ive quartic scaling. We also implement optimization procedures, which reduce the number of terms in the estimator by at least an order of magnitude, greatly expediting computation.
\end{itemize}

These estimators represent a significant advance compared to those used previously \citep[e.g.,][]{Planck:2019kim}, including (a) a much broader range of templates (rather than just local and EFT of inflation shapes), (b) the inclusion of polarization in all templates, (c) efficient optimization schemes for both contact and exchange templates, (d) full accounting for correlations between templates and mask-induced biases, (e) direct estimation, without having to first compute modal or binned spectra, and (f) a fast and flexible public-domain implementation. This will allow a wide variety of inflationary physics to be constrained for the first time, such as the collapsed limits of the cosmological collider scenario. In the next installment of this series \papertwo, we will implement the estimators in a public code, \href{https://github.com/oliverphilcox/PolySpec}{\textsc{PolySpec}}, and carefully test each aspect of the pipeline with simulated data, numerically validating the above conclusions. %, ensuring that the above features of the estimators are true in practice. 
Finally, in \paperthree, we will apply them to legacy temperature and polarization data from the \textit{Planck} satellite in order to obtain strong constraints on the trispectrum models discussed herein and their corresponding microphysical origins. 

Whilst we have discussed many theoretical models in this work, there are many more that have not been considered in detail. In some cases, obtaining separable trispectrum templates represents a difficult technical hurdle that must be overcome before their amplitudes can be constrained -- a notable example of this is the equilateral limit of the cosmological collider. For others, including isocurvature, tachyonic states and partially massless fields, there are no such limitations, and it would be straightforward to adapt the estimators presented herein to these models. This is an exciting avenue for future work and will be necessary to fully explore the quartic inflationary Universe.

\acknowledgments
{\small
\noindent We thank Giovanni Cabass, William Coulton, Adriaan Duivenvoorden, Sam Goldstein, Colin Hill, and Maresuke Shiraishi for insightful discussions, as well as Nicola Bartolo for comments on the draft. We are particularly indebted to Kendrick Smith, Matias Zaldarriaga and Leonardo Senatore for writing the sacred texts \citep{2011MNRAS.417....2S,2015arXiv150200635S}. \resub{Finally, we thank the anonymous referee for their careful reading of the manuscript and useful comments.}\begingroup\hypersetup{hidelinks} OHEP is a Junior Fellow of the Simons Society of Fellows, and thanks the \href{https://www.flickr.com/photos/198816819@N07/54238199131/}{Dallai Llama} for spiritual guidance.\endgroup OHEP would also like to thank the Center for Computational Astrophysics for their hospitality across the multiple years this set of papers took to write. The computations in this work were run at facilities supported by the Scientific Computing Core at the Flatiron Institute, a division of the Simons Foundation.
}

\appendix
\section{Direction-Dependent Trispectra}\label{app: direc-simp}
\noindent In this appendix, we derive the relation between the parity-odd trispectrum template and the general direction-dependent form. We start from the correlation function definition, written in terms of $\tau_{\rm NL}^{N, \rm odd}$ and $\tau_{\rm NL}^{n_1n_3n}$:
\beq
    \av{\zeta(\vk_1)\zeta(\vk_2)\zeta(\vk_3)\zeta(\vk_4)}'_c &\supset& \sum_{N\geq 0}\tau_{\rm NL}^{N, \rm odd}\Phi_N(\hk_1,\hk_3,\hK)P_\zeta(k_1)P_\zeta(k_3)P_\zeta(K)+\text{23 perms.}\\\nonumber
    &=& \sum_{n_1n_3n}\tau_{\rm NL}^{n_1n_3n}\Psi_{n_1n_3n}(\hk_1,\hk_3,\hK)P_\zeta(k_1)P_\zeta(k_3)P_\zeta(K)+\text{23 perms.}
\eeq
where $\Phi$ and $\Psi$ are the relevant angular basis functions:
\beq
    \Phi_N(\hk_1,\hk_3,\hK)&\equiv&-\frac{i}{6}\left[\mathcal{L}_N(\hk_1\cdot\hk_3)+(-1)^N\mathcal{L}_N(\hk_1\cdot\hK)+\mathcal{L}_N(\hk_3\cdot\hK)\right](\hk_1\times\hk_3\cdot\hK)\\\nonumber
    \Psi_{n_1n_3n}(\hk_1,\hk_3,\hK) &\equiv&\frac{1}{2}\sum_{m_1m_3m}\tj{n_1}{n_3}{n}{m_1}{m_3}{m}Y_{n_1m_1}(\hk_1)Y_{n_3m_3}(\hk_3)Y_{nm}(\hK),
\eeq
and we wish to obtain $\tau_{\rm NL}^{n_1n_3n}$ as a function of $\tau_{\rm NL}^{N,\rm odd}$. Noting the relations
\beq
    i(\hk_1\times\hk_3\cdot\hK) &=& \frac{\sqrt{2}}{3}(4\pi)^{3/2}\sum_{M_1'M_3'M'}\tj{1}{1}{1}{M_1'}{M_3'}{M'}Y_{1M_1'}(\hk_1)Y_{1M_3'}(\hk_3)Y_{1M'}(\hK)\\\nonumber
    \mathcal{L}_N(\hk\cdot\hk') &=& \frac{4\pi}{2N+1}\sum_{\mu}(-1)^{\mu}Y_{N\mu}(\hk)Y_{N-\mu}(\hk'),
\eeq
\citep[cf.][]{Shiraishi:2016mok}, we can expand $\Phi_N$ as
\beq
    \Phi_N(\hk_1,\hk_3,\hK)&=&-\frac{\sqrt{2}}{18}\frac{1}{2N+1}(4\pi)^{5/2}\sum_{M_1'M_3'M'\mu}(-1)^\mu \tj{1}{1}{1}{M_1'}{M_3'}{M'}Y_{1M_1'}(\hk_1)Y_{1M_3'}(\hk_3)Y_{1M'}(\hK)\\\nonumber
    &&\,\times\,\left[Y_{N\mu}(\hk_1)Y_{N-\mu}(\hk_3)+(-1)^NY_{N\mu}(\hk_1)Y_{N-\mu}(\hK)+Y_{N\mu}(\hk_3)Y_{N-\mu}(\hK)\right].
\eeq
The product of two spherical harmonics can be expressed in terms of the Gaunt factor:
\beq
    Y_{nm}(\hk)Y_{n'm'}(\hk) = \sum_{n''m''}(-1)^{m''}Y_{n''m''}(\hk)\G^{nn'n''}_{mm'-m''}
\eeq
yielding
\beq
    \Phi_N(\hk_1,\hk_3,\hK)&=&-\frac{\sqrt{2}}{18}\frac{1}{2N+1}(4\pi)^{5/2}\sum_{M_1'M_3'M'\mu}\sum_{n_1n_3nm_1m_3m}\tj{1}{1}{1}{M_1'}{M_3'}{M'}Y_{n_1m_1}(\hk_1)Y_{n_3m_3}(\hk_3)Y_{nm}(\hK)\\\nonumber
    &&\,\times\,\bigg[\G^{1n_1N}_{M_1'-m_1\mu}\G^{Nn_31}_{-\mu-m_3M_3'}(-1)^{m_1+m_3+\mu}\delta^{\rm K}_{n1}\delta^{\rm K}_{mM'}\\\nonumber
    &&\qquad\,+\,\G^{1n_1N}_{M_1'-m_1\mu}\G^{Nn1}_{-\mu -mM'}(-1)^N(-1)^{m_1+m+\mu}\delta^{\rm K}_{n_31}\delta^{\rm K}_{m_3M_3'}\\\nonumber
    &&\qquad\,+\,\G^{1n_3N}_{M_3'-m_3\mu}\G^{Nn1}_{-\mu-m M'}(-1)^{m_3+m+\mu}\delta^{\rm K}_{n_11}\delta^{\rm K}_{m_1M_1'}\bigg].
\eeq
Next, we note that the $3j$ symbols can be simplified via the Wigner $6j$ definition:
\beq
    \sum_{M_1'M_3'\mu}(-1)^{\mu+m_1+m_3+N}\tj{1}{1}{1}{M_1'}{M_3'}{m}\tj{1}{n_1}{N}{M_1'}{-m_1}{\mu}\tj{N}{n_3}{1}{-\mu}{-m_3}{M_3'}%\\\nonumber
    %&=& -\sum_{M_1'M_3'\mu}(-1)^{\mu+M_1'+M_3'}\tj{1}{1}{1}{-M_3'}{-m}{-M_1'}\tj{1}{n_1}{N}{M_1'}{-m_1}{\mu}\tj{N}{n_3}{1}{-\mu}{-m_3}{M_3'} \\\nonumber
    &=&-\begin{Bmatrix} 1 & n_1 & n_3\\ N & 1 & 1\end{Bmatrix}\tj{n_1}{n_3}{1}{m_1}{m_3}{m}\nonumber\\
    \sum_{M_1'M'\mu}(-1)^{m_1+m+\mu+N+1+n_1+n}\tj{1}{1}{1}{M_1'}{m_3}{M'}\tj{1}{n_1}{N}{M_1'}{-m_1}{\mu}\tj{N}{n}{1}{\mu}{-m}{M'} &=& %\nonumber\\
    %\sum_{M_1'M_3'\mu}(-1)^{M_1'+M'+\mu}\tj{1}{1}{1}{-M'}{-M_3'}{-M_1'}\tj{1}{n_1}{N}{M_1'}{-m_1}{\mu}\tj{N}{n}{1}{\mu}{-m}{M'} &=& 
    \begin{Bmatrix}1 & n_1 & n\\ N & 1 & 1\end{Bmatrix}\tj{n_1}{1}{n}{m_1}{m_3}{m}\\\nonumber
    \sum_{M_3'M'\mu} (-1)^{m_3+m+\mu+N}\tj{1}{1}{1}{m_1}{M_3'}{M'}\tj{1}{n_3}{N}{M_3'}{-m_3}{\mu}\tj{N}{n}{1}{-\mu}{-m}{M'} &=&%\nonumber\\
    %-\sum_{M_3'M'\mu} (-1)^{M_3'+M'+\mu}\tj{1}{1}{1}{-M'}{-M_1'}{-M_3'}\tj{1}{n_3}{N}{M_3'}{-m_3}{\mu}\tj{N}{n}{1}{-\mu}{-m}{M'} &=& 
    -\begin{Bmatrix}1 & n_3 & n\\ N & 1 & 1\end{Bmatrix}\tj{1}{n_3}{n}{m_1}{m_3}{m}.
\eeq
When the dust settles, we find
\beq
    \Phi_N(\hk_1,\hk_3,\hK)&=&\frac{\sqrt{2}}{3}(4\pi)^{3/2}(-1)^N\sum_{n_1n_3n}\Psi_{n_1n_3n}(\hk_1,\hk_3,\hK)\sum_{mm'}\sqrt{(2m+1)(2m'+1)}\tjo{1}{m}{N}\tjo{1}{m'}{N}\nonumber\\
    &&\,\times\,\begin{Bmatrix} 1 & m & m'\\ N & 1 & 1\end{Bmatrix}\bigg[\delta^{\rm K}_{n_1m}\delta^{\rm K}_{n_3m'}\delta^{\rm K}_{n1}+(-1)^{N}\delta^{\rm K}_{n_1m}\delta^{\rm K}_{n_31}\delta^{\rm K}_{nm'}+\delta^{\rm K}_{n_11}\delta^{\rm K}_{n_3m}\delta^{\rm K}_{nm'}\bigg].
\eeq
This implies the relation
\beq
    \left.\tau_{\rm NL}^{n_1n_3n}\right|_{N,\rm odd}&=&\frac{\sqrt{2}}{3}(4\pi)^{3/2}(-1)^N\tau_{\rm NL}^{N,\rm odd}\sum_{m,m'=N\pm 1}\sqrt{(2m+1)(2m'+1)}\tjo{1}{m}{N}\tjo{1}{m'}{N}\begin{Bmatrix} 1 & m & m'\\ N & 1 & 1\end{Bmatrix}\nonumber\\
    &&\,\times\,\bigg[\delta^{\rm K}_{n_1m}\delta^{\rm K}_{n_3m'}\delta^{\rm K}_{n1}+(-1)^{N}\delta^{\rm K}_{n_1m}\delta^{\rm K}_{n_31}\delta^{\rm K}_{nm'}+\delta^{\rm K}_{n_11}\delta^{\rm K}_{n_3m}\delta^{\rm K}_{nm'}\bigg],
\eeq
allowing the direction-dependent coefficients to be computed from a parity-odd trispectrum specified by $\tau_{\rm NL}^{N,\rm odd}$.

\section{Angular Dependence of the Collapsed Collider Trispectrum}\label{app: angle-simplifications}
\noindent Below, we rewrite the angular factor of the collapsed spin-$s$ collider trispectrum in separable form, which allows for efficient practical implementation. Starting from \eqref{eq: theta-function}, we can rewrite the associated Legendre polynomials in terms of spherical harmonics with respect to the axis $\hK$:
\beq\label{eq: angle-simp1}
    \Theta(\hk_1,\hk_3;\hK,\mu_s) &=& \frac{4\pi}{2s+1}\sum_{\lambda}Y_{s\lambda}(\hk_1;\hK)Y^*_{s\lambda}(\hk_3;\hK)W_\lambda(s,\mu_s).
\eeq
The rotated harmonics $Y_{sm}(\hk;\hK)$ can be related to those about some arbitrary axis $\hz$ via (inverse) Wigner $D$-matrices:
\beq
    Y_{s\lambda}(\hk;\hK) &=&\sum_{\lambda'}Y_{s\lambda'}(\hk;\hz)\mathcal{D}_{\lambda'\lambda}^{(s)*}(\alpha,\beta,\gamma) = (-1)^s\sqrt{\frac{4\pi}{2s+1}}\sum_{\lambda'}Y_{s\lambda'}(\hk;\hz){}_{\lambda}Y^*_{s\lambda'}(\beta,\alpha)e^{i\lambda\gamma},
\eeq
where $(\alpha,\beta,\gamma)$ are the Euler angles of $\hK$ about $\hz$, and we have expressed the Wigner $D$-matrix in terms of a spin-weighted spherical harmonic in the second expression. Inserting into \eqref{eq: angle-simp1}, we find two spin-weighted spherical harmonics: these can be contracted via the identity
\beq
    {}_{s_1}Y_{\ell_1m_1}(\hn){}_{s_2}Y_{\ell_2m_2}(\hn) = \sum_{\ell_3m_3s_3}\sqrt{\frac{(2\ell_1+1)(2\ell_2+1)(2\ell_3+1)}{4\pi}}\tj{\ell_1}{\ell_2}{\ell_3}{m_1}{m_2}{m_3}\tj{\ell_1}{\ell_2}{\ell_3}{-s_1}{-s_2}{-s_3}{}_{s_3}Y_{\ell_3m_3}^*(\hn).
\eeq
In combination, we find 
\beq
    \Theta(\hk_1,\hk_3;\hK,\mu_s) &=& \frac{(4\pi)^{3/2}}{2s+1}\sum_{S=0}^{2s}\sqrt{2S+1}\sum_{\lambda}(-1)^{S+\lambda}W_\lambda(s,\mu_s)\tj{s}{s}{S}{\lambda}{-\lambda}{0}\\\nonumber
    &&\,\times\,\left(\sum_{\lambda_1\lambda_3\Lambda}\tj{s}{s}{S}{\lambda_1}{\lambda_3}{\Lambda}Y_{s\lambda_1}(\hk_1)Y_{s\lambda_3}(\hk_3)Y_{S\Lambda}(\hK)\right),
\eeq
dropping the $\hz$ argument and noting that ${}_0Y_{\ell m}(\beta,\alpha)$ is a spin-weighted spherical harmonic in $\hK$. 

\section{Primordial Fisher Forecasts}\label{app: forecasts}
\noindent Given two primordial trispectra, $T_\zeta$, $T'_\zeta$, we define the inner product following the theoretical (primordial) trispectrum Fisher matrix outlined in \citep{2012PhRvD..86f3511F,Regan:2010cn}:
\beq
    \langle T_\zeta |T_\zeta' \rangle = \int_{\mathcal{V}_T}dk_1dk_2dk_3dk_4dsdt\,\frac{k_1k_2k_3k_4st}{\sqrt{g_1}P_\zeta(k_1)P_\zeta(k_2)P_\zeta(k_3)P_\zeta(k_4)}T_\zeta(k_1,k_2,k_3,k_4,s,t)T_\zeta'(k_1,k_2,k_3,k_4,s,t)
\eeq
where $\mathcal{V}_T$ is the tetrahedral domain specified by triangle conditions, as outlined in \citep{2012PhRvD..86f3511F,Regan:2010cn,Chen:2009bc,Floss:2022grj}, $g_1=s^2t^2u^2-s^2\kappa_{23}\kappa_{14}+t^2\kappa_{12}\kappa_{34}-(k_1^2k_3^2-k_2^2k_4^2)\kappa_{12}\kappa_{34}$, $\kappa_{ij}\equiv k_i^2-k_j^2$ and $s,t,u$ are (square-roots of) Mandelstam variables. This motivates the following cosine between templates:
\beq\label{eq: primordial-cosine}
    \mathrm{cos}(T_\zeta,T_\zeta') = \frac{\langle T_\zeta|T_\zeta'\rangle}{\sqrt{\langle T_\zeta | T_\zeta\rangle\langle T_\zeta'|T_\zeta'\rangle}}.
\eeq
In the below, we will evaluate the integrals numerically (based on the code of \citep{Floss:2022grj}) using a wavenumber-range $k_i\in[2,2000]/\chi_{\rm rec}$, with the lower-limit set by the distance to last scattering, and the upper (roughly) by the \textit{Planck} noise properties. 

\begin{table}[]
    \centering
    \begin{tabular}{c|ccc}
    & $\gnldotdot$ & $\gnldotdel$ & $\gnldeldel$\\\hline
    $\gnldotdot$ & 1 & 0.95 & 0.66\\
    $\gnldotdel$ & & 1 & 0.81\\
    $\gnldeldel$ & & & 1\\
    \end{tabular}
    \vskip 8pt
    \begin{tabular}{c|cccccc}
    & $\taunl$ & $\gnldotdot$ & $\tau_{\rm NL}^{\rm light}(0,3/2)$ & $\tau_{\rm NL}^{\rm light}(0,1)$ & $\tau_{\rm NL}^{\rm light}(0,1/2)$ & $\tau_{\rm NL}^{\rm light}(0,0)$\\\hline
    $\taunl$ &1&0.00&1.00&0.84&0.06&0.00\\
    $\gnldotdot$ &&1&0.00&0.01&0.27&0.33\\
    $\tau_{\rm NL}^{\rm light}(0,3/2)$ &&&1&0.85&0.06&0.00\\
    $\tau_{\rm NL}^{\rm light}(0,1)$ &&&&1&0.24&0.05\\ $\tau_{\rm NL}^{\rm light}(0,1/2)$ &&&&&1&0.84\\
    $\tau_{\rm NL}^{\rm light}(0,0)$ &&&&&&1
    \end{tabular}
    \vskip 8pt
    \begin{tabular}{c|cccccc}
    & $\taunl$ & $\gnldotdot$ & $\tau_{\rm NL}^{\rm heavy}(0,0)$ & $\tau_{\rm NL}^{\rm heavy}(0,1)$ & $\tau_{\rm NL}^{\rm heavy}(0,2)$ & $\tau_{\rm NL}^{\rm heavy}(0,3)$\\\hline
    $\taunl$ &1&0.00&0.00&0.00&0.00&0.00\\
    $\gnldotdot$ &&1&-0.33&-0.30&-0.13&0.08\\
    $\tau_{\rm NL}^{\rm heavy}(0,0)$ &&&1&0.86&0.28&-0.30\\
    $\tau_{\rm NL}^{\rm heavy}(0,1)$ &&&&1&0.39&-0.40\\ $\tau_{\rm NL}^{\rm heavy}(0,2)$ &&&&&1&-0.21\\
    $\tau_{\rm NL}^{\rm heavy}(0,3)$ &&&&&&1
    \end{tabular}
    \caption{Correlation between various primordial trispectrum templates, estimated via the cosine defined in \eqref{eq: primordial-cosine}. The top panel shows results for the EFT of inflation templates, whilst the middle (lower) panel shows results for the light (heavy) spin-zero collider, analyzed jointly with the local and equilateral shapes. We assume a $k$-range of $[2,2000]/\chi_{\rm rec}$ in all cases and evaluate the multi-dimensional integrals using the \textsc{vegas} integrator, following \citep{Floss:2022grj}. We conclude that (a) the three EFT of inflation templates are highly correlated \citep[cf.][]{2015arXiv150200635S}, (b) the light spin-zero collider templates asymptote to $\taunl$ for $\nu_0\to 3/2$ and have non-trivial correlations with equilateral shapes as $\nu_0\to 0$, and (c) the heavy templates correlate both with the equilateral templates (unless $\mu_0$ is large) and themselves (for similar $\mu_0$).}
    \label{tab: primordial-cosines}
\end{table}

In Tab.\,\ref{tab: primordial-cosines}, we give the correlations between various primordial templates, obtained from \eqref{eq: primordial-cosine}. The first panel shows the cosine between the three EFT of inflation templates (\S\ref{subsubsec: EFTI}); as noted in \citep{2015arXiv150200635S}, $\gnldotdot$ and $\gnldotdel$ are highly correlated, thus $\gnldotdel$ can be dropped from the analysis without appreciable loss of constraining power. For the low-mass collider templates of \S\ref{subsubsec: collider} (assuming spin zero and restricting to $K\leq k_{1,3}/2$ as in \eqref{eq: light-exchange-template}), we find that collider templates with dissimilar $\nu_0$ are clearly distinguishable; furthermore, there is limited overlap with the local template except for $\nu_0 = 3/2$, matching expectations. As $\nu_0\to 0$, we find increasing overlap with equilateral configurations (due to the reduced divergences in the collapsed limit), though this is somewhat tempered by our restriction to $K\leq k_{1,3}/2$. For the oscillatory high-mass templates, we find (a) some correlation with the equilateral shape (again reduced by our restrictions on $K$), (b) negligible correlation with the local shape (as expected, due to differing collapsed limits), and (c) some correlation with close $\mu_0$ templates. Whilst we find limited correlations with the light templates (which coincide at $\mu_0=\nu_0=0$), property (c) implies that similar frequency oscillations may be tricky to distinguish in practice. These conclusions echo those obtained in `dark ages' 21cm forecasts \citep{Floss:2022grj}.

Finally, the primordial cosine can be used to assess the validity of separable approximations to the $K\lesssim k/\alpha_{\rm coll}$ restriction present in the collider templates. 
%In particular, we wish to avoid the explicit and non-separable factors of $\Theta_{\rm H}(k_{1,3}-\alpha_{\rm coll} K)$ in the integrals.
As discussed in \S\ref{subsubsec: collider}, a simple option is to enforce $k_{1,3}\geq \alpha_{\rm coll}K_{\rm coll}$ and $K\leq K_{\rm coll}$ (or more simply $k\geq \alpha_{\rm coll}K_{\rm coll}$, given that the $k_{2,4}$ leg is not divergent); this ensures that the above condition is always satisfied. Noting that the constraints on scalar templates are usually dominated by small scales, the restriction on $k$ is safe; if the templates have significant power in the collapsed regime, we expect also little loss of information from the condition on $K$. Restricting to $K_{\rm coll}=k_{\rm max}/4$ and $\alpha_{\rm coll}=1$, we find that the low-mass templates are correlated with the full forms at $60\%$ ($\nu_0 = 0$) to $99\%$ ($\nu_0 = 3/2$) or around $60\%$ for the heavy particles. This implies that (a) we can use the above approximation to perform collider analyses; (b) this leads to only minor distortions to the templates. As mentioned in \S\ref{subsubsec: collider}, an alternative approach is to simply restrict the values of $\ell$ and $L$ entering the analysis; this has approximately the same result (since $\ell\sim k\chi_{\rm rec}$), but adds survey-dependence to the constraints and can be biased by projection effects and the finite width of the last-scattering surface.

\section{Lensing Estimators}\label{app: lensing}
\noindent Here, we derive the polarized lensing estimator discussed in \S\ref{subsec: lensing-estimator}. Our starting point is the correlation between lensed and unlensed fields induced by the lensing potential $\phi$:
\beq
    \av{\delta a_{\ell_1m_1}^{X_1}[\phi]a^{X_3}_{\ell_3m_3}}_a &=& \frac{1}{2}\sum_{LM}\phi_{LM}(-1)^M\left[L(L+1)+\ell_3(\ell_3+1)-\ell_1(\ell_1+1)\right]\sqrt{\frac{(2\ell_1+1)(2\ell_3+1)(2L+1)}{4\pi}}\\\nonumber
    &&\,\times\,\tj{\ell_1}{\ell_3}{L}{m_1}{m_3}{-M}\tj{\ell_1}{\ell_3}{L}{s_{X_1}}{-s_{X_1}}{0}\left[\epsilon_{\ell_1\ell_3L}C_{\ell_3}^{X_1X_3}-\beta_{\ell_1\ell_3L}C_{\ell_3}^{\bar{X}_1X_3}\right].
\eeq
Here we have averaged over the unlensed CMB fields with power spectra $C_\ell^{XY}$ and exchanged columns of the $3j$ symbol. As shown in \citep{Hanson:2010rp}, a reduced variance estimator can be wrought by setting $C_\ell^{XY}$ equal to the \textit{lensed} CMB power spectra. Next, we note two useful identities:
\beq
    \left[L(L+1)+\ell'(\ell'+1)-\ell(\ell+1)\right]\tj{\ell}{\ell'}{L}{s}{-s}{0}&\equiv&-\sqrt{L(L+1)}\sum_{\lambda=\pm 1}\sqrt{(\ell'+\lambda s)(\ell'-\lambda s+1)}\tj{\ell}{\ell'}{L}{s}{-s+\lambda}{-\lambda}
\eeq
\beq
    \sqrt{\frac{(2\ell_1+1)(2\ell_2+1)(2\ell_3+1)}{4\pi}}\tj{\ell_1}{\ell_2}{\ell_3}{m_1}{m_2}{m_3}\tj{\ell_1}{\ell_2}{\ell_3}{-s_1}{-s_2}{-s_3}&\equiv&\int d\hn\,{}_{s_1}Y^*_{\ell_1m_1}(\hn){}_{s_2}Y^*_{\ell_2m_2}(\hn){}_{s_3}Y^*_{\ell_3m_3}(\hn)
\eeq
from \citep[Eq.\,34.3.14]{nist_dlmf}. In combination, we find
\beq\label{eq: lensing-2pt}
    \av{\delta a_{\ell_1m_1}^{X_1}[\phi]a^{X_3}_{\ell_3m_3}}_a &=& -\frac{1}{2}\sum_{LM}\sum_{\lambda=\pm1}\sqrt{L(L+1)}\phi^*_{LM}\left[\epsilon_{\ell_1\ell_3L}C_{\ell_3}^{X_1X_3}-\beta_{\ell_1\ell_3L}C_{\ell_3}^{\bar{X}_1X_3}\right]\\\nonumber
    &&\,\sqrt{(\ell_3+\lambda s_{X_1})(\ell_3-\lambda s_{X_1}+1)}\int d\hn\,{}_{-s_{X_1}}Y^*_{\ell_1 m_1}(\hn){}_{s_{X_1}-\lambda}Y^*_{\ell_3m_3}(\hn){}_{\lambda}Y^*_{LM}(\hn).
\eeq
To form the estimator numerator, we project the two-point function onto two copies of the data, which involves terms of the form
\beq
    \sum_{\ell_1\ell_3m_1m_3X_1X_3}\av{\delta a_{\ell_1m_1}^{X_1}[\phi]a^{X_3}_{\ell_3m_3}}_a[\Si \alpha]^{X_1*}_{\ell_1m_1}[\Si \gamma]^{X_3*}_{\ell_3m_3} \equiv \sum_{LM}\sqrt{L(L+1)}\Phi_{LM}[\Si\alpha,\Si\gamma]\phi^*_{LM},
\eeq
defining $\Phi_{LM}$, which is an unnormalized quadratic estimator for $\phi_{LM}$. Using this notation, the lensing estimator numerator can be written succinctly
\beq
    \widehat{\mathcal{N}}_{A_{\rm lens}}[\alpha,\beta,\gamma,\delta] &=& \frac{1}{24}\sum_{LM}L(L+1)(-1)^M\Phi_{L(-M)}[\Si\alpha,\Si\gamma]\Phi_{LM}[\Si\beta,\Si\delta]C^{\phi\phi}_L+\text{11 perms.},
\eeq
introducing a lensing amplitude $A_{\rm lens}$ with $A_{\rm lens}^{\rm fid}=1$. 

To facilitate practical implementation, we must rewrite the $\Phi_{LM}$ functions in terms of nested harmonic transforms, starting from the explicit definition
\beq
    \Phi_{LM}[x,y] &=&-\frac{1}{2}\sum_{\ell_1\ell_3m_1m_3X_1X_3}\sum_{\lambda=\pm1}\left[\epsilon_{\ell_1\ell_3L}C_{\ell_3}^{X_1X_3}-\beta_{\ell_1\ell_3L}C_{\ell_3}^{\bar{X}_1X_3}\right]\sqrt{(\ell_3+\lambda s_{X_1})(\ell_3-\lambda s_{X_1}+1)}\\\nonumber
    &&\,\int d\hn\,{}_{-s_{X_1}}Y^*_{\ell_1 m_1}(\hn){}_{s_{X_1}-\lambda}Y^*_{\ell_3m_3}(\hn){}_{\lambda}Y^*_{LM}(\hn)B_{\ell_1}^{X_1*}B_{\ell_3}^{X_3*}x^{X_1*}_{\ell_1m_1}y^{X_3*}_{\ell_3m_3}.
\eeq
This can be split into two pieces differing by a phase factor $(-1)^{\ell_1+\ell_3+L}$, which can be absorbed into the spherical harmonics via ${}_sY_{\ell m}(-\hn) = (-1)^\ell{}_{-s}Y_{\ell m}(\hn)$; this leads to
\beq\label{eq: BLM-simp}
    \Phi_{LM}[x,y] &=&-\frac{1}{4}\sum_{\ell_3m_3X_1X_3}(-1)^{s_{X_1}}\sum_{\lambda=\pm1}\sqrt{(\ell_3+\lambda s_{X_1})(\ell_3-\lambda s_{X_1}+1)}\\\nonumber
    &&\,\bigg\{\left[C_{\ell_3}^{X_1X_3}+iC_{\ell_3}^{\bar{X}_1X_3}\right]\int d\hn\,{}_{+s_{X_1}}U^{X_1}[x](\hn){}_{s_{X_1}-\lambda}Y^*_{\ell_3m_3}(\hn){}_{+\lambda}Y^*_{LM}(\hn)B_{\ell_3}^{X_3*}y^{X_3*}_{\ell_3m_3}\\\nonumber
    &&\,+\,\left[C_{\ell_3}^{X_1X_3}-iC_{\ell_3}^{\bar{X}_1X_3}\right]\int d\hn\,{}_{-s_{X_1}}U^{X_1}[x](\hn){}_{\lambda-s_{X_1}}Y^*_{\ell_3m_3}(\hn){}_{-\lambda}Y^*_{LM}(\hn)B_{\ell_3}^{X_3*}y^{X_3*}_{\ell_3m_3}\bigg\},
\eeq
where we have defined the spin-$s$ maps
\beq
    {}_sU^X[x](\hn) = \sum_{\ell m}{}_sY_{\ell m}(\hn)x^X_{\ell m},
\eeq
which satisfy the conjugation relation ${}_sU^{X*}[x](\hn) = (-1)^s{}_{-s}U^{X}[x](\hn)$.\footnote{We hereafter drop the $(-1)^s$ factor, noting that $s\in\{0,2\}$.} We can further simplify using the following gradient-filtered maps:
\beq
    {}_{s_X}V^{X}_\lambda[x](\hn) = \sum_{\ell m Z}{}_{s_X-\lambda}Y_{\ell m}(\hn)\sqrt{(\ell+\lambda s_X)(\ell-\lambda s_X+1)}\left(C_{\ell}^{XZ}-iC_{\ell}^{\bar{X}Z}\right)x^{Z}_{\ell m};
\eeq
these are given explicitly by 
\beq
    {}_0V^{T}_\lambda[x](\hn) &=& \sum_{\ell m Z}{}_{-\lambda}Y_{\ell m}(\hn)\sqrt{\ell(\ell+1)}\left[C_{\ell}^{TT}x^{T}_{\ell m}+C_{\ell}^{TE}x^{E}_{\ell m}\right]\\\nonumber
    {}_2V^{E}_\lambda[x](\hn) = i{}_2V^B_\lambda[x](\hn) &=& \sum_{\ell m Z}{}_{2-\lambda}Y_{\ell m}(\hn)\sqrt{(\ell+2\lambda)(\ell-2\lambda+1)}\left[C_{\ell}^{ET}x_{\ell m}^T+C_{\ell}^{EE}x_{\ell m}^E+iC_{\ell}^{BB}x^{B}_{\ell m}\right],
\eeq
where ${}_0V_\lambda^{T*} = (-1)^{\lambda}{}_0V_{-\lambda}^{T}$. Note that the estimator involves Wiener-filtered maps, as in \citep{Carron:2022edh}. Collecting results, we obtain the final form
\beq\label{eq: PhiLM-simp-app}
    \Phi_{LM}[x,y] &=&-\frac{1}{4}\sum_{X_1,\lambda=\pm1}\bigg\{\int d\hn\,{}_{s_{X_1}}U^{X_1}[x](\hn){}_{s_{X_1}}V_\lambda^{X_1*}[y](\hn){}_{+\lambda}Y^*_{LM}(\hn)\\\nonumber
    &&\qquad\qquad\qquad\,-\,\int d\hn\,{}_{s_{X_1}}U^{X_1*}[x](\hn){}_{s_{X_1}}V^{X_1}_\lambda[y](\hn){}_{-\lambda}Y^*_{LM}(\hn)\bigg\},
\eeq
which can be evaluated using a single pair of spin-weighted harmonic transforms. Notably, $\Phi_{LM}^* = (-1)^M\Phi_{L-M}$, \textit{i.e.}\ $\Phi_{LM}$ are the harmonic coefficients of a real field (proportional to the lensing distortion map). If one considers only temperature anisotropies, the estimators simplify considerably. We find
\beq\label{eq: PhiLM-simp-app-T}
    \Phi_{LM}[x,y] &\to&\frac{1}{2}\sum_{\lambda=\pm1}\int d\hn\,U[x](\hn)V_\lambda[y](\hn){}_{-\lambda}Y^*_{LM}(\hn),
\eeq
with 
\beq
    U[x](\hn) &=& \sum_{\ell m}Y_{\ell m}(\hn)x^T_{\ell m}, \qquad
    V_\lambda[x](\hn) = \sum_{\ell m}{}_{-\lambda}Y_{\ell m}(\hn)\sqrt{\ell(\ell+1)}C_{\ell}^{TT}x^T_{\ell m}.
\eeq
This matches \citep[Eq.\,63]{2015arXiv150200635S} and the temperature-only limit of the \textit{Planck} lensing estimators \citep{Planck:2018lbu}, up to higher-order bias corrections discussed in \S\ref{sec: estimator-comparison}.

For the Fisher matrix, we start from the polarization/harmonic-space $Q$ definition \eqref{eq: Q-def-harmonic} and insert the lensing trispectrum, which yields
\beq
    Q_{\ell m,A_{\rm lens}}^{X}[x,y,z] &=& \sum_{L'M'}\sum_{\ell_im_iX_i}\frac{\partial\av{\left[\av{\delta a_{\ell m}^{X}[\phi]a_{\ell_3m_3}^{X_3}}_a+\av{\delta a_{\ell_3m_3}^{X_3}[\phi]a_{\ell m}^{X}}_a\right]\phi^*_{L'M'}}_\phi}{\partial A_{\rm lens}}\\\nonumber
    &&\,\times\,\sqrt{L'(L'+1)}\Phi_{L'M'}[x,z]y^{X_3*}_{\ell_3m_3}\,+\,\text{5 perms.},
\eeq
where we explicitly distinguish between two types of permutations and insert our definition of $\Phi$. Inserting \eqref{eq: lensing-2pt} and simplifying as before leads to
\beq\label{eq: Qlm-lens-app}
    Q_{\ell m,A_{\rm lens}}^{X}[x,y,z] &=&\frac{1}{4}\sum_{\lambda}\int d\hn\,\bigg\{{}_{-s_X}Y_{\ell m}^*(\hn){}_{s_X}V_\lambda^{X*}[y](\hn){}_{-\lambda}W[x,z](\hn)-{}_{+s_X}Y_{\ell m}^*(\hn){}_{s_X}V_\lambda^{X}[y](\hn){}_{+\lambda}W[x,z](\hn)\bigg\}\nonumber\\
    &&\,\,+\frac{1}{4}\sum_{\lambda X_3}\sqrt{(\ell+\lambda s_{X_3})(\ell-\lambda s_{X_3}+1)}\nonumber\\
    &&\,\times\,\bigg\{\left(C_{\ell}^{X_3X}+iC_\ell^{\bar{X}_3X}\right)\int d\hn\,{}_{+s_{X_3}}U^{X_3}[y](\hn){}_{s_{X_3}-\lambda}Y_{\ell m}^*(\hn){}_{-\lambda}W[x,z](\hn)\nonumber\\
    &&\qquad\,+\,\left(C_{\ell}^{X_3X}-iC_\ell^{\bar{X}_3X}\right)\int d\hn\,{}_{-s_{X_3}}U^{X_3}[y](\hn){}_{\lambda-s_{X_3}}Y_{\ell m}^*(\hn){}_{+\lambda}W[x,z](\hn)\bigg\}\,+\,\text{5 perms.},
\eeq
defining the spin-$\lambda$ map
\beq
    {}_{\lambda}W[x,y](\hn) = \sum_{LM}{}_{\lambda}Y_{LM}(\hn)L(L+1)\Phi_{LM}[x,y]C_L^{\phi\phi},
\eeq
which satisfies ${}_{\lambda}W^* = (-1)^{\lambda}{}_{-\lambda}W$, \textit{i.e.}\ $\{{}_{+1}W,-{}_{-1}W\}$ form a spin$\pm1$ pair. Although this may look a little monstrous, it can be efficiently computed using nested spin-weighted spherical harmonic transforms, first to compute ${}_{\pm}W$, then to assemble $VW$ and $UW$ products, and finally to return to polarization/harmonic-space for $Q_{\ell m}$. Again, this simplifies in the temperature-only limit, matching \citep{2015arXiv150200635S} (though going beyond the isotropic normalization used in, for example, \citep{Carron:2022edh}); we find
\beq\label{eq: Qlm-lens-T}
    Q^T_{\ell m,A_{\rm lens}}[x,y,z] &\to&\,\,-\frac{1}{2}\sum_\lambda\bigg(\int d\hn\,Y_{\ell m}^*(\hn)V_\lambda[y](\hn){}_{\lambda}W[x,z](\hn)\\\nonumber
    &&\qquad\qquad\qquad\,-\,\sqrt{\ell(\ell+1)}C_\ell^{TT}\int d\hn\,U[y](\hn){}_{+\lambda}Y_{\ell m}^*(\hn){}_{\lambda}W[x,z](\hn)\bigg)\,+\,\text{5 perms.}
\eeq
which is significantly simpler, due to the symmmetric behavior under $\lambda=\pm 1$.

\section{Analytic Fisher Matrices}\label{app: analytic-fisher}
\subsection{Contact Trispectra}\label{app: analytic-fisher-contact}
\noindent Under ideal scenarios (unit mask, translation-invariant noise and optimal weighting), the Fisher matrix for contact trispectra can be computed analytically (following \citep{2015arXiv150200635S}, but incorporating polarization). The resulting expressions are used when implementing the optimization scheme of \S\ref{sec: optim}. As noted in \S\ref{sec: optim}, we consider each template independently, which does not lead to loss of information (though may slightly reduce the efficiency of our optimization scheme).

To begin, we write the Fisher matrix as a Gaussian expectation over the estimator numerator:
\beq
    \mathcal{F} = \av{\widehat{\mathcal{N}}[d,d,d,d]\widehat{\mathcal{N}}^*[d,d,d,d]}_{d,\rm fc},
\eeq
where `fc' indicates that we take only the fully-connected part of the correlator. Explicitly, for the local shape:
\beq
     \mathcal{F}_{\gnl} &=& \left(\frac{9}{25}\right)^2\int_0^\infty r^2dr\int_0^\infty r'^2dr'\int d\hr\,d\hr'\,\av{P^3[d](\hr,r)Q[d](\hr,r)P^3[d](\hr',r')Q[d](\hr',r')}_{d,\rm fc}\\\nonumber
     &=& \left(\frac{9}{25}\right)^2\int_0^\infty r^2dr\int_0^\infty r'^2dr'\int d\hr\,d\hr'\,\av{P[d](\hr,r)P[d](\hr',r')}_d^3\av{Q[d](\hr,r)Q[d](\hr',r')}_d+\text{23 perms.}
\eeq
Since
\beq\label{eq: av-PQ}
    \av{P[d](\hr,r)[d]Q(\hr',r')}_d = \sum_{\ell XX'}\frac{2\ell+1}{4\pi}p_\ell^X(r)B_\ell^XS^{-1,XX'}_\ell B_\ell^{X'} q_\ell^{X'}(r')L_\ell(\mu)\equiv \zeta_{PQ}(r,r',\mu)
\eeq
for Legendre polynomial $L$ and $\mu\equiv\hr\cdot\hr'$ (and similar for $\zeta_{PP}$ and $\zeta_{QQ}$), this can be written as
\beq
     \mathcal{F}_{\gnl} &=& 6\left(\frac{36\pi}{25}\right)^2\int_0^\infty r^2dr\int_0^\infty r'^2dr'\int_{-1}^1\frac{d\mu}{2}\\\nonumber
     &&\,\times\,\bigg[\zeta^3_{PP}(r,r',\mu)\zeta_{QQ}(r,r',\mu)+3\zeta_{PQ}(r,r',\mu)\zeta_{QP}(r,r',\mu)\zeta_{PP}^2(r,r',\mu)\bigg],
\eeq
where the $\mu$ integral can be computed exactly using Gauss-Legendre quadrature with $(2\ell_{\rm max}+1)$ points. The procedure for the other local terms is analogous (though the permutation structure becomes somewhat arduous), leading to:
\beq\label{eq: analytic-fishes}
     \mathcal{F}_{g_{\rm NL}^{\rm con}} &=& 24\left(\frac{36\pi}{25}\right)^2\int_0^\infty r^2dr\int_0^\infty r'^2dr'\int_{-1}^1\frac{d\mu}{2}\zeta^4_{RR}(r,r',\mu)\\\nonumber
     \mathcal{F}_{\gnldotdot} &=& 24\left(\frac{1536\pi}{25}\right)^2\int_0^\infty r^2dr\int_{-\infty}^0d\tau\,\tau^4\int_0^\infty r'^2dr'\int_{-\infty}^0d\tau'\,\tau'^4\int_{-1}^1\frac{d\mu}{2}\zeta_{AA}^4(r,r',\tau,\tau',\mu)\\\nonumber
     \mathcal{F}_{\gnldotdel} &=& 4\left(\frac{13824\pi}{325}\right)^2\int_0^\infty r^2dr\int_{-\infty}^0d\tau\,\tau^2\int_0^\infty r'^2dr'\int_{-\infty}^0d\tau'\,\tau'^2\\\nonumber
     &&\,\times\,\int_{-1}^1\frac{d\mu}{2}\bigg\{\left[\zeta_{AA}^2\zeta_{BB}^2+4\zeta_{AA}\zeta_{AB}\zeta_{BA}\zeta_{BB}+\zeta_{AB}^2\zeta_{BA}^2\right]\\\nonumber
     &&\qquad\qquad\quad\,+\,\left[\zeta_{AA}^2\zeta_{BC}^2+4\zeta_{AA}\zeta_{AC}\zeta_{BA}\zeta_{BC}+\zeta_{AC}^2\zeta_{BA}^2\right]\,+\,\left[\zeta_{AA}^2\zeta_{CB}^2+4\zeta_{AA}\zeta_{AB}\zeta_{CA}\zeta_{CB}+\zeta_{AB}^2\zeta_{CA}^2\right]\\\nonumber
      &&\qquad\qquad\quad\,+\,\left[\tfrac{1}{2}\zeta_{AA}^2\left(\zeta_{CC^*}^2+\zeta_{CC}^2\right)+2\zeta_{AA}\zeta_{AC}\zeta_{CA}\left(\zeta_{CC}+\zeta_{CC^*}\right)+\zeta_{AC}^2\zeta_{CA}^2\right]\bigg\}\\\nonumber
      \mathcal{F}_{\gnldeldel} &=& 4\left(\frac{41472\pi}{2575}\right)^2\int_0^\infty r^2dr\int_{-\infty}^0d\tau\,\int_0^\infty r'^2dr'\int_{-\infty}^0d\tau'\\\nonumber
     &&\,\times\,\int_{-1}^1\frac{d\mu}{2}\bigg\{6\left[\zeta_{BB}^4+\zeta_{BC}^4+\zeta_{CB}^4\right]\,+\,12\zeta_{BB}^2(\zeta_{BC}^2+\zeta_{CB}^2)+4(\zeta_{BC}^2+\zeta_{CB}^2)(\zeta_{CC^*}\zeta_{CC}+\zeta_{CC}^2+\zeta_{CC^*}^2)\\\nonumber
     &&\qquad\qquad\,+2\,\zeta_{BB}^2(\zeta_{CC^*}^2+\zeta_{CC}^2)+8\zeta_{BB}\zeta_{BC}\zeta_{CB}(\zeta_{CC*}+\zeta_{CC})+4\zeta_{BC}^2\zeta_{CB}^2+\zeta_{CC}^4+\zeta_{CC^*}^4+4\zeta_{CC^*}^2\zeta_{CC}^2\bigg\},
\eeq
where $\zeta_{RR}$, $\zeta_{AA}$, $\zeta_{BB}$, $\zeta_{AB}$ are defined analogously to \eqref{eq: av-PQ}, and we drop the $(r,r',\tau,\tau',\mu)$ arguments in the final two expressions for clarity. For the spin-$\pm1$ fields $C$, the averages take the form (building on \citep{2015arXiv150200635S})
\beq
    \av{A[d](\hr,r,\tau){}_{\pm 1}C[d](\hr',r',\tau')}_d &=& \mp\sum_{\ell XX'}\frac{2\ell+1}{4\pi}a_{\ell}^X(r,\tau)B_\ell^XS_\ell^{-1,XX'}B_\ell^{X'}c_{\ell}^{X'}(r,\tau)d^{\ell}_{0(-1)}(\theta) \equiv \pm\zeta_{AC}(r,r',\tau,\tau',\mu)\nonumber\\
    \av{{}_{\pm 1}C[d](\hr,r,\tau)A[d](\hr',r',\tau')}_d &=& \mp\sum_{\ell XX'}\frac{2\ell+1}{4\pi}c_{\ell}^X(r,\tau)B_\ell^XS_\ell^{-1,XX'}B_\ell^{X'}a_{\ell}^{X'}(r',\tau')d^\ell_{10}(\theta) \equiv \pm\zeta_{CA}(r,r',\tau,\tau',\mu)\nonumber\\
    \av{{}_{\pm1}C[d](\hr,r,\tau){}_{\pm1}C[d](\hr',r',\tau')}_d &=& \sum_{\ell XX'}\frac{2\ell+1}{4\pi}c_{\ell}^X(r,\tau)B_\ell^XS_\ell^{-1,XX'}B_\ell^{X'}c_{\ell}^{X'}(r',\tau')d^\ell_{1(-1)}(\theta) \equiv \zeta_{CC}(r,r',\tau,\tau',\mu)\nonumber\\
    \av{{}_{\pm1}C[d](\hr,r,\tau){}_{\mp1}C[d](\hr',r',\tau')}_d &=& \sum_{\ell XX'}\frac{2\ell+1}{4\pi}c_{\ell}^X(r,\tau)B_\ell^XS_\ell^{-1,XX'}B_\ell^{X'}c_{\ell}^{X'}(r',\tau')d^\ell_{11}(\theta) \equiv -\zeta_{CC^*}(r,r',\tau,\tau',\mu)
\eeq
for $\theta=\arccos\mu$ and $d_{ss'}^\ell(\theta)$ are Wigner $d$-functions. Note that these expressions are all real, which simplifies \eqref{eq: analytic-fishes} considerably. As in \citep{2015arXiv150200635S}, the Wigner $3j$ symbols can be generated recursively from
\beq
    \alpha^{\ell+1}_{ss'}(\theta)d^{\ell+1}_{ss'}(\theta)-(2\ell+1)\left[\cos\theta-\frac{ss'}{\ell(\ell+1)}\right]d^\ell_{ss'}(\theta)+\alpha^{\ell}_{ss'}(\theta)d^{\ell-1}_{ss'}(\theta) = 0
\eeq
(for $\ell\geq |s|,|s'|$) with $\alpha^\ell_{ss'} = \sqrt{(\ell^2-s^2)(\ell^2-s'^2)}/\ell$, starting from
\beq
    d_{10}^1(\theta) = \frac{1}{\sqrt{2}}\sin\theta, \qquad d_{11}^1(\theta) = \frac{1}{2}(1+\cos\theta), \qquad d_{1(-1)}^1(\theta) = \frac{1}{2}(1-\cos\theta)
\eeq
and the $\ell=s$ term (with $0\leq |s'|<s$):
\beq    
    \alpha_{ss'}^{s+1}d_{ss'}^{s+1}(\theta)=(2s+1)\left[\cos\theta-\frac{s'}{s+1}\right]d^{s}_{ss'}(\theta).
\eeq

\subsection{Exchange Trispectra}\label{app: analytic-fisher-exchange}
\noindent One may additionally attempt to compute the Fisher matrix analytically for exchange-factorizable trispectra. Here, we will demonstrate the difficulties of this with the simplest template: $\taunl$. Starting from
\beq
    \mathcal{F}_{\taunl} = \av{\widehat{\mathcal{N}}_{\taunl}[d,d,d,d]\widehat{\mathcal{N}}^*_{\taunl}[d,d,d,d]}_{d,\rm fc},
\eeq
as before, we can insert the definition of $\widehat{\mathcal{N}}_{\taunl}$, finding
\beq
     \mathcal{F}_{\taunl} &=& \frac{1}{4}\left[\prod_{i=1}^4\int r_i^2dr_id\hr_i\right]\sum_{LL'MM'}F_L(r_1,r_2)F_{L'}(r_3,r_4)Y_{LM}^*(\hr_1)Y_{LM}(\hr_2)Y_{L'M'}(\hr_3)Y^*_{L'M'}(\hr_4)\\\nonumber
     &&\,\quad\,\times\,\av{\prod_{i=1}^4\left[P[d](\vr_i)Q[d](\vr_i)\right]}_{d,\rm fc}.
\eeq
This contains a lot of terms, each involving contractions of the form \eqref{eq: av-PQ}. The simplest involve:
\beq
     \mathcal{F}_{\taunl} &\supset& \frac{1}{4}\left[\prod_{i=1}^4\int r_i^2dr_i\right]\sum_{L}(2L+1)F_L(r_1,r_2)F_{L}(r_3,r_4)V^{L}_{ppqq}(r_1,r_3)V^{L}_{ppqq}(r_2,r_4),
\eeq
simplifying the spherical harmonics via Wigner $3j$ algebra and defining
\beq
    U^\ell_{pq}(r,r') &=& \sum_{XX'}p_\ell^X B_\ell^XS_\ell^{-1,XX'}B_\ell^{X'}q_\ell^{X'},\\\nonumber
    V^L_{ppqq}(r,r') &=& \sum_{\ell_1\ell_2}\frac{(2\ell_1+1)(2\ell_2+1)}{4\pi}\tjo{\ell_1}{\ell_2}{L}^2U^{\ell_1}_{pp}(r,r')U^{\ell_2}_{qq}(r,r').
\eeq
This term can be computed relatively efficiently, since the radial integrations can be written as a trace, \textit{i.e.}\ $\sim\mathrm{Tr}\left(F_L\cdot V^L\cdot F_L \cdot V^L\right)$. However, a number of non-trivial cross-terms also arise, such as
\beq
     \mathcal{F}_{\tau_{\rm NL}^{\rm loc}} &\supset& \frac{1}{4}\left[\prod_{i=1}^4\int r_i^2dr_i\right]\sum_{LL'}F_L(r_1,r_2)F_{L'}(r_3,r_4)U_{pp}^{\ell_1}(r_1,r_3)U^{\ell_2}_{qq}(r_1,r_4)U_{pp}^{\ell_3}(r_2,r_4)U_{qq}^{\ell_4}(r_2,r_3)\\\nonumber
     &&\,\times\,(-1)^{L+L'}\frac{(2\ell_1+1)(2\ell_2+1)(2\ell_3+1)(2\ell_4+1)(2L+1)(2L'+1)}{(4\pi)^2}\\\nonumber
     &&\,\times\,\tjo{\ell_1}{\ell_2}{L}\tjo{\ell_3}{\ell_4}{L}\tjo{\ell_1}{\ell_4}{L'}\tjo{\ell_3}{\ell_2}{L'}\begin{Bmatrix}L & \ell_1 & \ell_2\\ L' & \ell_3 &\ell_4\end{Bmatrix}.
\eeq
Due to the presence of the Wigner $6j$ symbol (in curly parentheses), these are significantly more expensive to compute, even if one restricts to low $L,L'$ in the estimators. This can be equivalently written in terms of Legendre polynomials using \eqref{eq: av-PQ}; however, there remains a coupled `square' integral in both $r$ and $\mu$, which makes direct computation infeasible. For these reasons, we compute the exchange trispectrum Fisher matrices using Monte Carlo methods as discussed in \S\ref{sec: estimators-exchange}, even in the idealized limit.

\bibliographystyle{apsrev4-1}
\bibliography{refs}% Produces the bibliography via BibTeX.

\end{document}